\documentclass[times,twocolumn,final,authoryear]{elsarticle}

\usepackage{jasr}
\usepackage{framed,multirow}

\usepackage{amssymb}
\usepackage{latexsym}
\usepackage{adjustbox}
\usepackage{xspace}

\usepackage{url}
\usepackage{xcolor}
\definecolor{newcolor}{rgb}{.8,.349,.1}
\usepackage{soul}

\usepackage[citebordercolor=white]{hyperref}

\journal{Advances in Space Research}

\begin{document}

\verso{B. Bhat}

\begin{frontmatter}

\title{Radiative transfer modeling of the observed line profiles in G31.41+0.31}%

\author[1]{Bratati \snm{Bhat}}

\author[1,2]{Prasanta \snm{Gorai}}
\author[1]{Suman Kumar \snm{Mondal}}
\author[1]{Sandip K. \snm{Chakrabarti}}

\author[1]{Ankan \snm{Das}\corref{cor1}}
\ead{ankan.das@gmail.com}
\cortext[cor1]{Corresponding author: 
  Tel.: +91-9830469158;}

\address[1]{Indian Centre for Space Physics, 43 Chalantika, Garia Station Road, Kolkata 700084, India}
\address[2]{Department of Space, Earth \& Environment, Chalmers University of Technology, SE-412 96 Gothenburg, Sweden}

\received{}
\finalform{}
\accepted{}
\availableonline{}
\communicated{}

\begin{abstract}
An inverse P-Cygni profile of H$^{13}$CO$^+$ (1 $\rightarrow$ 0) in G31.41+0.31 was recently observed, which indicates the presence of an infalling gas envelope. Also, an outflow tracer, SiO, was observed. Here, exclusive radiative transfer modelings have been implemented to generate synthetic spectra of some key species (H$^{13}$CO$^+$, HCN, SiO, NH$_3$, CH$_3$CN, CH$_3$OH, CH$_3$SH, and CH$_3$NCO) and extract
the physical features to infer the excitation conditions of the surroundings where they observed. The gas envelope is assumed to be accreting in a spherically symmetric system towards the
central hot core region. Our principal intention was to reproduce the observed line profiles toward G31.41+0.31
and extract various physical parameters. The LTE calculation with CASSIS and non-LTE analysis with the RATRAN radiative transfer codes are considered for the modeling purpose. The best-fitted line
parameters are derived, which represents the prevailing physical condition of the gas envelope. Our results suggest that an
infalling gas could explain the observed line profiles of all the species mentioned above except SiO. An additional outflow component is required to confer the SiO line profile. Additionally, an astrochemical model is implemented to explain the observed abundances of various species in this source.
\end{abstract}


\end{frontmatter}



\section{Introduction}
Ultra-Compact HII (UC HII) regions are often used as a tracer of high-mass ($>$8 $M_\odot$) star-forming regions (HMSFRs). However, the evolutionary tracks of high-mass star formations are poorly constrained. Specifically, the effects of fragmentation and formation of multiple systems in HMSFRs and mass accumulation processes toward the cores are not yet well-established \citep{bosc19,pala14, beut18}. Generally, Hot Molecular Cores (HMCs) are characterized by high gas temperature ($\geq 100$ K), high density ($\geq 10^6$ cm$^{-3}$) and small size ($\leq 0.1$ pc), which show numerous simple and complex organic molecular emissions \citep{kur00}. 
The hot molecular phase is believed to occur in the early stage of high-mass star formation. 
G31.41+0.31 (hereafter G31) is an exciting HMC due to its kinematics and chemical diversity. The dust continuum emission of G31 revealed that it has two components: the Main core and NE core \citep{belt18}. It looks featureless and homogeneous, suggesting no hint of fragmentation in the Main core \citep{belt18}. But the dull nature of the central core may arise from the significant opacity of the dust emission, which prevents a precise observation of the central core. The presence of the red-shifted absorption, rotational spin-up, and the existence of two embedded free-free emitting sources at the center suggest that this core might be undergoing fragmentation with infall and differential rotation due to conservation of angular momentum.

An UC HII region is located in a $\sim5^{''}$ distance from the main-core of G31 \citep{ces10}. This HMC is situated at a distance of $7.9$ kpc away from the Sun. The luminosity of this source is $\rm{\sim 3\times10^{5}} \ $\(L_\odot\) \citep{chu90}. The new parallax measurements suggest that it is located at a distance of $3.7$ kpc and luminosity would be $\rm{\sim 4\times10^{4}} \ $\(L_\odot\) \citep{reid19,belt19}. The systematic velocity 
($\rm{V_{LSR}}$) of this source is $\sim$ $97$ km s$^{-1}$ \citep{ces10,riv17}.

\cite{Gira09} observed inverse P-Cygni profile of low excitation line (C$^{34}$S, J=7-6) in G31 and discussed the infalling matter in it.
\cite{cesa11} observed C$^{17}$O and CH$_3$CN, and also gave a hint of infall in this source. Later, \cite{belt18} observed this source with high angular and spatial resolution using Atacama Large Millimeter/submillimeter Array (ALMA) and discussed the associated kinematics. They identified
J=5-4 transitions of SiO and discussed the possible presence of molecular outflow and their directions in this source.
They also observed the inverse P-Cygni profile in H$_2$CO and CH$_3$CN lines with various upper state energy (E$_{up}$) and found the accelerating infall
nature and the rotation present in the HMC. Recently, \cite{gora21} reported different transitions of SiO, HCN, and H$^{13}$CO$^+$ in G31, which trace outflow and infall signatures in this source.

It is not straightforward to explain the time variability of the infalling activity. The infall may proceed inside-out \citep{shu77} or outside-in \citep{fost93}. \cite{oso09} modeled the observed spectral energy distribution (SED) in detail for G31. They fitted the SED using different models and extracted the fundamental physical parameters responsible for the obtained spectral profile. They modeled the source using density and temperature profiles obtained from the singular logatropic sphere solution. Along with the
other relevant physical properties, they estimated the core mass $\sim25M_\odot$, the outer radius of the envelope $\sim30000$ AU by considering a distance of $7.9$ kpc (continuum model of \cite{van13} used an outer radius $\sim 119000$ AU). 
\cite{may14} identified the presence of infall motion in G31. This infall is identified from the spectral signatures of the red and blue asymmetries present in blue-shifted and red-shifted regions concerning the rest of the frame. Along with the infall motion in the source, \cite{may14} also reported
the inversion transitions of ammonia ((2,2), (3,3), (4,4), (5,5), (6,6)). They introduced a new signature called `central blue spot' in first-order moment maps, confirming the infalling motion. Recently, using the central blue spot signature, \cite{esta19} estimated the range of the central masses of G31 $\sim70-120M_\odot$. 

Radiative transfer code is frequently used to explain the observed line profiles. For example, \cite{wyr16} modeled the observed $\rm{HCO^+}$ line profile by using the RATRAN radiative transfer code. 
They targeted nine massive molecular clumps (region of a molecular cloud where density is high, and there reside much dust and many gas cores)
of the ATLASGAL submillimeter dust continuum survey of our Galaxy with the Stratospheric Observatory for Infrared Astronomy (SOFIA) and identified multiple $\rm{NH_3}$
transitions and some transitions of $\rm{HCO^+, HCN}$, and HNC. 
They extracted
several physical parameters and obtained the infall signature from red-shifted absorption. Absorption lines of ammonia ($\rm{NH_3}$) were identified in all nine sources, which signify the presence of infall with an infall rate between 
$3 - 10\times10^{-3} M_\odot/yr$. 
 \cite{her19}  used the one-dimensional radiative transfer code, MOLLIE \citep{keto10} to explain their massive molecular outflow in G331.512-0.103  (a compact radio source in the center of an energetic molecular outflow and an active and extreme high-mass star-forming environment in our Galaxy). In this source, \cite{her19} reported SiO with the band 7 data of ALMA.
They used the radiative transfer model to reproduce the observed SiO spectral feature and obtain the source's outflow parameters. They classified the observed emission lines into two categories: the relatively narrow component tracing
the emission from the core and the comparatively broad component that traces the massive outflow. 

Earlier, many observations were carried out to explore the chemical composition of G31. Identifying numerous complex organic molecules (COMs) in this source makes it a fascinating target for astronomers. Using the IRAM facility, \cite{belt09} identified glycolaldehyde (CH$_2$OHCHO), methanol (CH$_3$OH), and methyl formate (HCOOCH$_3$). Recently, \cite{riv17} spotted dimethyl ether ($\rm{CH_3OCH_3}$), glycolaldehyde, methyl formate, ethylene glycol ((CH$_2$OH)$_2$), and ethanol ($\rm{C_2H_5OH}$) using Green Bank Telescope (GBT), IRAM $30$ m telescope and Submillimeter Array (SMA) interferometric observations toward the main molecular core of G31. 
Subsequently, \cite{gora21} observed peptide bond containing molecule (methyl
isocyanate, CH$_3$NCO) and complex sulfur-bearing species (methanethiol, CH$_3$SH) for the first time in G31.

This paper reports a modeling attempt to explain the line profiles observed in G31, using chemical and radiative transfer codes. 
A chemical model is prepared to explain the observed abundances and radiative transfer codes to explain the observed spectral signatures. The physical properties (infall, outflow, temperature, etc.) associated with G31 are also deduced by examining the observed line profiles of H$^{13}$CO$^+$, HCN, SiO, and NH$_3$. The line profiles of some COMs such as $\rm{CH_3SH}$, $\rm{CH_3NCO}$, and $\rm{CH_3OH}$ are also modeled. The same observed spectral data that were presented in \cite{gora21} is used. For the NH$_3$, the spectral data from \cite{oso09} is used.
The paper is organized as follows: Section \ref{sec:compu-detail} 
describes the computational methodology used to obtain the physical and chemical parameters for our radiative transfer model. 
Results and discussions related to our models are presented 
in Section \ref{sec:result-discussions}. Finally, in Section \ref{sec:conclusion}, we draw our conclusions.

\section{Computational Details and Methodology}
\label{sec:compu-detail}
\subsection{Observations}
\label{sec:obs}
\begin{figure}
    \includegraphics[width=8cm]{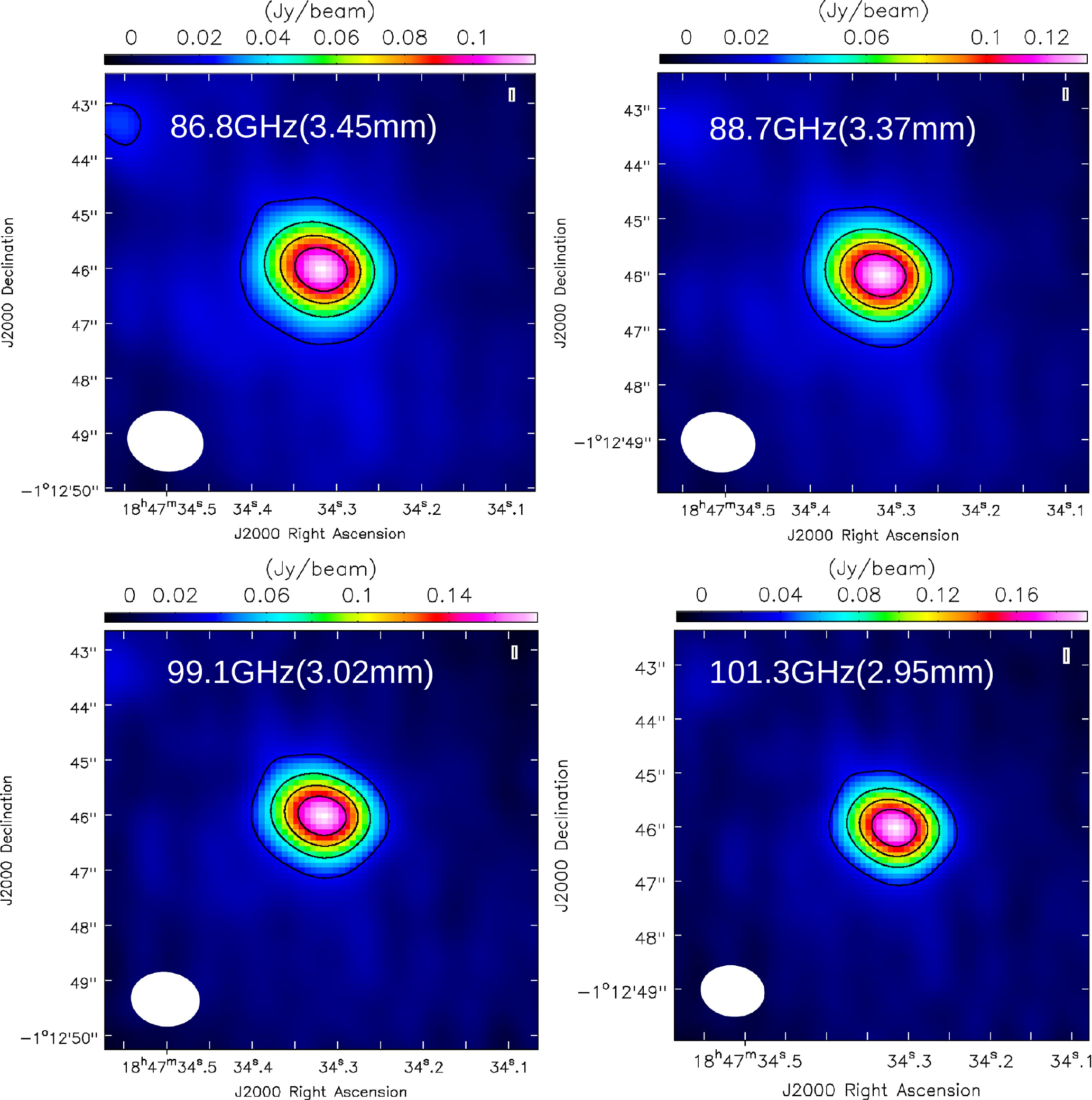}
    \caption{Continuum emission observed towards G31 with ALMA band 3 at (i) 86.8 GHz (peak intensity = 118.34 mJy/beam), (ii) 88.7 GHz (peak intensity = 133.51 mJy/beam),  (iii) 99.1 GHz (peak intensity 179.2 mJy/beam), and (iv) 101.3 GHz (peak intensity = 196.2 mJy/beam). Contour levels are at 20\%, 40\%, 60\%, and 80\% of the peak flux. The synthesized beam is shown in the lower left-hand corner of each figure.}
    \label{fig:cont}
  \end{figure}

 Here, the cycle 3 archival data of the Atacama Large millimeter/submillimeter Array (ALMA) ($\#$2015.1.01193.S) is analyzed. The dataset consists of four spectral windows (86.559-87.028 GHz, 88.479-88.947 GHz, 98.977-99.211 GHz, and 101.076-101.545 GHz). Figure \ref{fig:cont} shows the continuum emission images at these four spectral windows. The data cube has a spectral resolution of $244$ kHz ($\sim 0.84$ km s$^{-1}$ ) with a synthesized beam of  $\sim 1.19^{''} \times 0.98^{''}$ ( $9402 \times 7743$ AU considering 7.9 kpc distance and $4403 \times 3636$ AU considering 3.7 kpc distance) with a position angle 76$^\circ$. Unless otherwise stated, 3.7 kpc distance is used for the calculations. The systematic velocity of this source is
$97$ km s$^{-1}$. \cite{gora21} reported a peak intensity of dust continuum $\sim 158.4$ mJy/beam, the integrated flux $\sim 242.4$ mJy, and the RMS noise of the continuum map  $\sim 5$ mJy/beam. They obtained the molecular hydrogen column density of $\sim 1.53 \times 10^{25}$ cm$^{-2}$. 

\cite{oso09} modeled the observed spectral energy distribution and ammonia emission of the G31 hot core. They considered the hot core has an infalling envelope in an intense accreting phase. They derived the physical properties of the envelope and stellar part by fitting the observed spectral energy distribution. There are numerous observations of the G31 region. Still, most of the observations are indistinguishable from the G31 HMC and  UCHII region, whose peaks are only differed by 5$^{''}$. \cite{cesa94,cesa98} had the VLA data where the combined spectral resolution was $\sim 0^{''}.63$. \cite{oso09} used this high spatial resolution data to carry out the spectral line data analysis. \cite{oso09} also used low angular resolution data (40$^{''}$) of \cite{chu90} to test their model prediction. They obtained a best-fit when they used a central mass $\sim 25 M_{\odot}$ by considering a distance of 7.9 kpc, a mass accretion rate $\sim 3 \times 10^{-3} M _{\odot}$, and index of the power-law that describes the dust absorption coefficient ($\beta$) $\sim 1$.

\subsection{Physical condition}
\label{sec:phys}
\begin{figure}
\centering
\includegraphics[height=6cm]{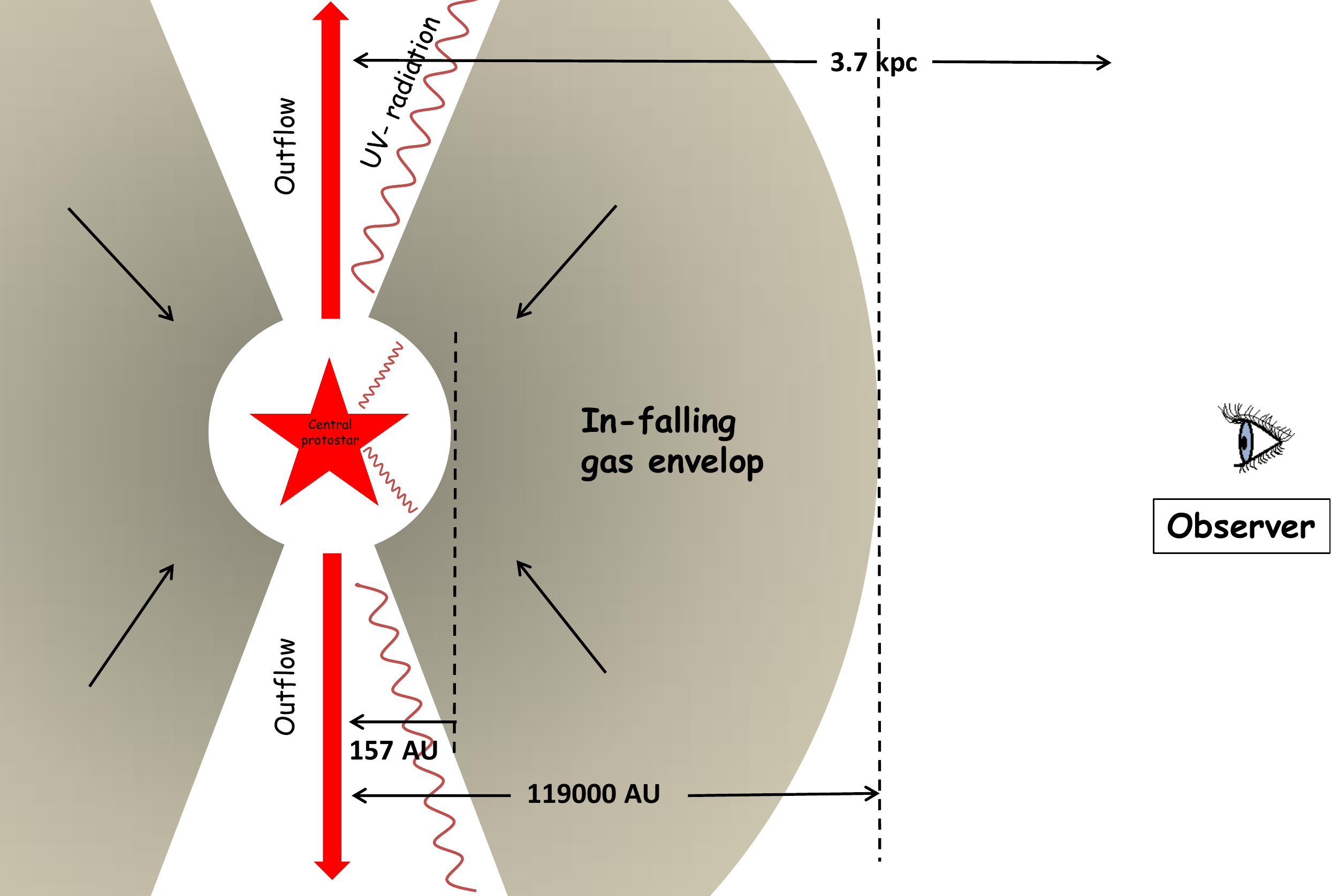}
  \caption{Cartoon diagram to represent the modeled region of the cloud.}
  \label{fig:cartoon}
  \end{figure}

\begin{figure}
\centering
\includegraphics[height=8cm]{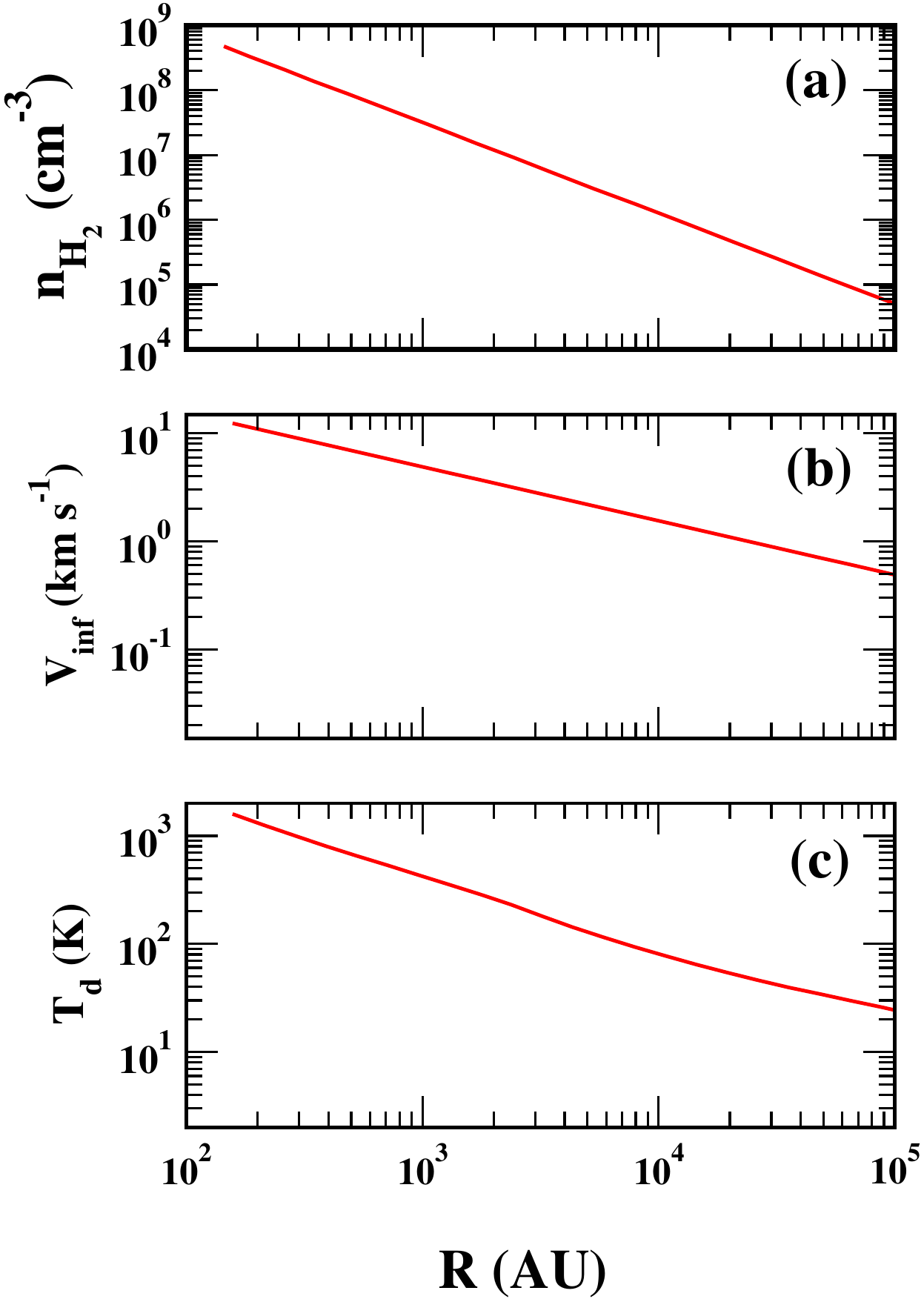}
  \caption{Spatial distribution of the physical parameters (a. H$_2$ density, b. infall velocity, and c. dust temperature) considered here are shown. The density and temperature variation are taken from \protect\cite{van13} and infall velocity variation is derived from the Equation \ref{eqn:infall} by using $v_{1000}= 4.9$ km s$^{-1}$}.
  \label{fig:physical_profiles}
\end{figure}

The infalling region of the cloud envelope is divided into $23$ grids. \cite{van13} considered the envelope beyond $156$ AU. The reason behind this choice is that \cite{oso09} obtained a dust temperature $\sim 1200$ K around this radius. It is the sublimation temperature of the dust. So our model has considered the infalling envelope region ranging between 156 AU - 119000 AU ($7.56 \times 10^{-4} - 5.769 \times 10^{-1}$ pc).
A cartoon diagram of the modeled region is shown in Figure \ref{fig:cartoon}. It depicts an infalling gas envelope towards the central protostar. An outflow component is also present in the cloud. \cite{mars10} identified a cold foreground cloud in the line of sight of G31. But these foreground clouds are at entirely different velocities. Thus, any foreground cloud component is not considered to be associated with this source.
A logarithmic spacing is applied to select the grid spacing. 
The density distribution of the envelope is defined by using a density exponent ($p$)
$\sim 1.40$ obtained with the dust continuum model of \cite{van13}. 
The spatial distribution of the temperature is also taken from \cite{van13}. The kinetic gas temperature and dust temperature are considered equal, assuming they are well coupled. 
The spatial distribution of the physical parameters is shown in Figure \ref{fig:physical_profiles}.

 The infalling matter is in the free-fall state \citep{belt18}, which is estimated as:
\begin{equation}
v_{inf}(r)=v_{1000}\left(\frac{r}{1000 \ AU}\right)^{-0.5},
\label{eqn:infall}
\end{equation}
where r represents the radius and $v_{1000}$ is the infall velocity at $1000$ AU.
\cite{oso09} fitted the observed spectral energy distribution towards this hot core. As the resolution of most of the observations were $>5^{''}$, and an angular distance separates the central emitting region of G31 and nearby HII region $\sim 5^{''}$, the emission had contaminated with
the emission of the HII region.  Thus, \cite{oso09} considered
the SED fitting by considering many data points from archival data and predicted the upper limit of different physical parameters from their fitting. Unless otherwise stated, $v_{1000} \sim 4.9$ km s$^{-1}$ is used from \cite{oso09}.  The rotational motion of the infalling gas envelope is not considered here for simplicity, which might cause some discrepancies in the result.

\subsection{Chemical Modeling}
\label{sec:chemical}
Detailed chemical modeling has been carried out to study the abundances of the species observed in G31. 
Our Chemical Model for Molecular Cloud (CMMC) \citep{das15a,das15b,das16,gor17a,gor17b,sil18,gora20,sil21,das21} is considered for this modeling. 
The gas-phase pathways are mainly adopted from the UMIST database \citep{mce13}, whereas the ice phase pathways and binding energies (BEs) of the
surface species are considered from the KIDA database \citep{rua16} and \cite{das18,sil17}.
 The time evolution of the physical parameters (H$_2$ density and temperature) are considered in three significant steps: isothermal collapsing phase,
warm-up phase, and post-warmup phase. This type of straightforward model is best suited to study the chemical evolution of hot cores \citep{gora20,sil21}.

\begin{figure}
\hskip -1cm
\includegraphics[height=6.8cm]{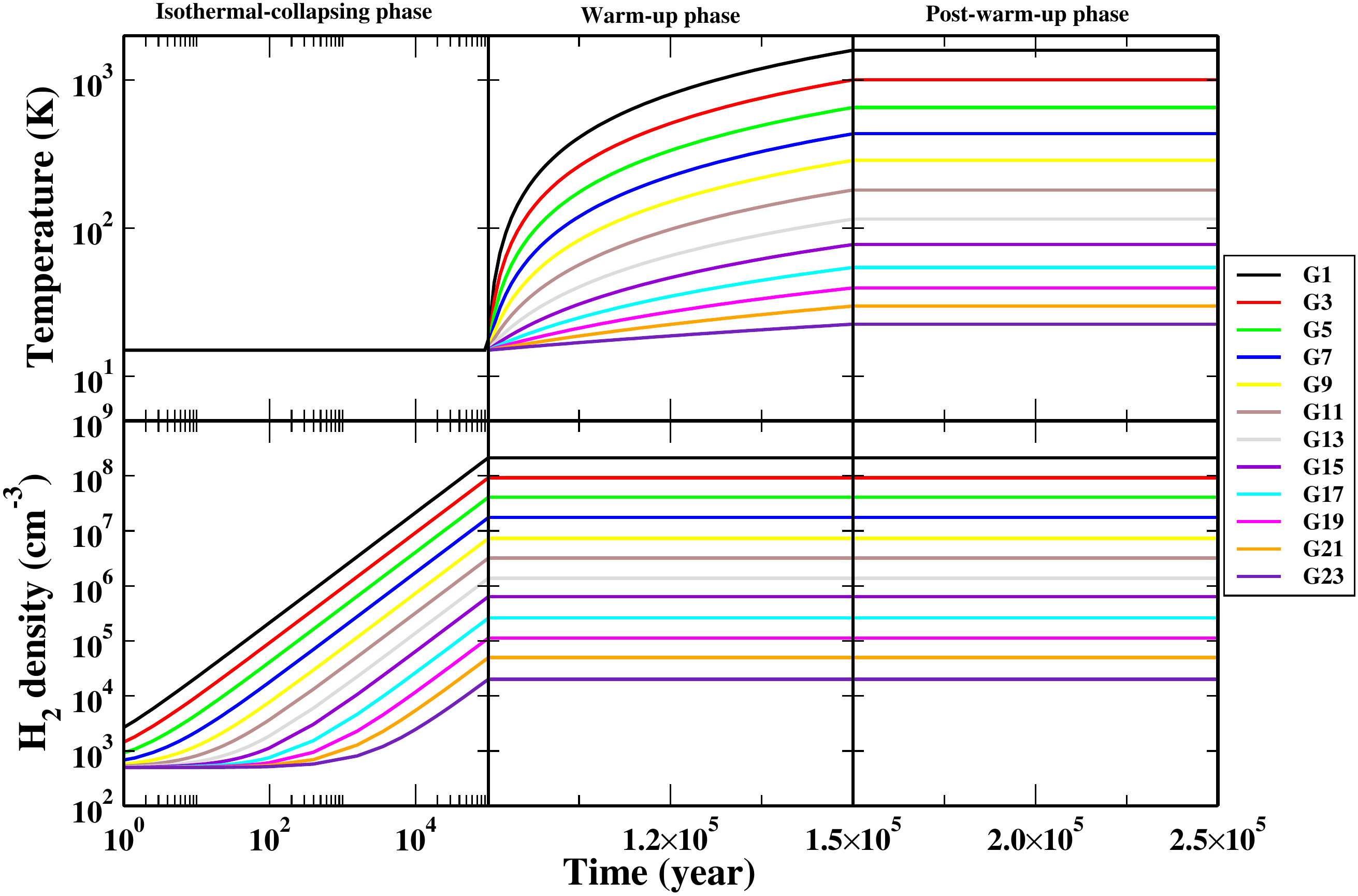}
\caption{Physical evolution considered in this simulation is shown. The cloud envelope is divided the cloud into $23$ spherical shells. The density and temperature evolution in the alternative shells (G1-G23) are shown for the representation.
The entire simulation time
scale is divided into three parts: The isothermal (gas and grain at $15$ K) and the collapsing phase, which extends up to $10^5$ years. The second
phase corresponds to the warm-up time scale, whose span is for $5 \times 10^4$ years, and finally, the third phase
corresponds to the post-warmup phase, whose span is for $10^5$ years. .
}
\label{fig:physical_model}
\end{figure}

\begin{figure*}
\centering
\includegraphics[height=16cm,width=10cm,angle=270]{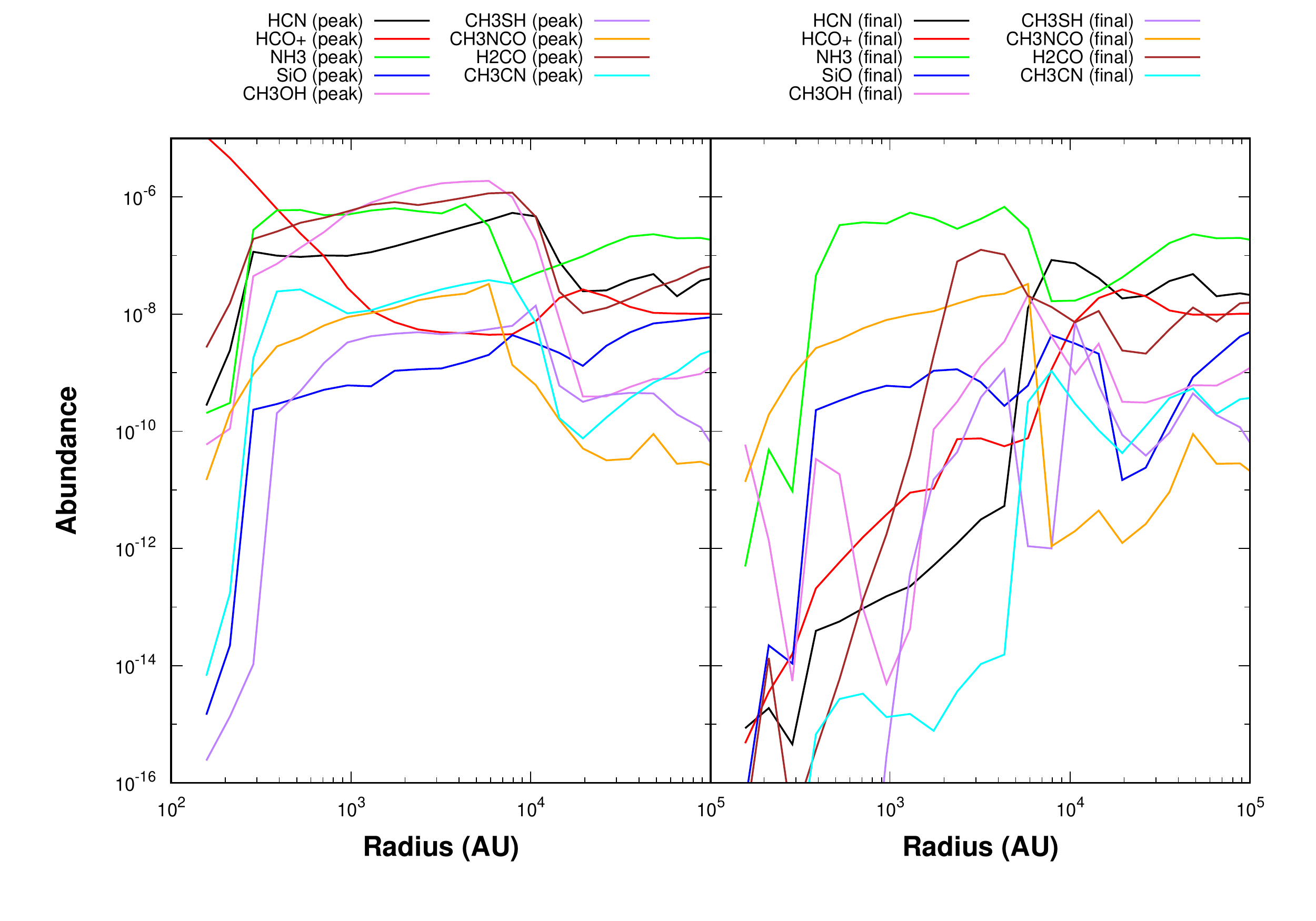}
\caption{The spatial distribution of the peak abundances (left panel) and final abundances (right panel) of some key interstellar species are shown. The peak abundances are taken beyond $10^5$ years, whereas the final abundances are taken at the end of the simulation time scale $\sim 2.5 \times 10^5$ years.}
\label{fig:abundance}
\end{figure*}

As discussed in Section \ref{sec:phys}, 
the gas cloud envelope is divided into $23$ spherical shells.
In the first phase (i.e., isothermal phase), all the shells are kept at a constant temperature $\sim 15$ K.
During this time interval ($\sim 10^5$ years), the $\rm{H_2}$ number density of the cloud is allowed to evolve from $ 5 \times 10^2$ cm$^{-3}$ to a final value. The shorter collapsing time scale is used to represent the high mass star formation scenario.
The final density of each shell is taken from the results presented in Figure \ref{fig:physical_profiles}.
For example, in this phase, the outermost
shell's density is allowed to evolve up to $4 \times 10^4$ cm$^{-3}$, and in the innermost shell, it is up to $4 \times 10^8$ cm$^{-3}$. 
All the shells are allowed to evolve from their initial temperature
to the final temperature within $5 \times 10^4$ years in the warm-up stage. This fast warm-up time scale is chosen to represent the high mass star formation. The final temperature of each shell is defined from the results presented in Figure \ref{fig:physical_profiles}. For example, the outermost
shell's temperature is assumed to evolve up to $23$ K, and in the innermost shell, it is up to $1593$ K. 
After the completion of the warm-up phase, the post-warmup phase starts. All the shells are allowed to remain at the same density and temperature of their earlier stage. A time scale of $10^5$ years is considered for the post-warmup phase. 
Thus, our total simulation time scale (i.e., collapsing time + warmup time + post-warmup time) is $\sim 2.5 \times 10^5$ years. 
The density and temperature variation of our three-phase model is shown in Figure \ref{fig:physical_model}.
The obtained abundances of various chemical species concerning H$_2$ molecules are shown in 
Figure \ref{fig:abundance}. 
 The left panel of Figure \ref{fig:abundance} depicts the peak abundance obtained beyond the collapsing time scale, whereas
the right panel displays the abundances obtained at the end of the simulation time scale ($\sim 2.5 \times 10^5$ years). 
For a better understanding, in Table \ref{table:abundances}, the final and peak abundances of these species are noted at different radius.
It is interesting to note that
deep inside the cloud, where the temperature is higher, abundances of the complex organic molecules
 (especially CH$_3$SH and CH$_3$NCO) achieve peak abundance during the end of the simulation. 
For the rest of the molecules (NH$_3$, HCO$^+$, HCN, and SiO), peak abundance appears during the intermediate time. Detailed discussions regarding the obtained abundance of these
species are discussed in the respective section.

\subsection{Radiative transfer models}
In G31, \cite{gora21} reported the identification of several COMs (CH$_3$SH, CH$_3$NCO, and CH$_3$OH). They further showed that  H$^{13}$CO$^+$, HCN, and SiO exhibited certain unique spectral features. Here, CASSIS spectrum analyzer (developed by IRAP-UPS/CNRS, \cite{vast15}, \url{http://cassis.irap.omp.eu})  
and 1D RATRAN program \citep{hog00}
together with the spectroscopic  database `Cologne Database for Molecular Spectroscopy' \citep[CDMS,][]{mull01,mull05,endr16} \url{(https://cdms.astro.uni-koeln.de)} and Jet Propulsion Laboratory (JPL) \citep{pick98} \url{(http://spec.jpl.nasa.gov/)} are used to model the observed line profiles.

The MCMC algorithm was developed in constraining the probabilistic behavior of a collection of atomic particles. It was challenging to do this analytically. This stochastic model describes a sequence of possible events in which each event's probability depends only on the state attained in the previous event \citep{gag17}. The MCMC method is an interactive process that goes through all of the line parameters (e.g., excitation temperature, source size, line width) with a random walk and heads into the solution's space, and $\chi^2$ minimization gives the final solution. Here, the MCMC fitting is mainly implemented for those species for which more than two transitions are identified. Like the rotational diagram method, the MCMC method is also helpful for species where multiple transitions
covering a wide range in energy have been detected \citep{vast18}.
Since the MCMC method is an iterative method that heads toward the solution by the $\chi^2$ minimization process over parameter space. Statistically, it will be beneficial to estimate the physical parameters from the fitting of multiple transitions.

Since H$^{13}$CO$^+$, HCN, and SiO show some unique line profiles and not many transitions are identified, Markov Chain Monte Carlo Method is not used. Two slab model of \cite{mye96} and further improved by \cite{difr01} is used to explain the unique characteristics (inverse P-Cygni profile, a representative of the infalling envelope) of H$^{13}$CO$^+$. 
This model considers the infalling gas envelope in two different spherically symmetric regions of
different excitation temperatures. Along the line of sight, it places two parallel slabs having different excitation temperatures. In the front
layer, the excitation temperature is $T_f$, and in the rear layer, it is $T_r$, respectively. It considers a peak optical depth for both the region, 
$\tau_0$. An infall velocity, $V_{in}$ is regarded throughout the contracting envelope. 
Some fraction of the beam is filled by the continuum temperature ($T_c$, the brightness temperature corresponding to the continuum image's peak flux density). This beam filling factor is denoted by $\Phi$. The rear layer is illuminated by the background
radiation of temperature ($T_b$ = $2.7$ K), and the core is considered a black body. \cite{mye96} defined the radiative transfer solution in terms of $T_f$, $T_r$, $T_b$, and T$_B$ (brightness temperature). Later \cite{difr01} introduced the continuum source into account.
There are two main limitations to the two slab model: the simplicity and a large number of free parameters. However, this model is beneficial for reproducing the line profile of the infalling envelope.
The CASSIS program is utilized to fit the observed spectral signature based on this two slab model.

1D RATRAN model is utilized to explain all the observed line profiles.
Various input parameters applied for RATRAN modeling (inner and outer radius of the envelope, 
dust opacity, distance from the Sun, dust to gas mass ratio, background radiation, and bolometric
temperature) are listed in Table \ref{table:RATRAN-input-parameters}. The  dust emissivity ($\kappa$) is considered by assuming a power-law emissivity model. The index $\beta$ can vary between $1-1.4$.
 \begin{table*}
\centering
{\scriptsize
 \caption{Key parameters used for our 1D-RATRAN modeling. \label{table:RATRAN-input-parameters}}
\begin{tabular}{|l|l|}
  \hline
  \hline
  Input parameters & Used \\
  \hline
  \hline
 Inner radius of the envelope& $156$ AU\\
Outer radius of envelope & $119000$ AU\\
T$_{cmb}$ & $2.73$ K\\
Gas to dust mass ratio& $100$\\
 $\kappa$ (dust emissivity) & $\kappa=\kappa_0(\frac{\nu}{\nu_0})^\beta$, where $\kappa_0= 19.8$ cm$^2$/gm \citep{oss94},
 $\nu_0= 6 \times 10^{11}$ Hz,
 and $\beta=1 - 1.4$ \\
Distance & $3700$ pc\\
Bolometric temperature & $55$ K \citep{mue02}\\
  \hline
  \hline
 \end{tabular}}
 \end{table*} 
It is noticed that the velocity components present in a collapsing gas envelope can strongly affect the line profiles. The thermal broadening of lines is automatically considered in RATRAN. The other factors affecting the line profile are infall motion, expansion, 
and non-thermal turbulent motion. These contributions are considered in our model. The Doppler `$b$' parameter (characterizes the spectral line's width) is regarded as a constant throughout the envelope to assess the effects of turbulence present in the envelope affecting the line profile. This parameter is related to the line broadening and directly related to the FWHM of 
the line transition by, 
$$
\frac{b}{FWHM}=\frac{1}{2\times\sqrt{ln(2)}}=0.60.
$$
 In our RATRAN model, a constant linewidth throughout the entire region of the cloud is considered. But in reality, this parameter would radially vary. Increasing this parameter would decrease the line intensity, whereas reducing this parameter would increase the line intensity.

\section{Results and discussions}
\label{sec:result-discussions}
\subsection{The Markov Chain Monte Carlo model \label{sec:MCMC}}
Recently, \cite{gora21} reported the observation of complex sulfur-bearing species, methanethiol ($\rm{CH_3SH}$) and methyl isocyanate ($\rm{CH_3NCO}$) in G31. They also reported various transitions of methanol ($\rm{CH_3OH}$) in G31. Their observations extracted the excitation temperature, column density, FWHM, etc., using a rotation diagram and LTE model analysis. Here, the MCMC method is implemented to estimate those parameters and compare them with the results of \cite{gora21}. 

\begin{figure}
\hskip -1cm
\includegraphics[width=11cm, height=12 cm, angle=270]{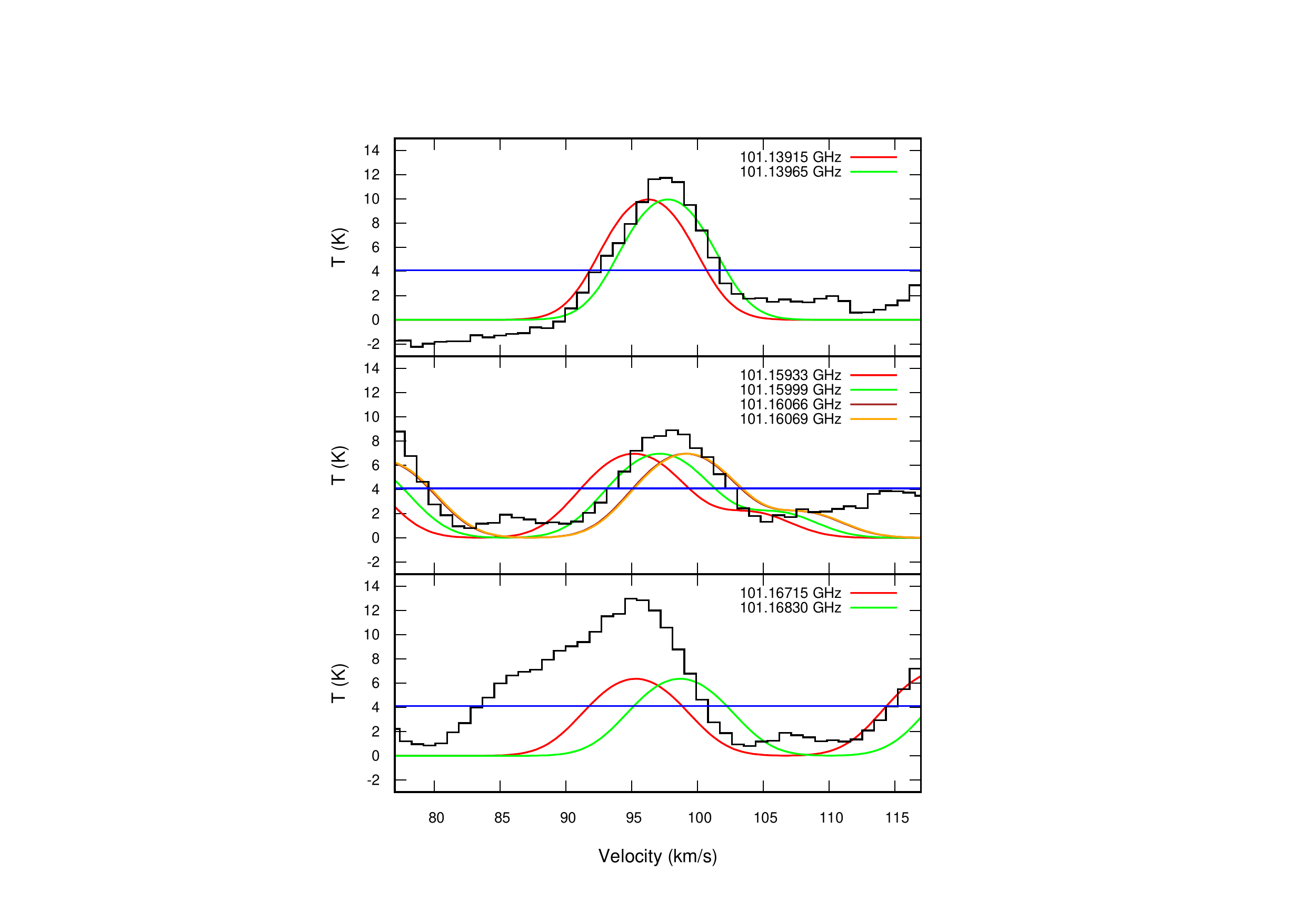}
\caption{The MCMC method is used to fit the observed transitions of $\rm{CH_3SH}$ in G31. The
observed spectral profile is shown with black lines, whereas the modeled profile is shown with the other colours. The horizontal blue line represents the 3$\sigma$($4.1$ K) RMS noise level.} 
\label{fig:ch3sh-mcmc}
\end{figure}

\begin{figure}
\hskip -3cm
\includegraphics[width=14cm,height=14cm,angle =270]{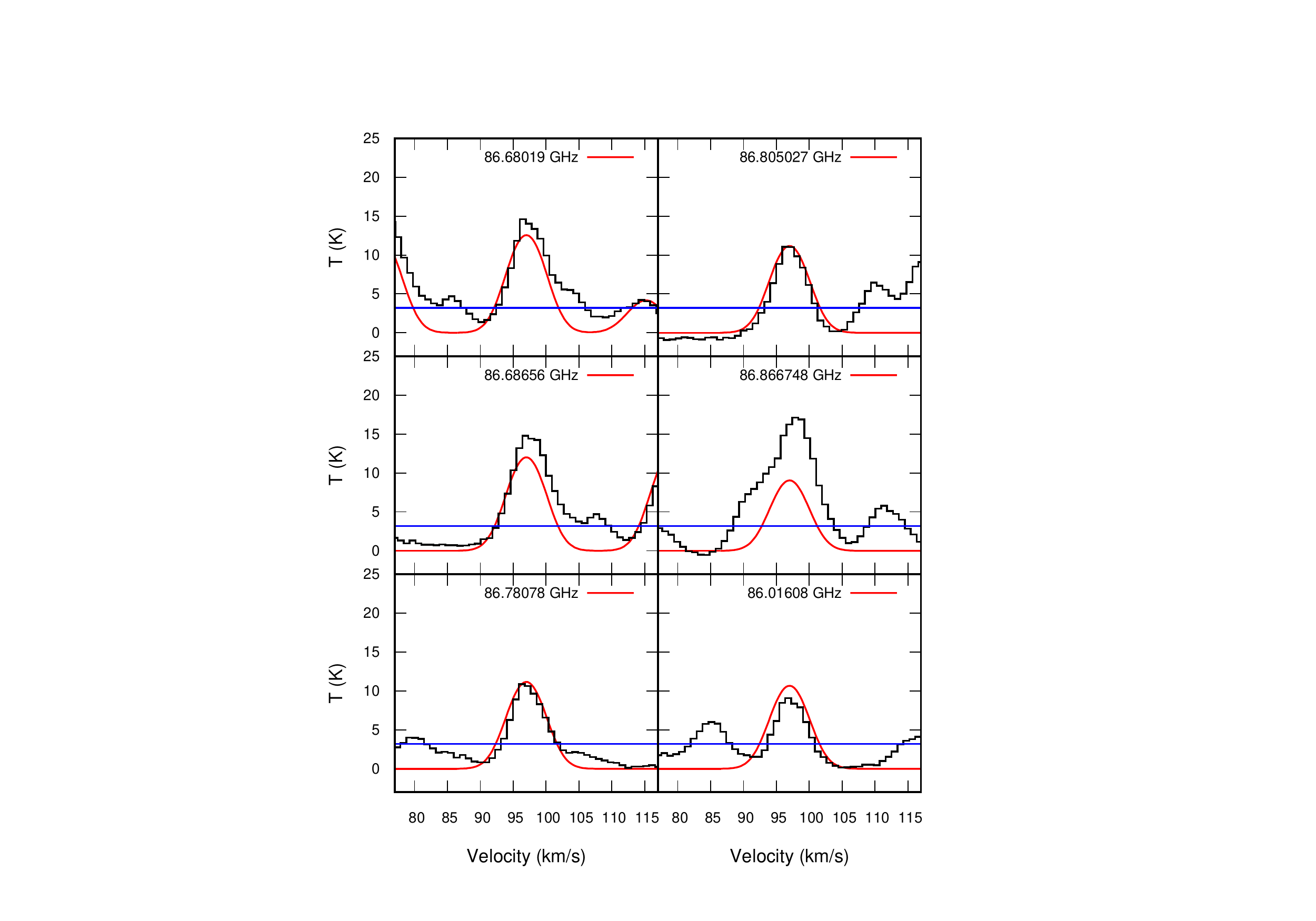}
\caption{The MCMC method used to fit the observed transitions of $\rm{CH_3NCO}$ in G31. The red lines represent the modeled spectral profile, whereas the observed spectra are shown in black. The horizontal blue line represents the 3$\sigma$(3.2K) RMS noise level.}
\label{fig:ch3nco-mcmc}
\end{figure}

\begin{figure}
\hskip -3cm
\includegraphics[width=14cm,height=15cm,angle=270]{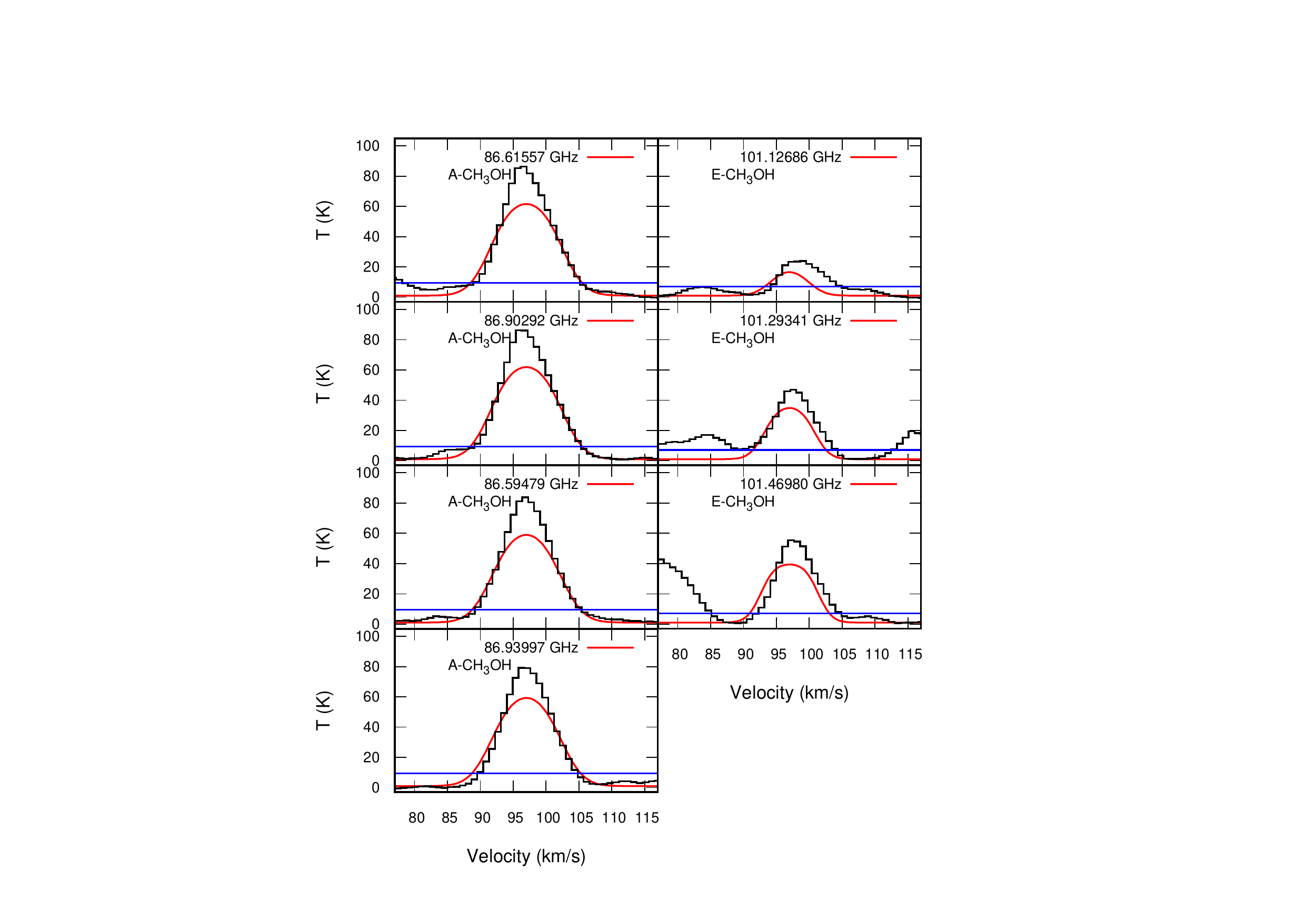}
\caption{MCMC method used to fit the observed transitions of methanol ($\rm{CH_3OH}$) towards G31. The red lines represent the modeled spectral profile to the observed spectra in black. The horizontal blue line represents the 3$\sigma$(9.4K for A-CH$_3$OH and 7K for E-CH$_3$OH) RMS noise level.}
\label{fig:ch3oh-mcmc}
\end{figure}

Initial constraints used for this fitting procedure and the best-fitted values of the physical parameters and the uncertainties obtained from the $\chi^2$ minimization are noted in Table 
\ref{table:mcmc_lte}. All the fitted spectra of $\rm{CH_3SH}$, $\rm{CH_3NCO}$, and $\rm{CH_3OH}$ 
are shown in Figures \ref{fig:ch3sh-mcmc}, \ref{fig:ch3nco-mcmc}, and \ref{fig:ch3oh-mcmc} respectively.
With the best-fitted line parameters shown in Table \ref{table:mcmc_lte}, it is noticed that all the transitions of CH$_3$SH and CH$_3$NCO are optically thin. In contrast, all the transitions of A-CH$_3$OH  and two transitions of E-CH$_3$OH are optically thick.
A column density of $1.4 \times 10^{17}$ cm$^{-2}$, $1.2 \times 10^{18}$ cm$^{-2}$, respectively are obtained for CH$_3$SH and CH$_3$NCO. The transitions of CH$_3$SH are fitted well with an excitation temperature $\sim 77$ K, whereas for CH$_3$NCO it is $\sim 205$ K.
For the methanol, it is not possible to fit all the transitions simultaneously. Thus following \cite{gora21}, we separate the transitions arising from A/E methanol. A column density of 
$2.1 \times 10^{19}$ cm$^{-2}$, and $8.2 \times 10^{19}$ cm$^{-2}$ are obtained for A-CH$_3$OH, and E-CH$_3$OH, respectively. In the case of A-CH$_3$OH, comparatively a higher temperature ($\sim 173$ K) is associated than E-CH$_3$OH ($\sim 82$ K). The peak abundance profile of CH$_3$OH shown in Figure \ref{fig:abundance} and noted in Table \ref{table:abundances} depicts that deep inside the envelope ($<5840$ AU), the abundance of methanol gradually decreases. Thus, obtaining a comparatively higher abundance with the cold component (i.e., transitions of E-CH$_3$OH) of methanol is valid. At the footnote of the table, these column densities with that was obtained by \cite{gora21} by the LTE analysis and rotational diagram method are also noted.

\begin{table*}
\tiny{
\caption{Summary of the best fitted line parameters obtained by using MCMC method.\label{table:mcmc_lte}} 
\begin{center}
\addtolength{\leftskip} {-2cm}
\addtolength{\rightskip}{-2cm}
\begin{tabular}{|c|c|c|c|c|p{1.4cm}|c|p{0.9cm}|c|}
\hline
\hline
Species&Quantum numbers&Frequency&E$_u$&Best fit FWHM&Best fitted column&Best fitted
&Best fitted&Optical depth \\
&&(GHz)&(K)&(Km s$^{-1}$)&density$^a$ (cm$^{-2}$) & T$_{ex}$ (K)& source size ($^{''}$)& ($\tau$)\\
\hline\hline
&4(0,4)-3(0,3), m=0&101.13915&12.14&&&&&0.526\\
CH$_3$SH&4(0,4)-3(0,3), m=1&101.13965&13.56&&&&&0.517\\
&4(-2,3)-3(-2,2), m=0&101.15933&31.26&&&&&0.308\\
&4(-3,2)-3(-3,1), m=1&101.15999&52.39&&&&&0.137\\
&4(3,1)-3(3,0), m=0&101.16066&52.55&6.5&(1.4$\pm$1.0)$\times$10$^{17}$&77.20$\pm$30.08&0.97$\pm$0.11&0.136\\
&4(-3,2)-3(-3,1), m=0&101.16069&52.55&&&&&0.136\\
&4(-2,3)-3(-2,2), m=1&101.16715&29.62&&&&&0.315\\
&4(2,2)-3(2,1), m=1&101.16830&30.27&&&&&0.312\\
\hline
&10(0,10)-9(0,9), m=0(v$_t$=0)&86.68019&22.88&&&&&0.568\\
&10(0,0)-9(0,0), m=1(v$_t$=0)&86.68656&34.97&&&&&0.536\\
$\rm{CH_3NCO}$&10(2,9)-9(2,8), m=0(v$_t$=0)&86.78078&46.73&6.5&(1.2$\pm$ 0.4)$\times$10$^{18}$&202.3$\pm$16.05&0.90$\pm$0.11&0.485\\
&10(2,8)-9(2,7), m=0(v$_t$=0)&86.80503&46.73&&&&&0.485\\
&10(-3,0)-9(-3,0), m=1(v$_t$=0)&86.86675&88.58&&&&&0.373\\
&10(2,0)-9(2,0), m=1(v$_t$=0)&87.01608&58.81&&&&&0.455\\
\hline
&$7(2)^{-}-6(3)^{-},V_{t}=0$&86.61557&102.70&&&&&1.576\\
A-$\rm{CH_3OH}$&$7(2)^{+}-6(3)^{+},V_{t}=0$&86.90292&102.72&8.8&$(2.1 \pm 0.82) \times 10^{19}$&$172.51 \pm 15.46$&$1.98 \pm 0.25$&1.582\\
&$15(3)^{+}-14(4)^{+},V_{t}=0$&88.59479&328.26&&&&&1.350\\
&$15(3)^{-}-14(4)^{-},V_{t}=0$&88.93997&328.28&&&&&1.356\\
\hline
&$5(-2)-5(1),V_{t}=0$&101.12686&60.73&&&&&0.478\\
E-$\rm{CH_3OH}$&$7(-2)-7(1),V_{t}=0$&101.29341&90.91&6.0&$(8.2 \pm 1.47) \times 10^{19}$&81.74 $\pm$ 7.74&$1.97 \pm 0.14$ &1.75\\
&$8(-2)-8(1),V_{t}=0$&101.46980&109.49&&&&&2.754\\
\hline
\hline
\end{tabular}
\end{center}}
\hskip 2cm {$^a$ With the LTE analysis, \cite{gora21} obtained the column density of $4.13 \times 10^{17}$ cm$^{-2}$, $7.22 \times 10^{17}$ cm$^{-2}$, and $1.84 \times 10^{19}$ cm$^{-2}$ respectively for  CH$_3$SH, CH$_3$NCO, and CH$_3$OH.} \\
{\noindent With the rotational diagram analysis, \cite{gora21} obtained it $2.85 \times 10^{16}$ cm$^{-2}$, $1.58 \times 10^{16}$ cm$^{-2}$ and $2.94 \times 10^{19}$ cm$^{-2}$ respectively.}\\
\end{table*}

\subsection{Two slab model}
\label{sec:twoslab}
An inverse P-Cygni profile shows a red-shifted absorption lobe along with a blue-shifted
emission lobe. The detection of $\rm{H^{13}CO^+}$ towards G31 suggest the infalling gas envelope towards the center of the source \citep{gora21}. To explain the emission and absorption nature of the inverse P-Cygni profile, here, two slab model is used which was introduced by \cite{mye96} and further modified by \cite{difr01}. 
  The input variables used to fit the inverse P-Cygni profile are listed in Table \ref{table:H13CO+_fit} along with the best-fitted parameters. These
parameters are extracted by minimizing the value of $\chi^2$.
$\rm{HCO^+}$ is widely used as a useful tracer of the envelope region.
\cite{gora21} recently reported an inverse P-Cygni profile for the $1-0$ transition of 
$\rm{H^{13}CO^+}$ in G31.
Their analysis of this profile had suggested a clear indication of the presence of infalling gas envelope towards 
the core of G31. 
Inverse P-Cygni line profiles ease extract the dynamic behavior 
(infall velocity, mass infall rate, velocity dispersion, etc.) of the infalling envelope.
Here, two slab model is utilized to explain the observed inverse P-Cygni profile of H$^{13}$CO$^+$. Altogether, there are eight parameters involved in the solution. This freedom can produce a lot of best-fit solutions. Some parameters are fixed to a realistic estimation to avoid any misleading results.  The best fitted parameters are noted in Table \ref{table:H13CO+_fit}. 

\begin{table}
{\scriptsize
 \caption{Best fitted parameters obtained with the two slab model for $\rm{H^{13}CO^+}$.\label{table:H13CO+_fit}}
 \centering
   \begin{tabular}{|l|l|l|}
  \hline
  \hline
  Input parameters & Range of values used as input & Best fitted parameters \\
  \hline
  \hline
 $T_r$ (K)& 10.0 - 200.0& 29.0\\
$\tau_0$ & 0.1 - 11.0& 1.19\\
$V_{in}$ (km s$^{-1}$) & 1.0 - 4.0 & 2.50 \\
$\sigma_v$ (km s$^{-1}$)& 0.1 - 3.0 & 0.97\\
$v_{lsr}$ (km s$^{-1}$) & 96.0 - 97.5 & 96.5 \\
$T_f$ (K) & 10.0 - 10.0 & 10.0\\
$T_c$ (K)& 36.0 & 36.0\\
$\Phi$ & 0.5 - 0.9 & 0.62\\
  \hline
  \hline
 \end{tabular}}
 \end{table}

 \begin{figure}
 \centering
\hskip -1.0cm
\includegraphics[height=5cm,width=6cm]{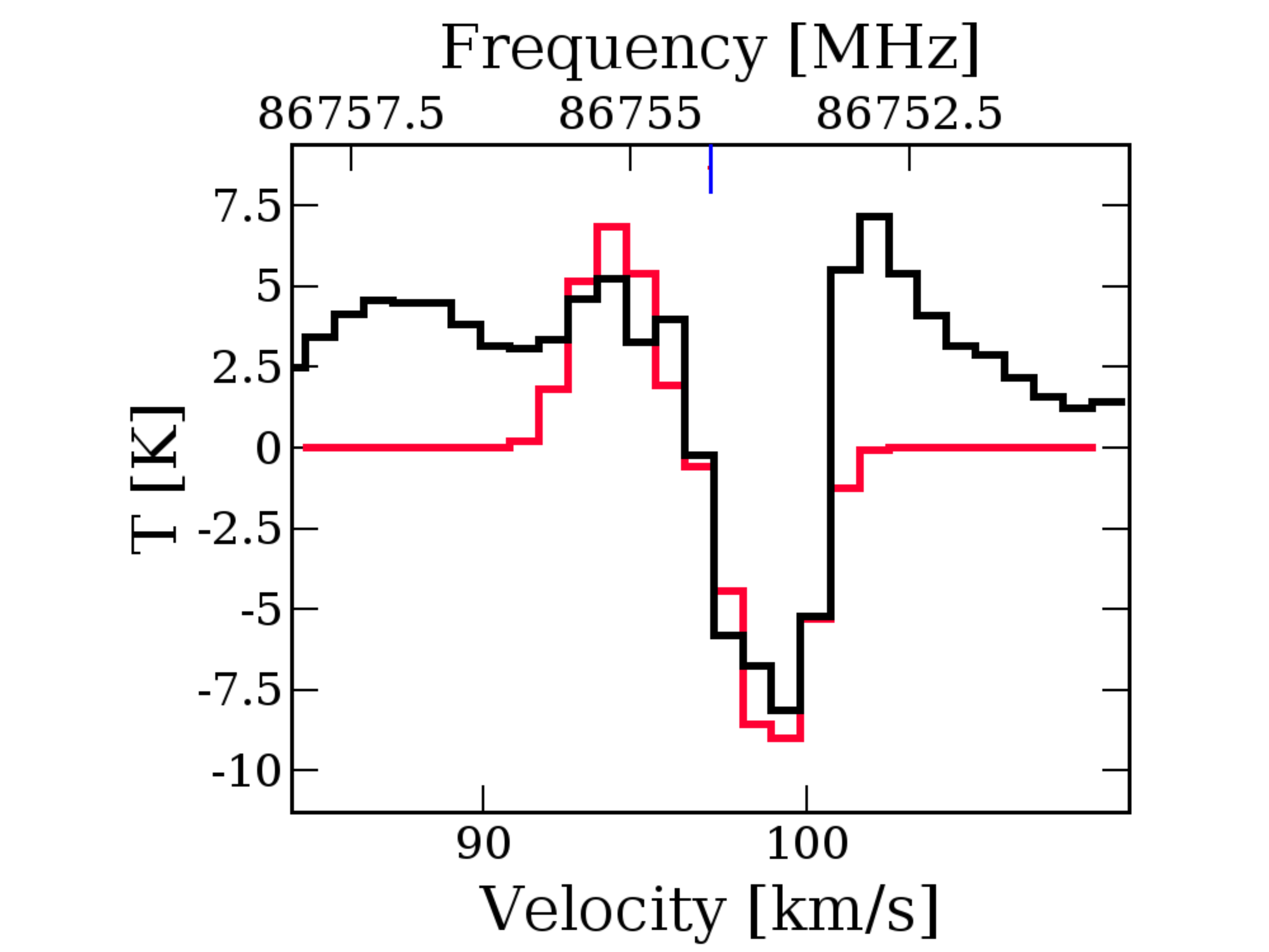}
\caption{Fitted inverse P-Cygni profile using two slab model is shown. Black line is the observed spectrum, whereas red line is the fitted spectrum.}
\label{fig:inP-cygni}
\end{figure}

From the best-fitted infall velocity,  the mass infall rate is further used by using the method 
described in \cite{klaa07}. Here, the source is considered to be spherically symmetric. This crude
approximation is often violated and it is required to consider the
outflow component for the computation of the mass infall rate. However, this approximation can 
present a good educated estimation of the
mass infall rate. The mass infall rate is described by the following relation taken from equation 3 of \cite{klaa07},
\begin{equation}
\dot{M}=\frac{dM}{dt}\simeq\frac{M}{t}=\frac{\rho V v_{in}}{r_{gm}}=\frac{4}{3}\pi n_{H_2}\mu m_H r^{2}_{gm} v_{in},
\end{equation}
where $\rm{v_{in}}$ is the infall velocity, $\rm{m_{H}}$ is the mass of a hydrogen atom, $n_{H_2}$ is the ambient source density (=$5.4\times10^6$cm$^{-3}$), $\mu$ represents the mean molecular weight of the gas ($=2.35$)and $r_{gm}$ is the geometric mean radius of $\rm{H^{13}CO^+}$ emitting region derived from the relation $r_{gm}=\sqrt{f} r_{beam}$. Here, $f$ is the beam filling factor (0.62 from Table \ref{table:H13CO+_fit}) and r$_{beam}$ is the beam radius (1.1$^{''}$ from \cite{gora21}). Here, a best-fitted infall velocity of $\rm{H^{13}CO^+}$ $\sim 2.5$ km s$^{-1}$ is obtained. Using this value in the above equation, 
a mass infall rate of $\sim 1.3\times10^{-3} \ M_\odot yr^{-1}$ with the 3.7 kpc distance consideration and $\sim 5.9\times10^{-3} \ M_\odot yr^{-1}$ with the 7.9 kpc distance consideration is used. These values are consistent with the previous prediction $\sim 3\times10^{-3} \ M_\odot yr^{-1}$ by \cite{oso09} for their best-fitted model with a central mass 25M$_\odot$.
The infall rate indicates that the source is in the state of a high accretion phase. The best-fitted spectra along with the reported inverse P-Cygni profile of \cite{gora21}
is shown together in Figure \ref{fig:inP-cygni}.

\subsection{1D spherically symmetric RATRAN model}
\label{sec:RATRAN}
The observed inverse P-Cygni profile ($\rm{H^{13}CO^+}$) and the feature of SiO, HCN, and NH$_3$ 
are further modeled with the 
RATRAN code to explain the physical properties of the envelope and surrounding medium in more detail. Moreover, the transitions of COMs are also modeled with this code.

\subsubsection{H$^{13}$CO$^+$ \label{sec:hco+}}
\begin{figure}
\centering
\includegraphics[height=5cm]{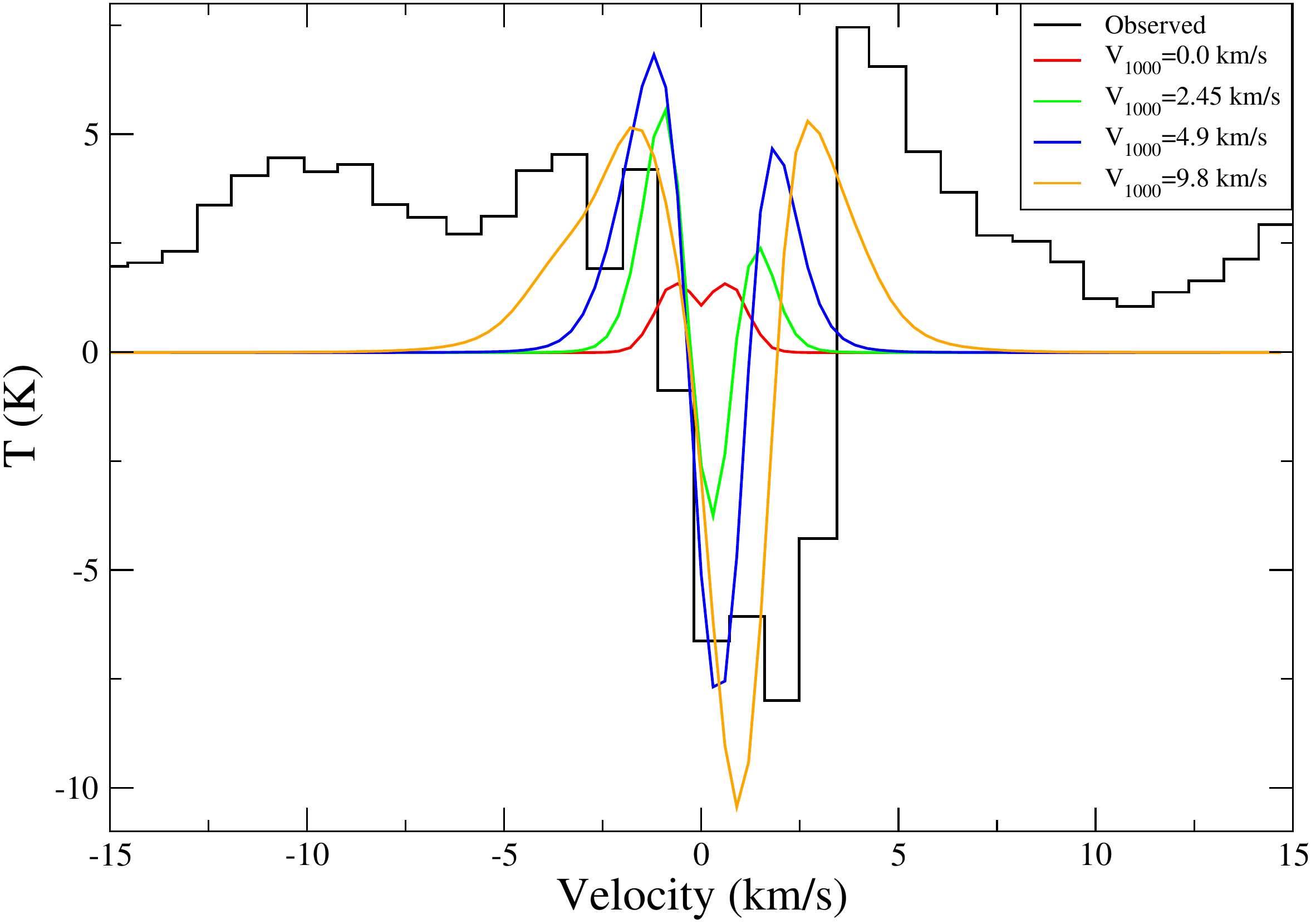}
\caption{A comparison between the observed line profile (black line) and modeled line profile with a constant abundance of H$^{13}$CO$^+$ ($7.07 \times 10^{-11}$) is shown. The best fit is obtained when $\beta=1$ is used. Different cases with the infall velocity are shown. It is evident that with the static envelope (red line), there is no asymmetry, and asymmetry increases with the increase in velocity. The best fit is obtained when an infall velocity of $4.9$ km/s at 1000AU is used.}
\label{fig:h13co+_best}
\end{figure}

The observed inverse P-Cygni profile of $\rm{H^{13}CO^+}$ is an indication of the collapsing envelope of a star-forming
region. The synthetic spectra of the $1-0$ transition ($86.753970$ GHz) 
of $\rm{H^{13}CO^+}$ is modeled using the 1D RATRAN code. The collisional
rates of $\rm{H^{13}CO^+}$ with H$_2$ are taken from the LAMDA database. 
The physical parameters of the model are described in Section \ref{sec:phys}. Initially, a constant abundance of H$^{13}$CO$^+$ is considered throughout the region.
Figure \ref{fig:h13co+_best} depicts the comparison between the observed (black) and modeled (orange) line profile.
 The constant abundance of H$^{13}$CO$^+$, FWHM, and $\beta$ are varied to find out the best fit line profile.
The best fit is obtained from the two slab model when an FWHM of $0.97$ km/s is used (see Table \ref{table:H13CO+_fit}). With the 1D RATRAN modeling, the best fit is obtained when an FWHM of $\sim 1.42$ km s$^{-1}$ is used. For the best fit line profile, $\beta=1$ and a constant abundance of $\sim 7.07 \times 10^{-11}$ is used.
A fractionation ratio of $\rm{H^{13}CO^+:HCO^+}=1:65$ \citep{hog00} is used, which yields the best-fitted HCO$^+$ abundance $4.6 \times 10^{-9}$.
The best fitted HCO$^+$ abundance is well within the limit of our modeled abundance noted in Table \ref{table:abundances} and shown in Figure \ref{fig:abundance}. 


 Figure \ref{fig:h13co+_best} shows that when the envelope is completely static (i.e., infall velocity is zero, red curve), it gives a symmetric line profile, whereas the inclusion of the infall velocity induces the asymmetry. Increasing the infall velocity shows an increase in the red-shifted absorption and blue-shifted 
emission nature. From Figure \ref{fig:h13co+_best}, it is noticed that the modeled line profile considering $v_{1000}= 2.45$ (green line), $4.9$ (blue line), and $9.8$ (orange line) km/s gradually increases the asymmetry. 
 The best fit is obtained when an infall velocity of $\sim 4.9$ km/s (blue line) at 1000 AU is considered, which agrees well with \cite{oso09}. 
It is noticed that the spectral profile is heavily dependent on the choice of $\beta$.
Since \cite{oso09} obtained $\beta=1$ from their analysis, here also $\beta=1$ is used.  

\begin{figure}
\centering
\includegraphics[height=5cm]{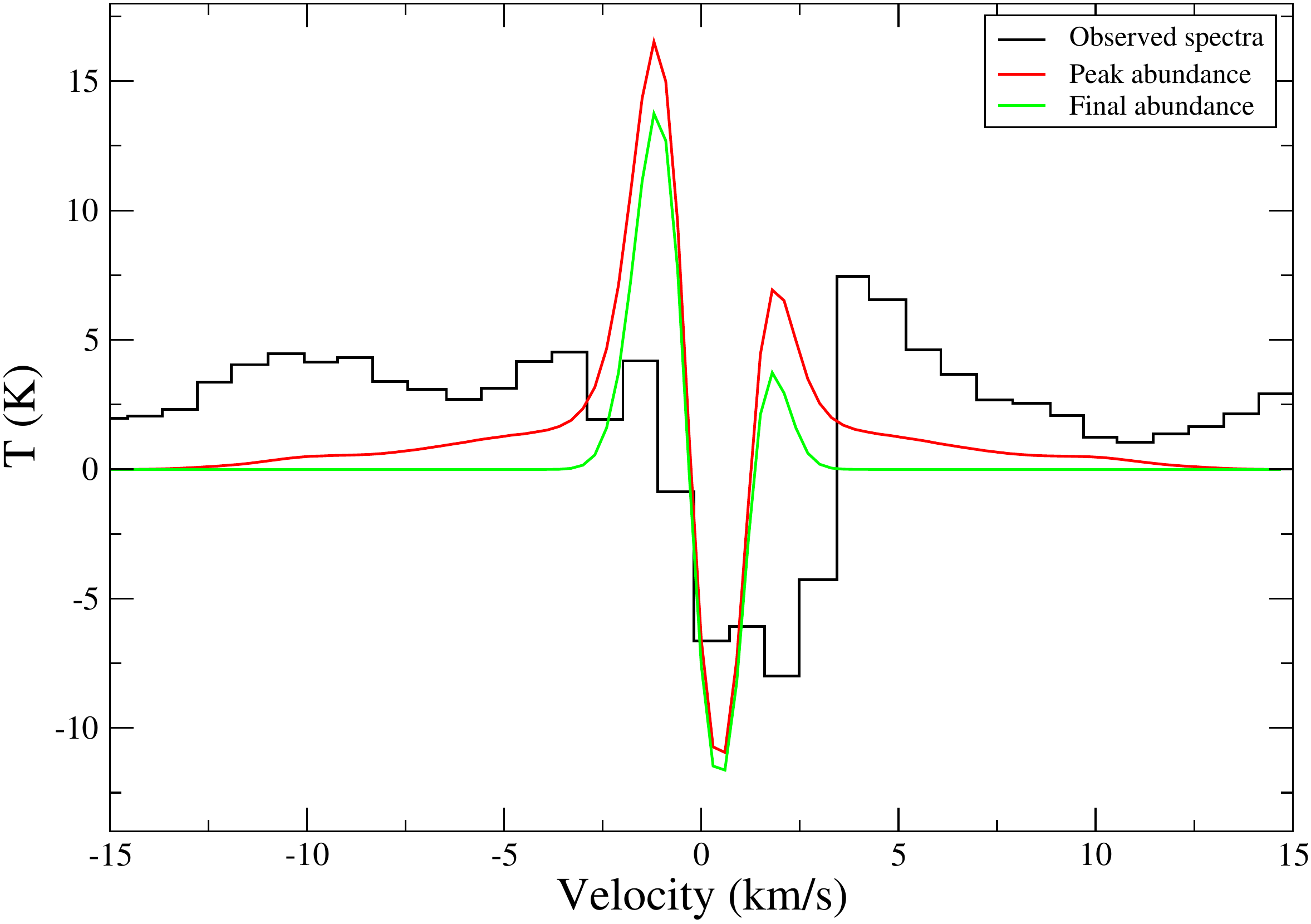}
\caption{ A comparison between the observed line and modeled line with the abundances obtained from our chemical modeling is shown. The abundance profile of HCO$^+$ obtained from our chemical modeling is used. Here, this abundance profile is further reduced by a factor of 65 to have the peak and final abundance profile of H$^{13}$CO$^+$. The red line shows the modeled line profile with the peak value, whereas the green line shows the modeled line profile with the final value obtained from our chemical model.}
\label{fig:h13co+_chemmodel}
\end{figure}

The obtained abundance distribution of HCO$^+$ from our chemical model discussed in 
Section \ref{sec:chemical} is further used.
Figure \ref{fig:abundance} shows that the peak abundance of HCO$^+$ varies in between ($\sim 7.6 \times 10^{-9}-10^{-8}$) beyond $10,000$ AU. In the innermost grid, the peak abundance of HCO$^+$ heavily increased.
Beyond $10,000$ AU, the peak abundance of HCO$^+$ and final abundance roughly matches. Inside the $10,000$ AU, the final abundance of HCO$^+$ drastically 
decreased.
Since our chemical code does not consider carbon fractionation, no abundance of 
H$^{13}$CO$^+$ is obtained from our code. However, as a guess,   an atomic fractionation ratio between 
$^{13}$C and $^{12}$C  $\sim 1:65$ is used to generate the spatial distribution of H$^{13}$CO$^+$.
Thus, the HCO$^+$ abundance is reduced by a factor of $65$ to have an educated estimation of the abundance of H$^{13}$CO$^+$.
Figure \ref{fig:h13co+_chemmodel} shows the comparison between our observed and modeled line profile by considering the abundances from our chemical model.  The red curve represents the modeled line profile when the peak abundance (taken during the warmup to post-warmup time scale) obtained from our chemical model is used. The green curve represents the modeled line profiles when considering the final abundance from our chemical model. Figure \ref{fig:h13co+_chemmodel} depicts that the observed absorption profile is well reproduced with our chemical model's abundance profile.
However, this abundance profile shows a strong emission feature compared to the observed blue-shifted emission nature. 
It is evident from the deuterium fractionation of molecule that the initial
atomic D/H ratio is not always reflected in the molecules \citep{case02,das13a,das15a,das16,maju14a}. So there might have some discrepancies in considering the abundances of H$^{13}$CO$^+$.
It is also noticed that our generated absorption nature in Figure \ref{fig:h13co+_best} and Figure \ref{fig:h13co+_chemmodel} is narrower compare to the observation. 
This inconsistency might be due to the absence of rotational motion in our radiative transfer model.

\begin{figure*}
  \centering
\begin{minipage}{0.42\textwidth}
    \includegraphics[height=7cm,width=9cm]{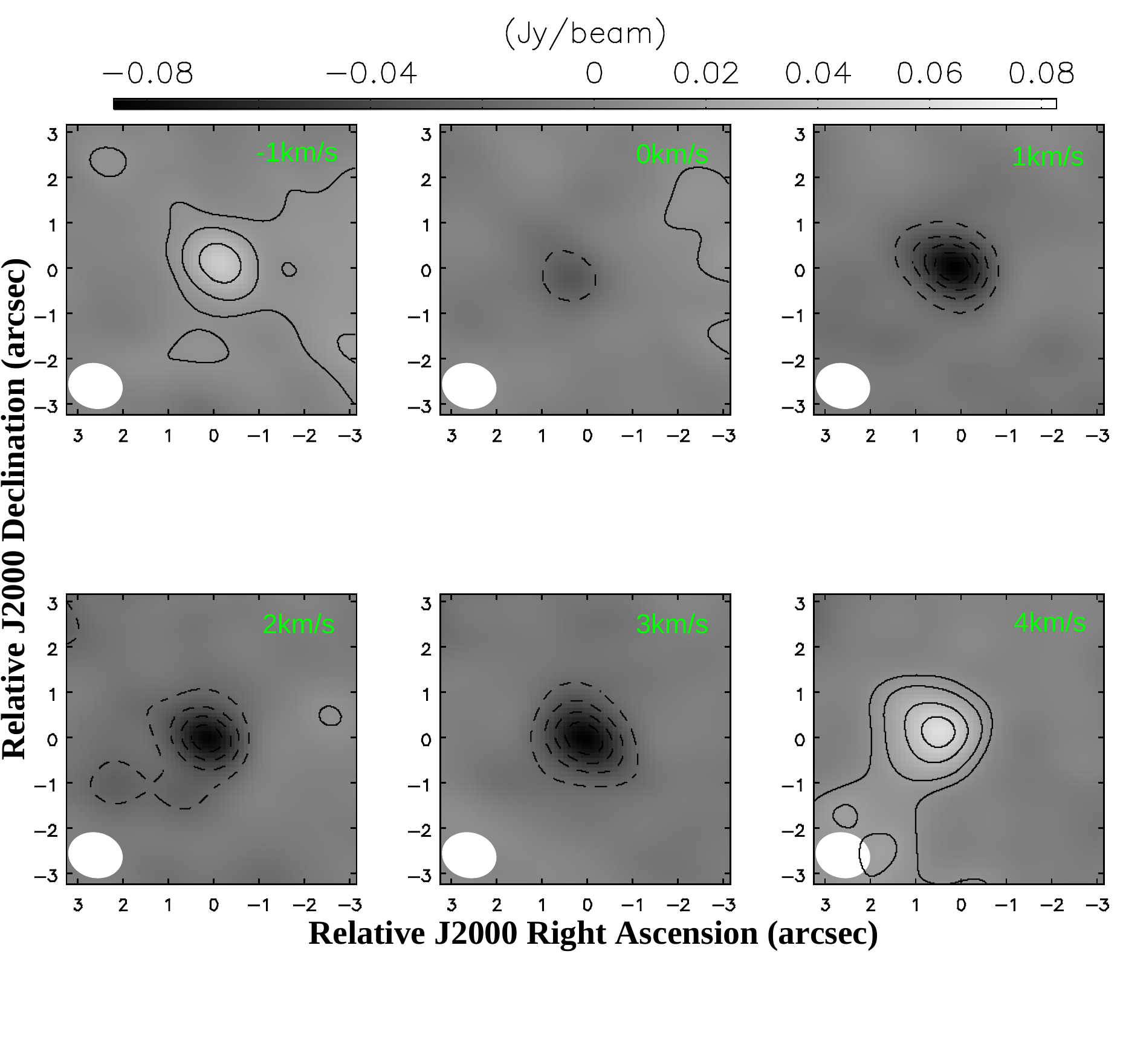}
  \end{minipage}
\hskip 2.5cm
\begin{minipage}{0.42\textwidth}
    \includegraphics[height=7cm,width=9cm]{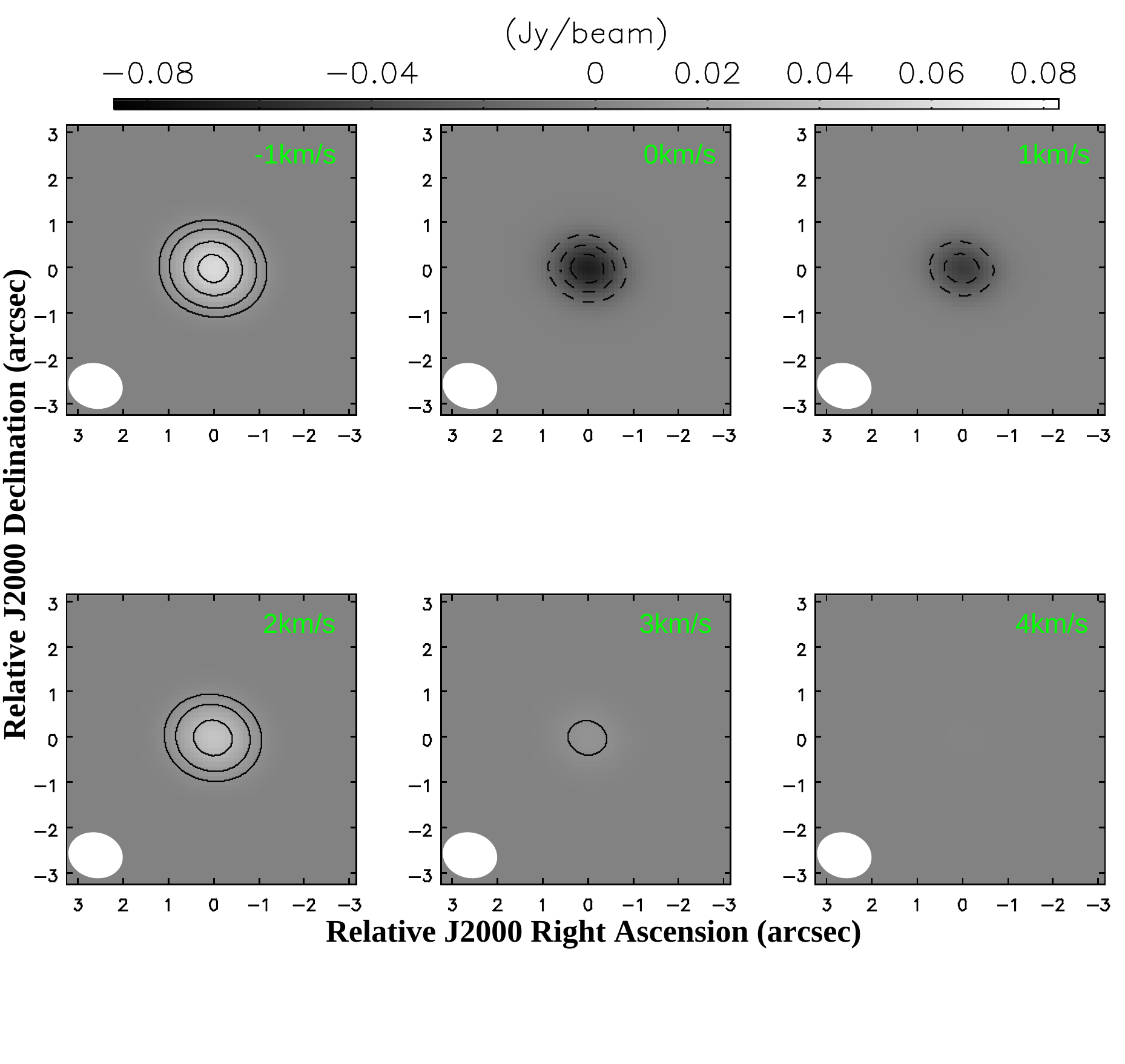}
  \end{minipage}
\caption{ A comparison between the observed channel map emission of  H$^{13}$CO$^+$ 
(left) and modeled channel map emission (right) by considering a constant abundance ($\sim 7.07 \times 10^{-11}$) profile is shown. The contour levels are drawn at -10\%, -20\%, -40\%, -60\%, 80\%, 10\%, 20\%, 40\%, 60\%, and 80\% of the  peck observed intensity ($0.086$ mJy/beam). The solid and dashed contours represent the emission and absorption, respectively. } 
\label{fig:channel-h13co+}
\end{figure*}

To visualize the difference between our generated line profile and the actual observation, our modeled `fits' file of H$^{13}$CO$^+$ is further processed to simulate the interferometric observations with the Common Astronomy Software Applications package \citep[hereafter, CASA,][]{mcmu07}. Here, instead of convolving the model with a Gaussian beam, it is simulated what ALMA would observe with the same array configuration as for the observation. 
Here, the model image is produced (for simplicity, here the line profile generated with the constant abundance $\sim 7.07 \times 10^{-9}$ is only shown) to represent the sky brightness distribution and have generated UV data using the `simobserve' task available in CASA. Then for the convolution, the `tclean' task in CASA is further used. 
The velocity channel maps obtained with the actual observation and the model are shown in Figure \ref{fig:channel-h13co+}. To understand its distribution in G31, in the left panel of Figure \ref{fig:channel-h13co+}, 
the observed spatial distribution with a few velocity channels are shown, which include both red-shifted 
(channels that are having velocity greater than the systematic velocity) and blue-shifted (channels that are having a velocity less than the systematic velocity) velocity channels together. The channel map of the
observed velocity channels  represent negative contour (i.e., $0$, $1$, $2$, and $3$, 
km s$^{-1}$), which mimics the absorption profile of the H$^{13}$CO$^+$. 
All other channels show emission signatures for the observed channel maps. 
Our simulated velocity channel maps (see right panel of Figure \ref{fig:channel-h13co+}) are 
in good agreement with the real observation.
Similar to the observed
channel maps, modeled channel maps at $0$ km s$^{-1}$ and $1$ km s$^{-1}$ show the absorption. 
However, the absorption is more extended in
the observation (i.e., absorption at {$2$ and $3$ km s$^{-1}$} is missing in our simulated image). It is 
also noted in the context of Figure \ref{fig:h13co+_best} and Figure \ref{fig:h13co+_chemmodel}. A more realistic physical model, including the
rotational motion and inclusion of outflow in the physical model may explain these mismatches.


Since H$^{13}$CO$^+$ traces the envelope region, with the present resolution, the inverse P-Cygni profile of H$^{13}$CO$^+$ is observed. Due to the higher resolution data ($0.22^{''}$), \cite{belt18} observed the source's more inner region, and thus no such profile of H$^{13}$CO$^+$ was expected.
Interferometric observations with too high angular resolution create the risk of filtering out the extended emission features. On the other hand, a shallow resolution will cause beam dilution. Infall and outflow present in an astrophysical source is a property of the envelope. So, studying the infall and outflow property using interferometric data is tricky. So molecules with extended emission are sensitive to the case that interferometric data filtered out the emission. \cite{cesa11} imaged the same transition using VLA (interferometer) and IRAM-30M (single dish) to recover the emission filtered out by the VLA data. But in our observation \citep{gora21}, the observation's angular resolution is approximately 1.1$^{''}$, which is not well resolved by the source or at best marginally resolved. For example, \cite{zhu11} reported the results on infall and outflow in the star-forming region W3-SE (it is not related to G31, but the linear resolution of these two sources is comparable), analyzing the line profile of HCN, HCO$^+$, and N$_2$H$^+$ using the interferometric (SMA-1, SMA-2) data of $\sim2.5^{''}$ resolution.

\begin{figure}
  \begin{minipage}{0.35\textwidth}
   \includegraphics[height=5.5cm,width=8cm]{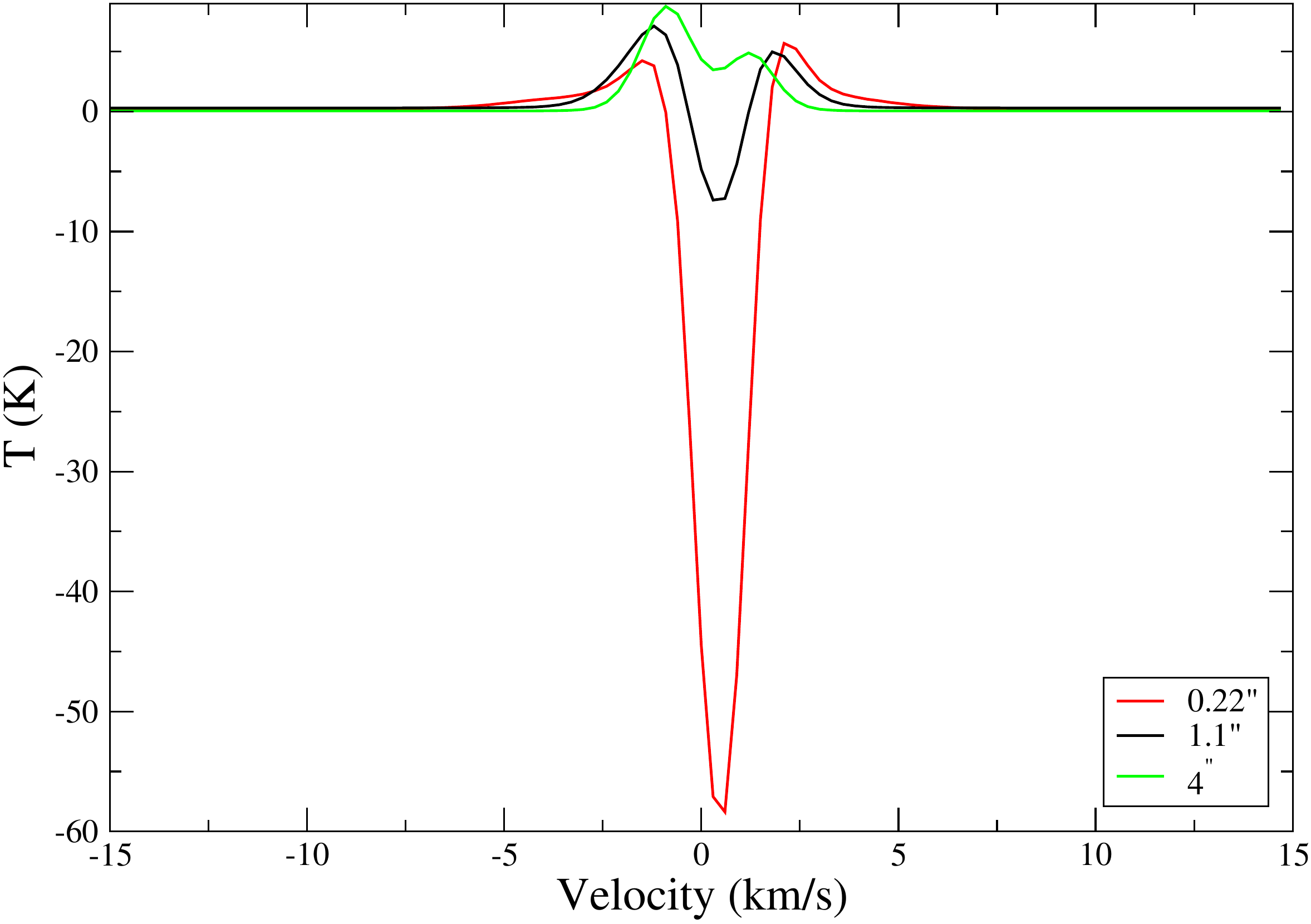}
  \end{minipage}
 \caption{The modeled line profiles of H$^{13}$CO$^+$ with the $0.22^{''}$ (red line), $1.1^{''}$ (black line), and $4^{''}$ (green line) resolution are shown. It depicts that the inverse P-Cygni profile is only visible when $1.1^{''}$ resolution is used.}
\label{fig:resolution}
\end{figure}

In Figure \ref{fig:resolution}, a comparison between the low resolution ($4^{''}$, green line and $1.1^{''}$, black line) and high resolution ($0.22^{''}$, red line) modeled spectra of H$^{13}$CO$^+$ are shown. It is noticed that the emission peak gradually diminishes with the increase in the resolution. Figure \ref{fig:resolution} depicts that the inverse P-Cygni nature is only visible when $1.1^{''}$ resolution is used. 
It justifies the relevance of considering the extended emission profile of $\rm{H^{13}CO^+}$ in our comparatively low-resolution data of ALMA.   

\begin{figure}
\includegraphics[height=5cm,width=8cm]{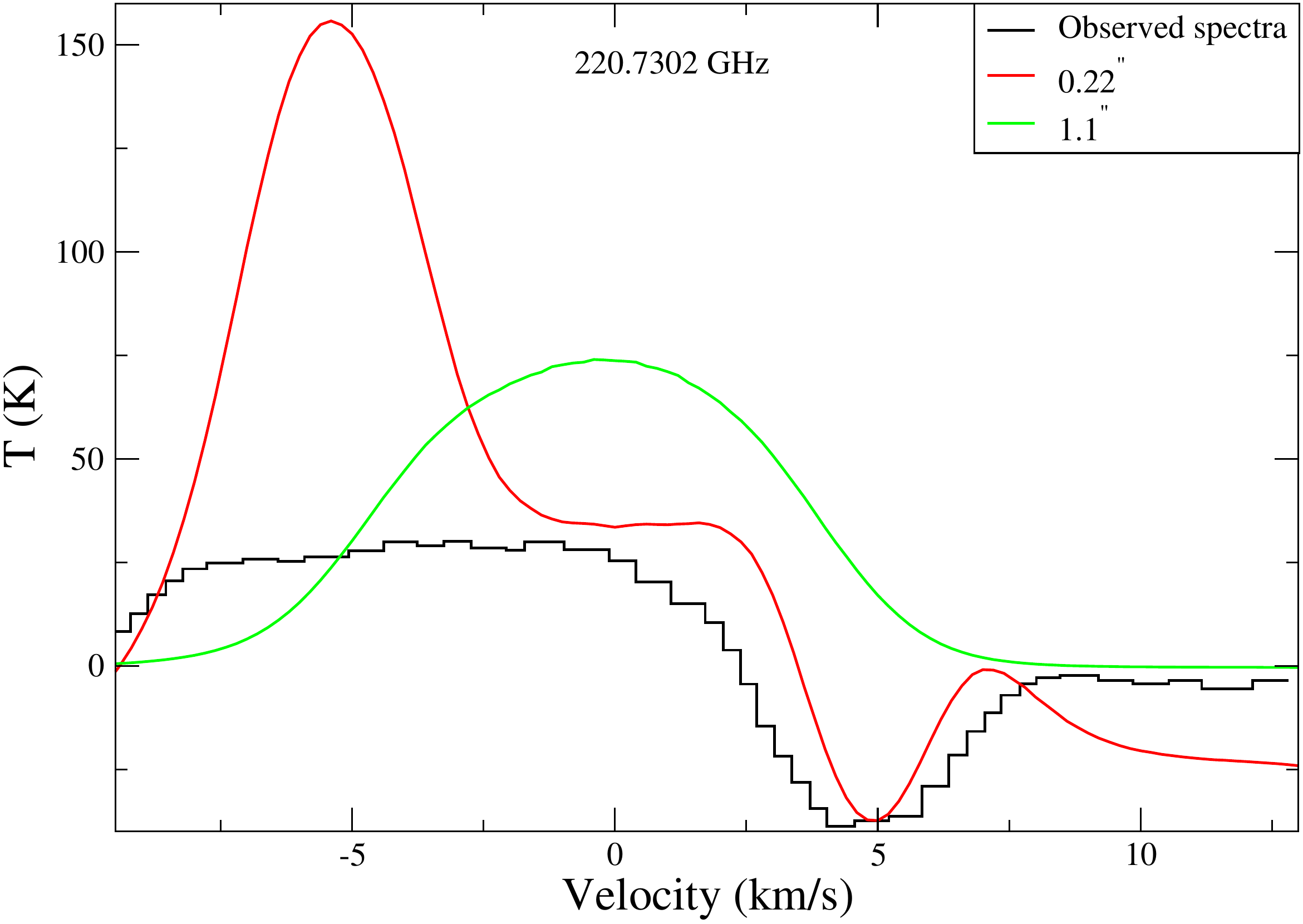}
\caption{ A comparison between the observed (0.22$^"$, black line) (data taken from \cite{belt18})} and modeled (0.22$^"$ in red and 1.1$^"$ in green) line profiles of CH$_3$CN is shown. It depicts that the inverse P-Cygni nature is not visible with our low-resolution data. For the best-fitted case, $\beta=1$, FWHM = $2.5$ km/s, V$_{1000}$=6.0 km/s and a constant abundance of $6 \times 10^{-8}$ are used.
\label{fig:ch3cn}
\end{figure}

With the higher resolution (0.22$^{''}$),
\cite{belt18} was able to observe various transitions of $\rm{CH_3CN}$ and its isotopologues having different upper state energies. CH$_3$CN is mainly formed on the grain surface and populates the gas phase when it is warmer. Thus, it mainly traces the hot inner region \citep{hung19}. 
The shift of the red-shifted absorption peak from the systematic velocity represents the infall velocity. \cite{belt18} noticed that this shift is increased with the increase in the up-state energy (K $<$ 10). It indicates the presence of an accelerated infall in this source. Though molecules with extended emission (like H$^{13}$CO$^+$) heavily suffer from interferometric filtering, in our case, the angular resolution is comparatively lower (1.1$^{''}$) than the resolution of \cite{belt18} (0.22$^{''}$). Previously, \cite{olm96a,olm96b,cesa11} observed $\rm{CH_3CN}$ transitions towards the 
main core of G31 using single dish (IRAM-30m) and interferometric (SMA). Here, the 1D RATRAN radiative transfer model is used to reproduce the observed line profile of CH$_3$CN for J=12 - 11 (K=2) transition (220.7302 GHz) by \cite{belt18}. It is important to note that our observational resolution greatly differs from \cite{belt18}. 
Figure \ref{fig:ch3cn} shows a comparison between the high resolution and low resolution modeled spectrum with the observed profile of CH$_3$CN. 
\cite{belt18} obtained the abundance of of CH$_3$CN $\sim 1.0 \times 10^{-8}$. 
The best-fit is obtained when a constant abundance of $\sim 6 \times10^{-8}$, FWHM of $2.5$ km/s, and $\beta$ $\sim 1$ is used. 
From our chemical modeling (Figure \ref{fig:abundance} and Table \ref{table:abundances}), it is noted that the peak abundance of CH$_3$CN can reach upto $\sim 3.8 \times 10^{-8}$.

From Figure \ref{fig:ch3cn}, it is evident that high resolution ($0.22^{''}$) modeled data show inverse P-Cygni profile, which is consistent with the observed profile of CH$_3$CN. However, in our model, a strong emission feature is obtained in the blue-shifted region. On the other hand, low resolution ($1.1^{''}$) modeled data has a significantly lower intensity than high resolution. It does not show any inverse P-Cygni profile because CH$_3$CN traces the infall materials toward the source's inner region.

\begin{figure}
  \begin{minipage}{0.35\textwidth}
   \includegraphics[height=7cm,width=8cm]{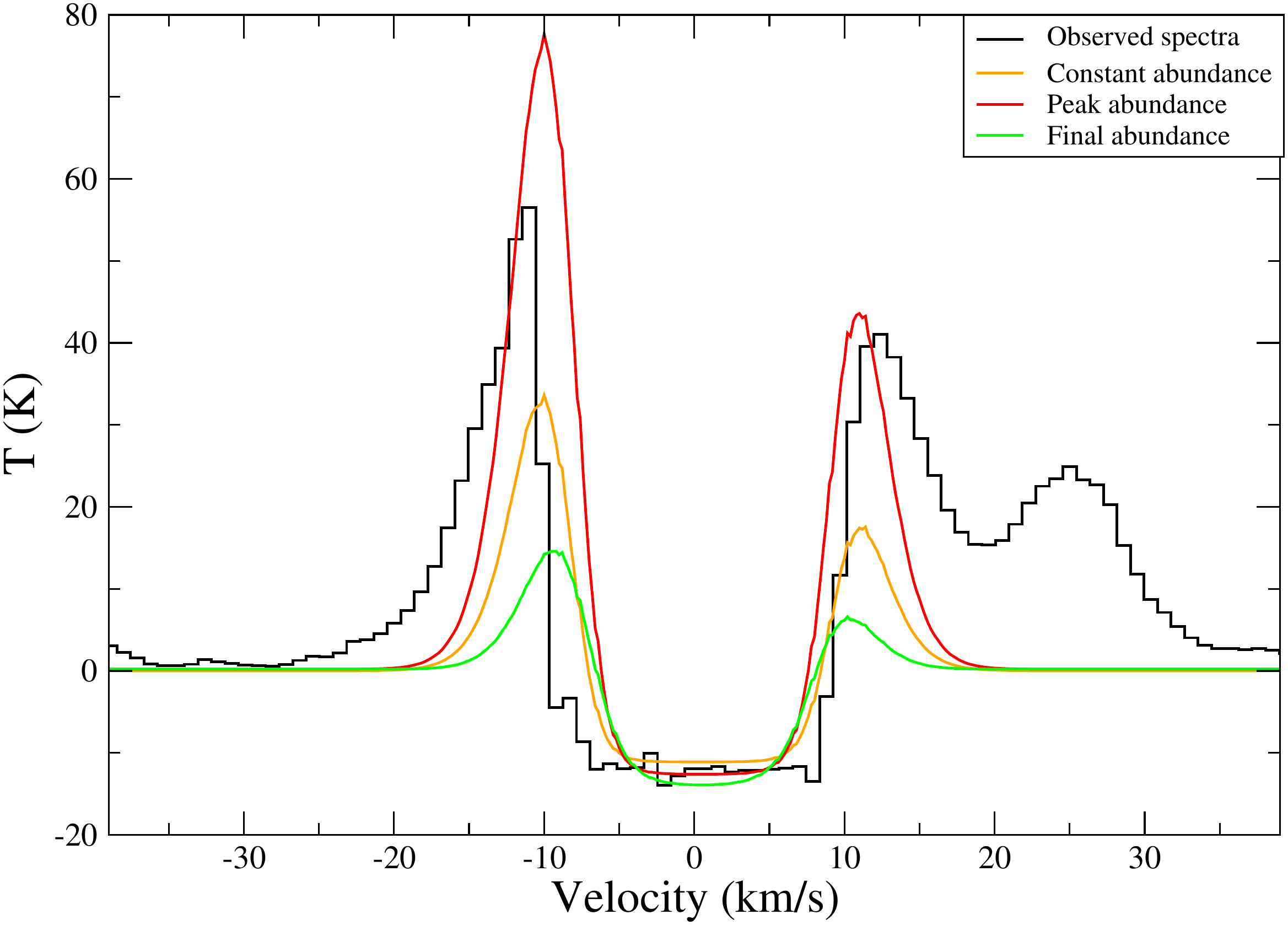}
  \end{minipage}
 \caption{ A comparison between the synthetic spectrum of HCN generated using 1D
RATRAN radiative transfer code and line profile observed towards G31 is shown. The solid black line represents the observed spectrum, whereas the orange line is the line profile of the transition of HCN obtained by using constant abundance ($7.6 \times 10^{-8}$) and red by using the peak abundance profile obtained from the chemical model, green by using the final abundance profile obtained from chemical model. We have obtained the best fit with $\beta=1.4$ and an FWHM of $10$ km/s.}
\label{fig:cassis_hcn}
\end{figure} 

\begin{figure*}
  \centering
\begin{minipage}{0.48\textwidth}
    \includegraphics[width=\textwidth]{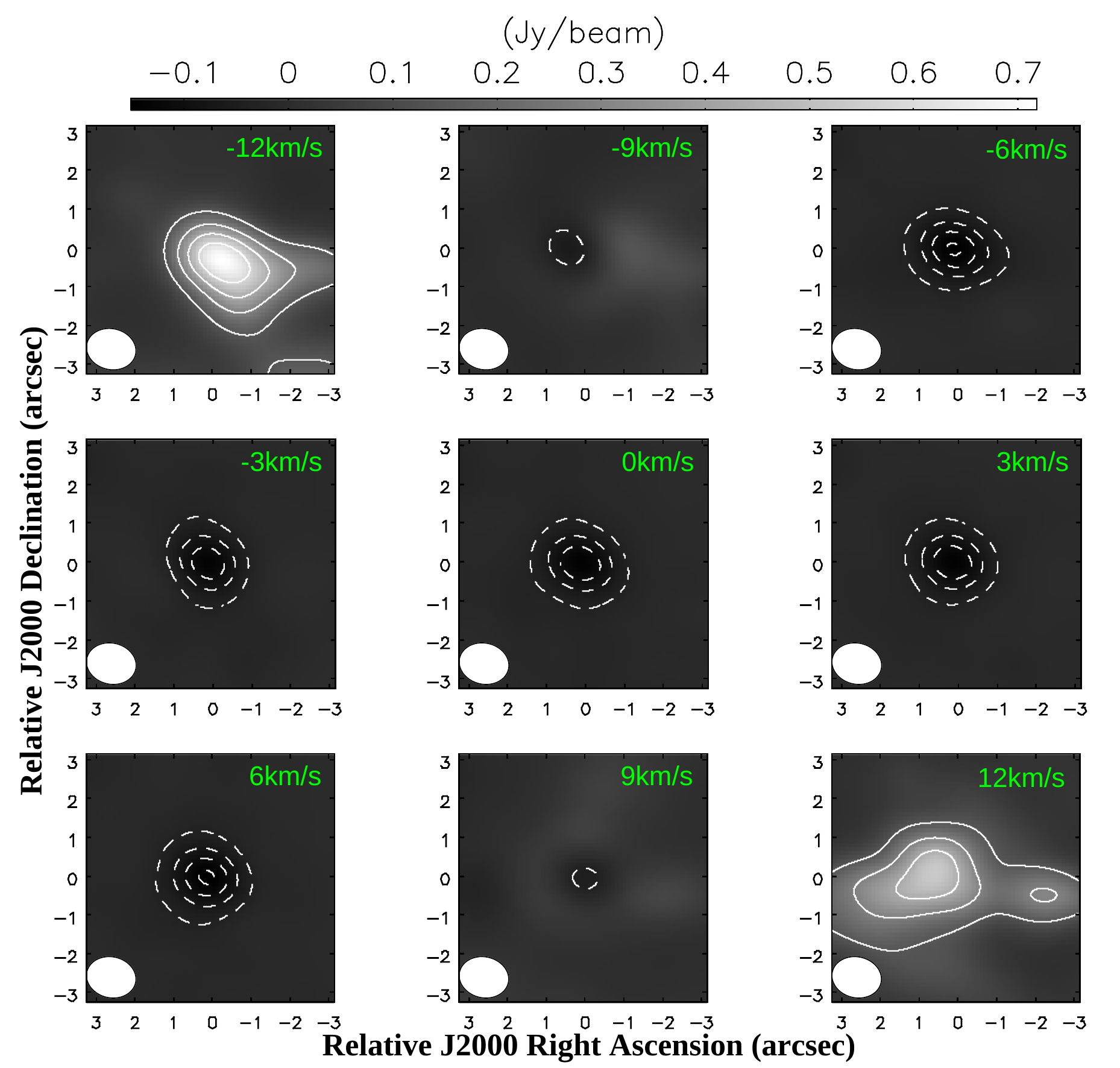}
  \end{minipage}
\hskip 0.5cm
\begin{minipage}{0.48\textwidth}
    \includegraphics[width=\textwidth]{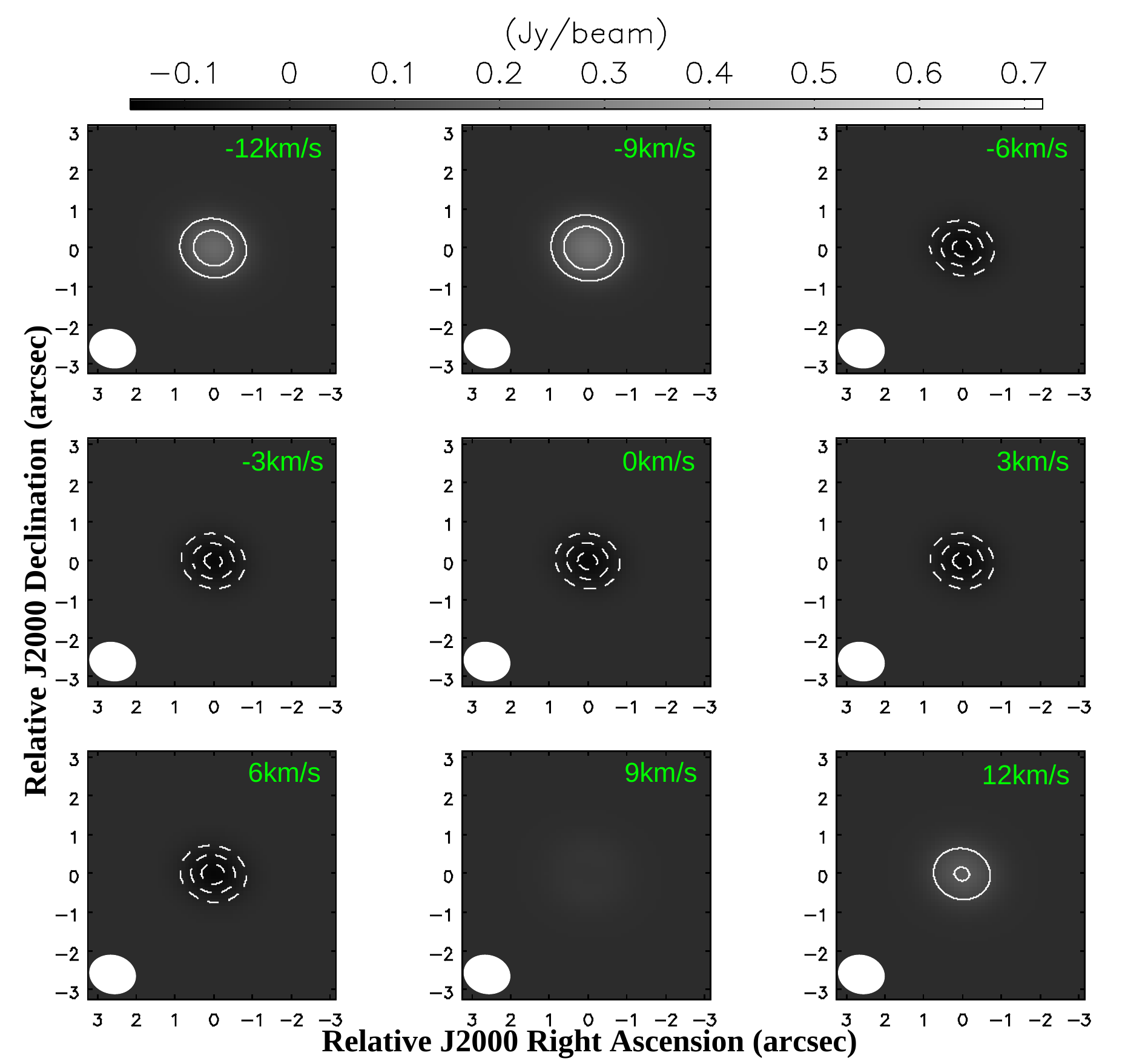}
  \end{minipage}
\caption{ Comparison of channel map emission of HCN between observation in left and model (constant abundance $\sim 7.6 \times 10^{-8}$) in right. Contour levels are at -5\%, -10\%, -15\%, -20\%, 20\%, 40\%, 60\%, and 80\% of the  peak observed intensity (0.719mJy/beam). Solid and dashed contours represent the emission and absorption, respectively.}
\label{fig:hcn-channel}
\end{figure*}

\subsubsection{HCN}
\cite{gora21} observed three hyperfine transitions of HCN, which appeared in absorption and were blended.
The observed spectral profile of HCN was broad.
They identified multiple transitions of HCN (F = 1$\rightarrow$1, 2$\rightarrow$1, and 1$\rightarrow$0 hyperfine components 
of the J=1$\rightarrow$0 transition). For the 1D RATRAN model, the physical input parameters described in Section 
\ref{sec:RATRAN} are considered. 
In Figure \ref{fig:cassis_hcn}, a good match between the modeled profile (orange line) and the observaed profile (black line) is obtained when an FWHM of $\sim 10$ km s$^{-1}$, constant abundance $\sim 7.6 \times 10^{-8}$, and $\beta=1.4$ are used.
The collisional rate between HCN and H$_2$ is used from the LAMDA database for the hyperfine transitions of HCN.
It is noticed that among the three hyperfine transitions of HCN, the F = 2$\rightarrow$1 (88.63184 GHz) transition is more robust. The obtained abundance distribution of HCN from our chemical model is further utilized to study. Figure \ref{fig:abundance} shows a slight jump in the abundance of HCN. Its peak abundance varies in between $2.8 \times 10^{-10}-5.4 \times 10^{-7}$. Our best-fitted constant abundance $\sim 7.6 \times 10^{-8}$ is well within this limit.
Inside the cloud, the final abundance of HCN shows a steady decreasing trend.
With the peak and final abundance obtained from the chemical modeling, the line profile of F = 2$\rightarrow$1 transition of HCN is generated. These are shown in Figure \ref{fig:cassis_hcn}  with the red and green curves, respectively. The red curve shows an excellent match with the observed line profile of HCN.
Figure \ref{fig:hcn-channel} shows the comparison of channel map emission between the
observation and modeling (by considering best-fitted constant abundance). A similar method is used in generating this simulated emission mentioned in section \ref{sec:hco+}. It depicts that our modeled channel map emissions are extended like the observed channel map emissions of HCN.

\begin{figure}
\includegraphics[height=6cm]{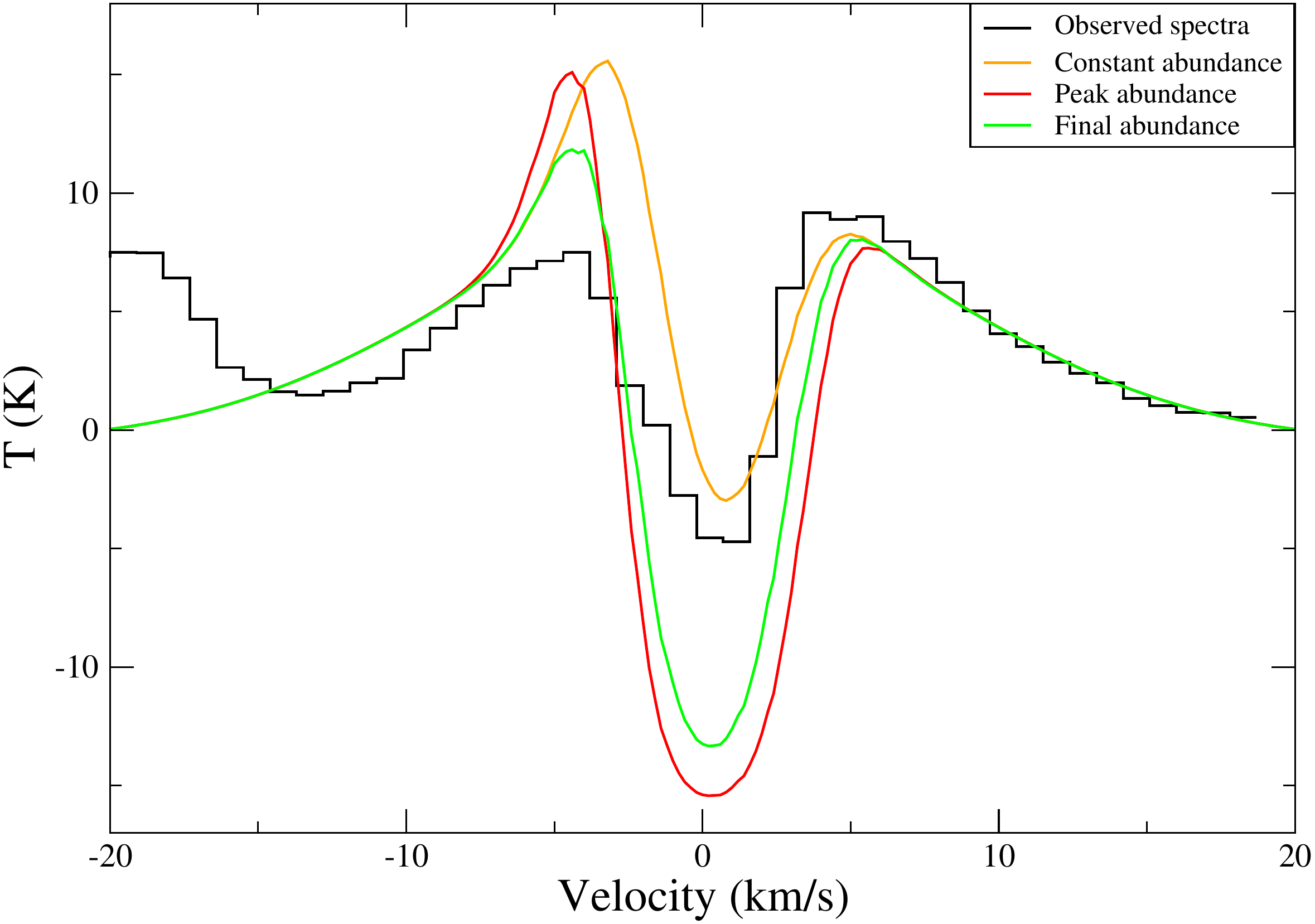}
\caption{A comparison between the observed and modeled SiO spectral profile in G31 is shown. The modeled line profile with the constant abundance ($9.5 \times 10^{-10}$) is shown in orange. The modeled line profiles with the peak and final abundance profiles are indicated with the red and green lines. The best fit was obtained when $\beta=1$ and an FWHM of $\sim 4.67$ km/s is used.}
\label{fig:sio-best}
\end{figure}

\subsubsection{SiO \label{sec:sio}}
The 5-4 \citep{belt18} and 2-1  \citep{maxi01,gora21} transitions of SiO were identified in G31.  
$\rm{SiO}$ is an excellent tracer of outflow present in the source.
 From the various SiO emissions, it is now known that there are at least three outflows associated with the G31 core:
E-W outflow to the south of the main core's dust continuum emission peak and is not associated with the two free-free embedded sources; N-S outflow, which could be related to the main core or related to one of the free-free embedded sources;
NE-SW outflow to the south of the main core, and it is not associated with it.
\cite{gora21} calculated the dynamical timescale from the velocity obtained of red-shifted and the blue-shifted lobe of $\rm{SiO}$ spectra. 
 They estimated the average dynamic timescale from the SiO observation $\sim 4.1 \times  10^3$ years (by considering $\sim 7.9$ kpc distance) and $1.92 \times 10^3$ years (by considering $\sim  3.7$ kpc distance).
Previously the molecular outflow present in the hot molecular core G31 is studied by \cite{olmi96} in $\rm{CO}$ 
molecule. Other studies \citep{ara08,belt18} in the literature hint at multiple outflow directions.

Here, the observed 2-1 transition \citep{gora21} of $\rm{SiO}$ is modeled by using the 1D RATRAN radiative transfer model.
The collisional rates of $\rm{SiO}$ with $\rm{H_2}$ is used from the LAMDA database.  
The physical structure considered for the modeling is described in Section \ref{sec:phys}. An additional outflow component in the modeling is also considered as the outflow highly influences siO spectra. It is to be noted that no outflow is considered in our physical model. Instead, a broad Gaussian
component is considered during the ray-tracing method when the ray passes from the
back half to the front half. The intensity in each velocity channel is then updated \citep{mot13}.
The constant abundance of $\rm{SiO}$ and the radial abundance profile obtained from our chemical model are both used in our model. With the constant abundance, the best fit is obtained when $\beta=1$, FWHM of $4.67$ km/s, and a constant abundance $9.5 \times 10^{-10}$ is used.

The intensity of the outflow component of $\sim 10$ K  and the FWHM of $\sim 20$ km s$^{-1}$ for the Gaussian outflow component are considered. 
In Figure \ref{fig:sio-best}, the observed line profile of the SiO (2-1) transition is shown in black, and modeled (constant abundance) line profile is shown in orange.
It depicts that the modeled spectra can successfully reproduce the observed absorption. However, the emission is a little stronger than this was observed.
The left panel of Figure \ref{fig:abundance} shows the radial distribution of the SiO abundance. 
It shows that the peak abundance of $\rm{SiO}$ is increased from 156 AU to 287 AU very rapidly, and beyond this, its peak abundance increased at a prolonged rate ($2.2 \times 10^{-10}-9.2 \times 10^{-9}$). Our best fitted constant abundance ($9.5 \times 10^{-10}$) of SiO is falling in between our modeled peak abundance range.
The final abundance shows ups and down in the abundance profile and significantly differs from its peak abundance. 
In Figure \ref{fig:sio-best}, modeled spectra with the peak abundance profile (red line) and final abundance profile (green line) are shown. Figure \ref{fig:sio-best} depicts that with the modeled peak abundance profile, the shape of the observed line profile can be reproduced. Our chemical model does not include shock, which may be vital for SiO-related chemistry.

\begin{figure*}
  \centering
\begin{minipage}{0.40\textwidth}
    \includegraphics[width=\textwidth]{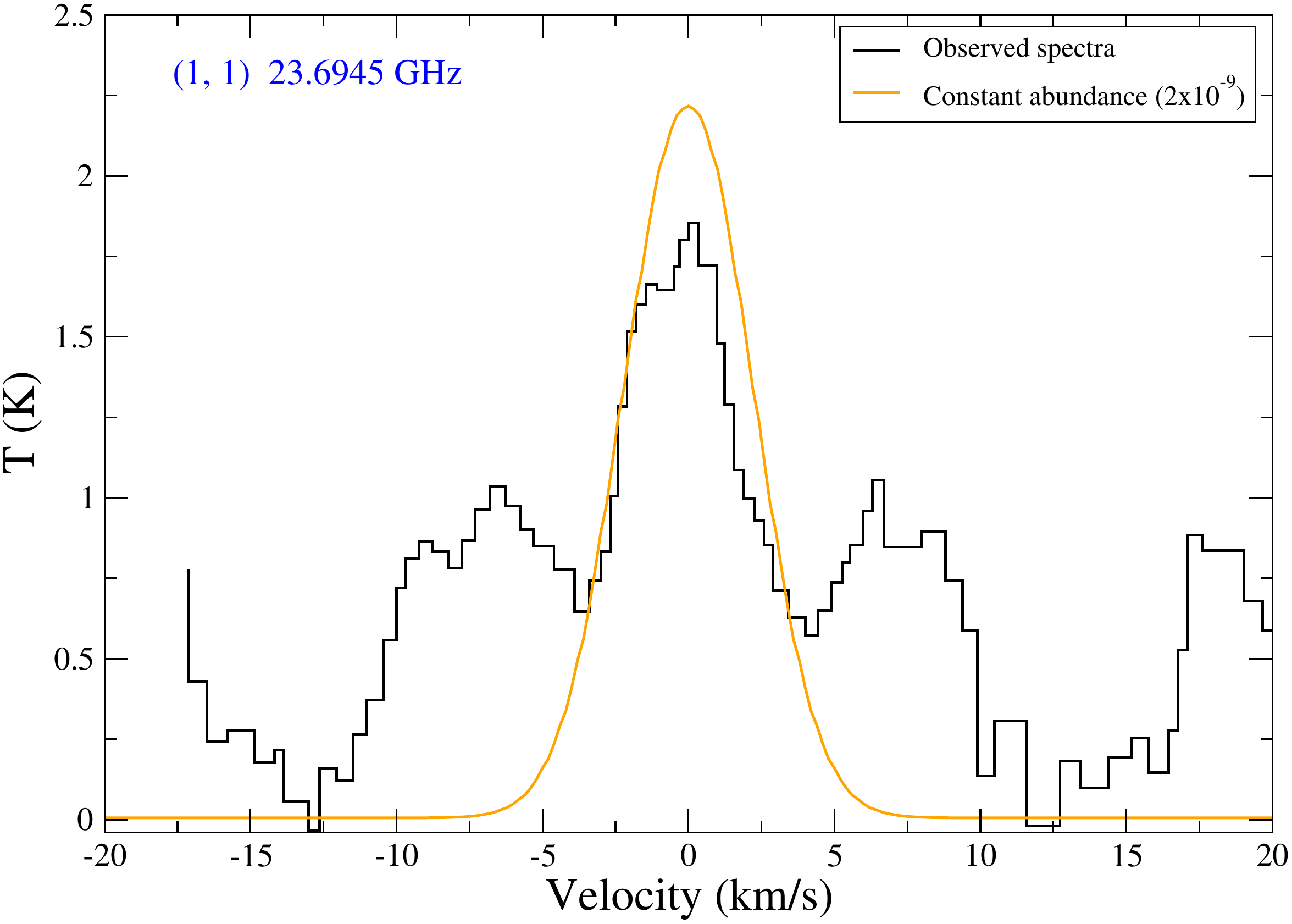}
  \end{minipage}
\hskip 0.5cm
  \begin{minipage}{0.40\textwidth}
    \includegraphics[width=\textwidth]{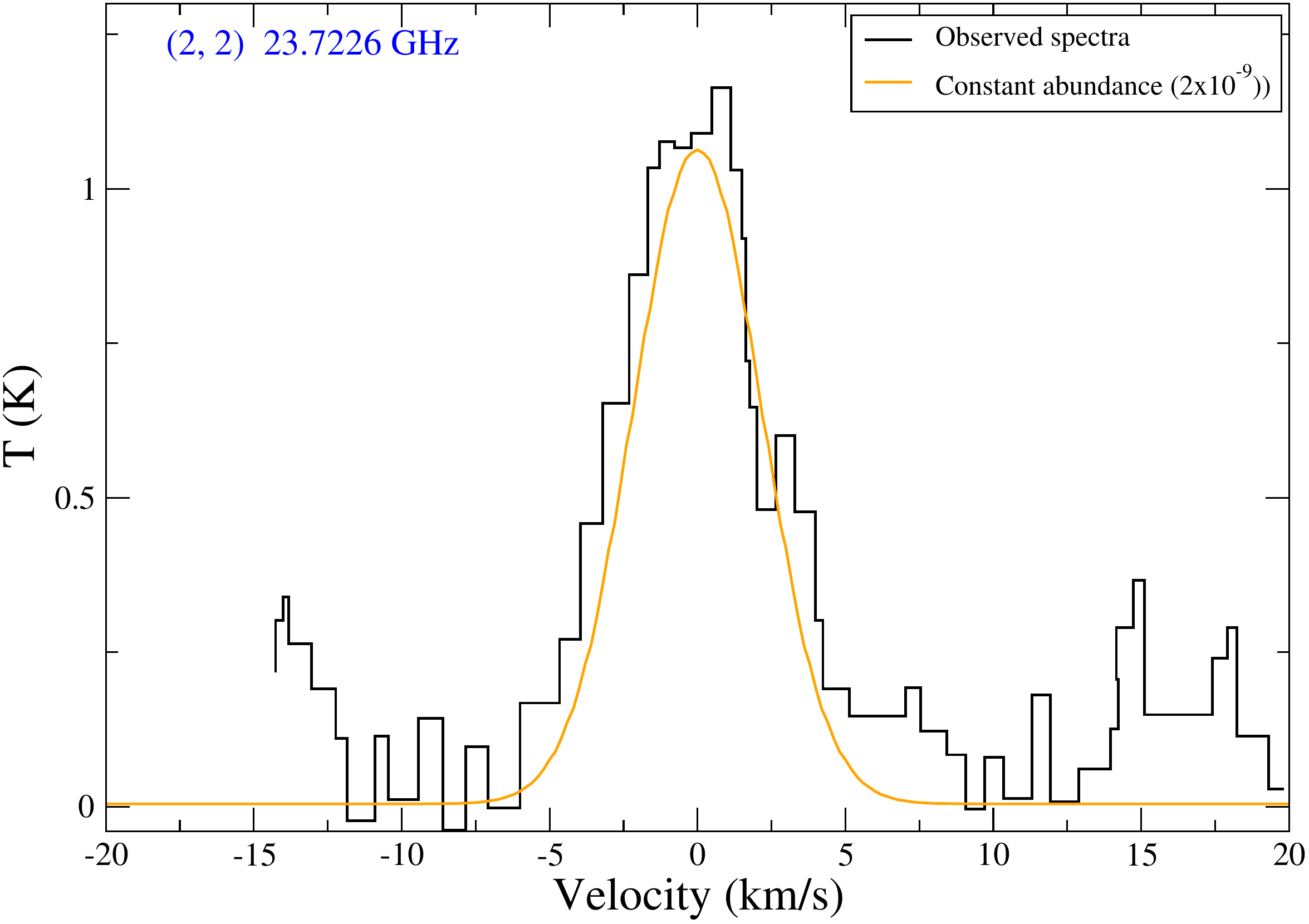}
  \end{minipage}
  \begin{minipage}{0.40\textwidth}
    \includegraphics[width=\textwidth]{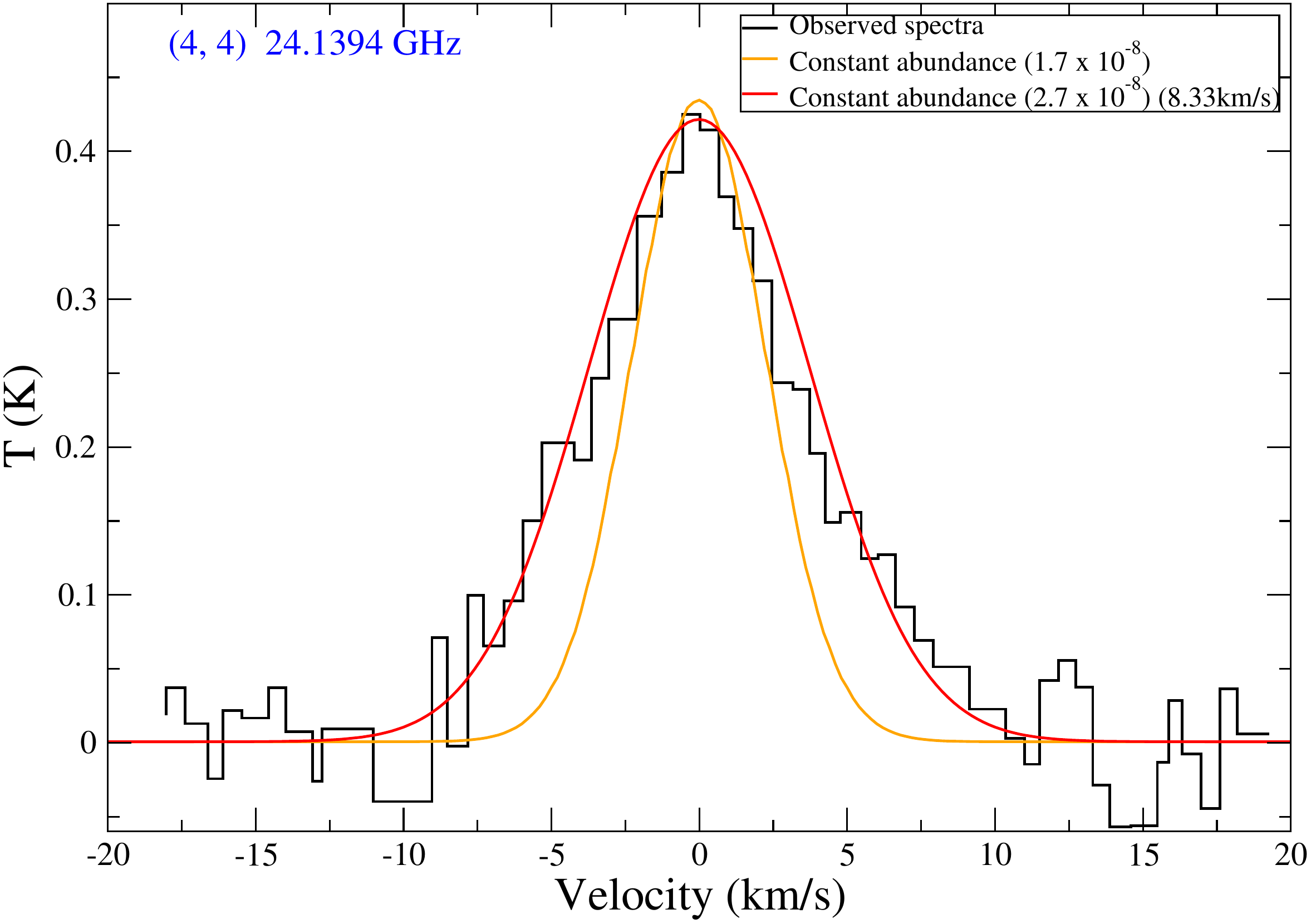}
  \end{minipage}
\hskip 0.6cm
  \begin{minipage}{0.40\textwidth}
    \includegraphics[width=\textwidth]{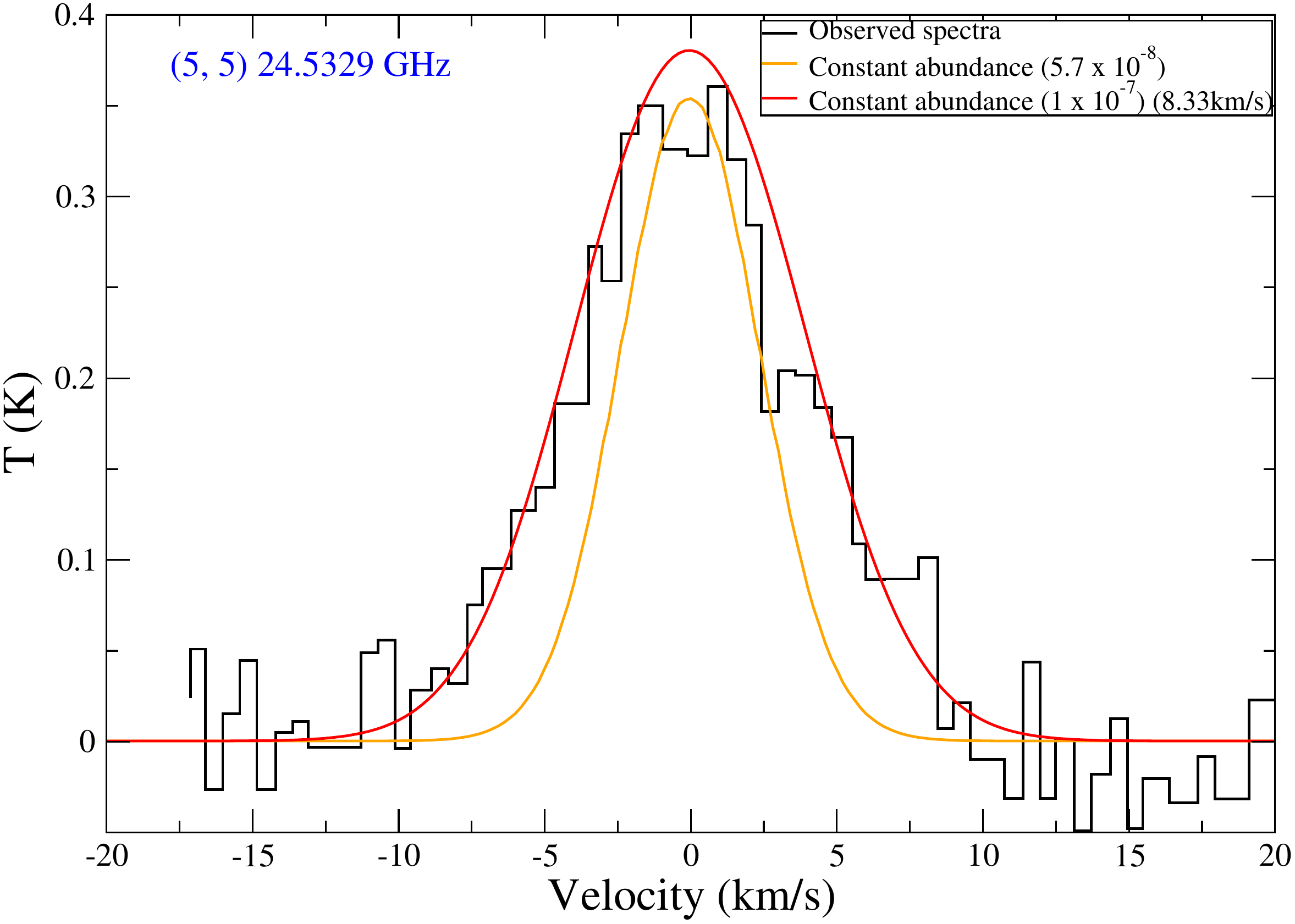}
  \end{minipage}
 \caption{ A comparison between the observed line profile (black) and synthetic line profiles (orange) of
 (a) 1,1 , (b) 2,2 , (c) 4,4 , (d) 5,5 transitions of $\rm{NH_3}$ are shown. These line profiles are generated with an FWHM of $\sim 4.9$ km/s, $\beta=1.4$ and a constant abundance of $\sim 2 \times 10^{-9}$ for 1,1 and 2,2 transitions, a constant abundance $\sim 1.7 \times10^{-8}$ for 4,4 transition and $ \sim 5.7\times10^{-8}$ for 5,5 transition. For 4,4 and 5,5, the best line profiles (red curves) with little higher FWHM ($8.33$ km/s) appeared with the constant abundance $2.7 \times 10^{-8}$ and $1.0 \times 10^{-7}$, respectively.}
 \label{fig:nh3_best}
\end{figure*}

\begin{figure}
  \centering
    \includegraphics[width=7cm]{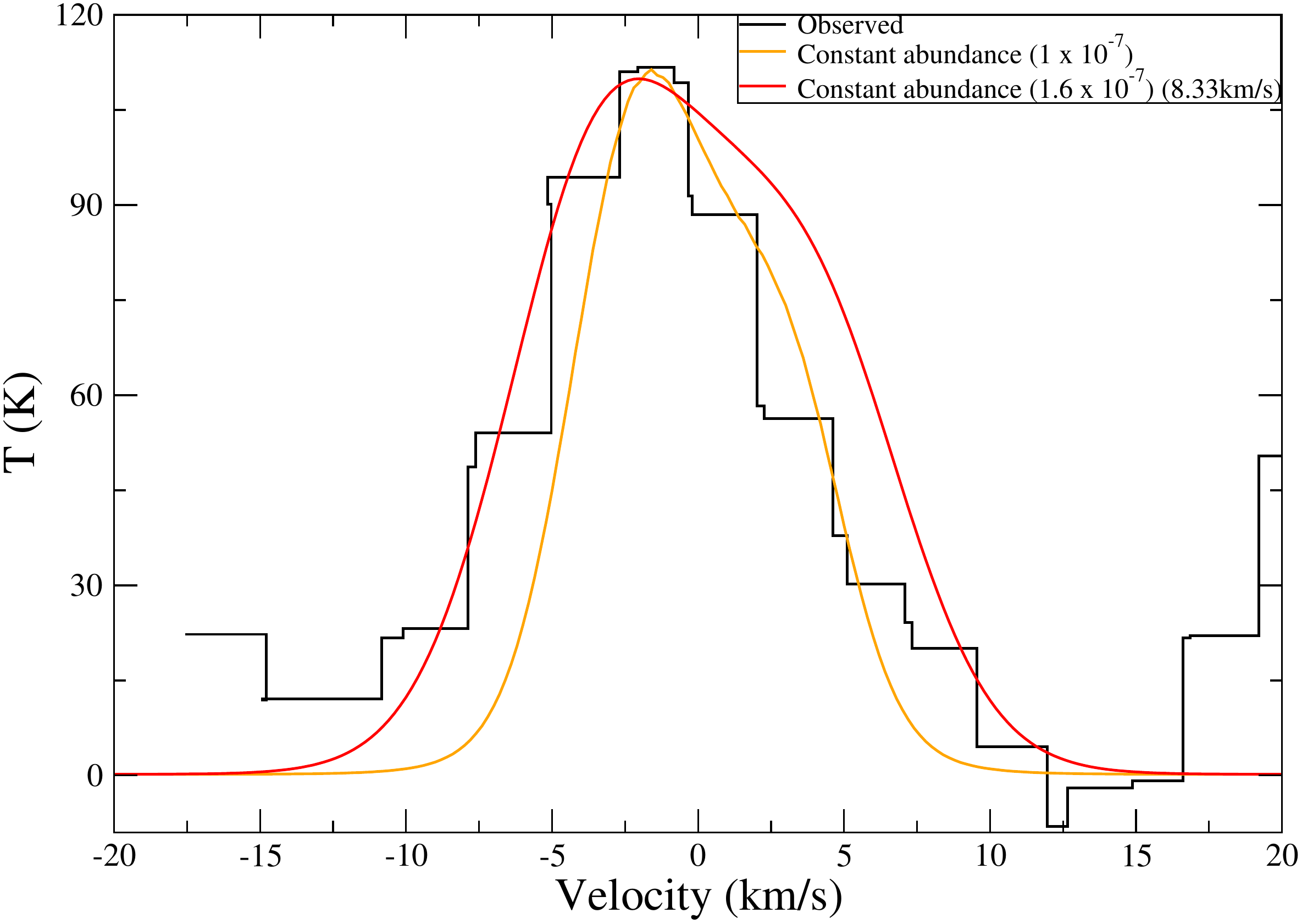}
 \caption{ A comparison between the high spatial resolution (0.63$^{''}$) VLA observation of the (4,4) transition (black) of NH$_3$ and modeled line profile
 (orange) of $\rm{NH_3}$ is shown. The synthetic line profile is generated with an FWHM of $\sim 4.9$ km/s, $\beta=1.4$, and a constant abundance $\sim 1 \times 10^{-7}$. With a little higher FWHM (8.33 km/s, red curve), the best fitted abundance is $1.6 \times 10^{-7}$.}
 \label{fig:nh3_best44}
 \end{figure}

\subsubsection{$\rm{NH_3}$}
\label{sec:NH3}
The UC HII region is situated beside the G31 core and is separated by an angular distance of $5^{''}$. Due to this small separation, 
the observations having angular resolution $>5^{''}$, often contaminated with the HII region. 
\cite{ces92} used a single dish $100$ m telescope (GBT) having a beam size of $\sim 40^{''}$ to study the G31 HMC region.
Multiple inversion transitions of ammonia were observed by \cite{ces92} and \cite{chu90}.
\cite{oso09} had modeled all these inversion transitions with the radiative transfer model. 
They first modeled these transitions starting with a constant abundance of $\rm{NH_3}$. 
However, they also parameterized the abundance profile depending upon the condensation and sublimation temperature of NH$_3$ and water. 
\cite{oso09} used the following relation to generate the abundance profile of $\rm{NH_3}$,
\begin{equation}
 X_{mol}=\frac{X_{max}-X_{min}}{1+\eta}+X_{min}.
\end{equation}
Here, $\eta$ represents the ratio of the species in the solid phase with that in the gas phase.
The total amount of the molecular species was assumed to be constant.
They considered that the variation of $\eta$ depends on the condensation and sublimation temperature and the density. 
They considered $\eta \gg 1$ for the minimum gas-phase abundance ($X_{min}$), whereas $\eta=0$ is considered for the maximum gas-phase abundance ($X_{max}$).
The sublimation effect of NH$_3$ was also considered.
\cite{oso09} obtained a good correlation with the observational results when they used 
$X_{max}=3\times10^{-6}$ and $X_{min}=2\times10^{-8}$.
Figure \ref{fig:abundance} shows a jump profile for the abundance of NH$_3$. 
A peak abundance variation in between $3.2 \times 10^{-8}-6.4 \times 10^{-7}$ is obtained from our model.
Its abundance
is increasing deep inside the cloud. \cite{oso09} used a minimum constant abundance (X$_{min}$) $\sim 2 \times 10^{-8}$ 
when the temperature is roughly below $90$ K. In the inner envelope, the temperature gradually increases, and
they used a constant maximum abundance ($X_{max}$) 
of $\sim 3 \times 10^{-6}$ for temperature greater than $100$ K. 
From our chemical model 
(see the left panel of Figure \ref{fig:abundance}), in the outer part of the envelope, peak NH$_3$ abundance varies in between $\sim 5.0 \times 10^{-8} - 2.3 \times 10^{-7}$. 
This abundance is obtained beyond $\sim 8000$ AU, corresponding to the region of temperature falling below $93$ K. 
Inside, a sudden jump in the
abundance profile is noticed and varies in between $\sim 3.4 \times 10^{-8} -6.0  \times 10^{-7}$. Inside $300$ AU, the peak abundance
of NH$_3$ heavily decreased.
The final abundance and peak abundance of NH$_3$ show more or less similar trends. 
It is exciting to note that a striking match between our obtained peak abundance profile and that used in \cite{oso09} is obtained when the gas and grain temperature at the initial stage of our model is kept at $20$K instead of $15$K 
reported here. 
When an initial temperature $\sim 20$ K is used, a peak abundance $\sim 1.7 \times 10^{-6}$  is obtained, which is more closer to the X$_{max}=3 \times 10^{-6}$ used in \cite{oso09}.  
\cite{oso09} did not carry out any chemical model to obtain this abundance. So a complete match with \cite{oso09} is not expected. 

\cite{chu90,ces92} used $100$ m telescope for the identification of
$\rm{NH_3}$(1,1), $\rm{NH_3}$(2,2), $\rm{NH_3}$(4,4), and $\rm{NH_3}$(5,5)
transitions. Here, all these transitions are modeled by considering a collapsing envelope.
Since \cite{ces92,chu90} observed the transitions of NH$_3$ with a $100$ m single dish, the obtained spectra is convolved by using a 40$^{''}$ beam size and have compared it with the observed one. 
Figure \ref{fig:nh3_best} shows the modeled (1,1),
(2,2), (4,4), and (5,5) transitions of NH$_3$ along with their observed (black) line profiles.
The collisional rate of $\rm{p-NH_3}$ is taken from the LAMDA database. 

The obtained line profile is convolved with the 40$^{''}$ beam size of the GBT $100$ m telescope to compare it directly with the observed spectra. 
In Figure \ref{fig:nh3_best}, the modeled spectra (with constant abundance, orange) along with the observed one (black) 
is shown.
It is noticed that a constant abundance can not able to explain the four observed line profiles simultaneously.  
For the $(4,4)$ transition, \cite{ces92} obtained an FWHM of $\sim 4.9\pm 0.1$ km s$^{-1}$. Here also, an FWHM of $\sim 4.9$ km s$^{-1}$ yields a good fit. 
 The best fit with the observed spectrum is obtained when a constant NH$_3$ abundance of $\sim 2 \times 10^{-9}$ for the (1,1) and (2,2) transitions is used. The good fit for the (4,4) and (5,5) transitions is obtained when a constant abundance of $1.7 \times 10^{-8}$ and $5.7 \times 10^{-8}$ is used, respectively.
It is noticed that with a higher FWHM ($\sim 8.33$ km/s), the (4,4) and (5,5) transitions shows a better fit with $2.7 \times 10^{-8}$ and $1.0 \times 10^{-7}$, respectively.
As in our case, \cite{oso09} were also unable to match all the peak intensities simultaneously with one abundance profile (see Figure 6 of \cite{oso09}). 
Here, comparatively, a higher constant abundance is needed to explain the (4,4) and (5,5) transitions. 
It is because the (1,1) and (2,2) transitions have comparatively lower upper state energies ($1.14$ K and $42.32$ K respectively) compared to the (4,4) and (5,5) transitions ($178.39$ K and $273.24$ K). Due to these differences in upper state energies, it is expected that the (4,4) and (5,5) transitions would originate comparatively from the warmer (i.e., inner) region of the envelope. From Figure \ref{fig:abundance}, we can see that NH$_3$ abundance is higher in the inner area of the envelope compared to the outer part. So the usage of the higher abundance for (4,4) and (5,5) transition is justified. 

In addition to the constant abundance, the peak abundance variation obtained from our chemical model is further used.  It is noticed that our modeled line profiles are unable to fit the observed transitions. The intensity of all the transitions is overproduced with our modeled abundance profile.
 
 In Figure \ref{fig:nh3_best44}, our modeled (4,4) transition is compared with that obtained with the VLA observation  (angular resolution 0.63$^{''}$) \citep{cesa98}. The best fit is obtained when a constant abundance of $\sim 1 \times 10^{-7}$, $\beta=1.4$, and FWHM $4.9$ km/s is used.
 With a little higher FWHM ($\sim 8.33$ km/s), the best fitted is obtained with an abundance $\sim 1.6 \times 10^{-7}$.

\begin{figure}
\includegraphics[height=5cm,width=8cm]{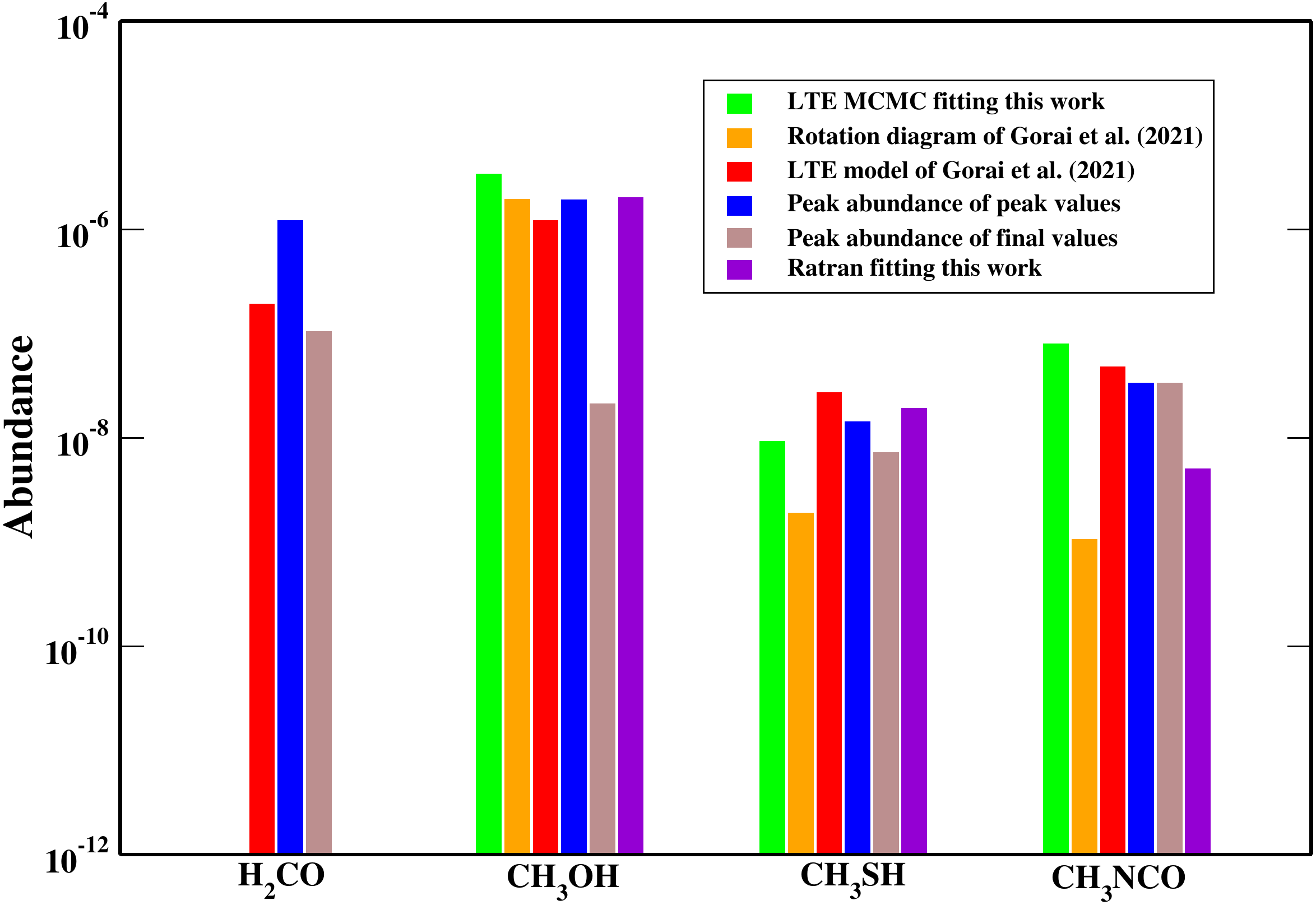}
\caption{Comparison between the observed (LTE and rotation diagram method of \cite{gora21} and MCMC fitting discussed in section \ref{sec:MCMC} (for CH$_3$OH average of A-CH$_3$OH and E-CH$_3$OH is taken from Table \ref{table:mcmc_lte}), RATRAN fitting) and simulated abundance is shown. Maximum values obtained from the peak abundance profile shown in Table \ref{table:abundances} is shown in blue, whereas the maximum abundance obtained from the final abundances is shown in brown.}
\label{fig:comp}
\end{figure}

\begin{figure*}
\begin{minipage}{0.30\textwidth}
\includegraphics[width=\textwidth]{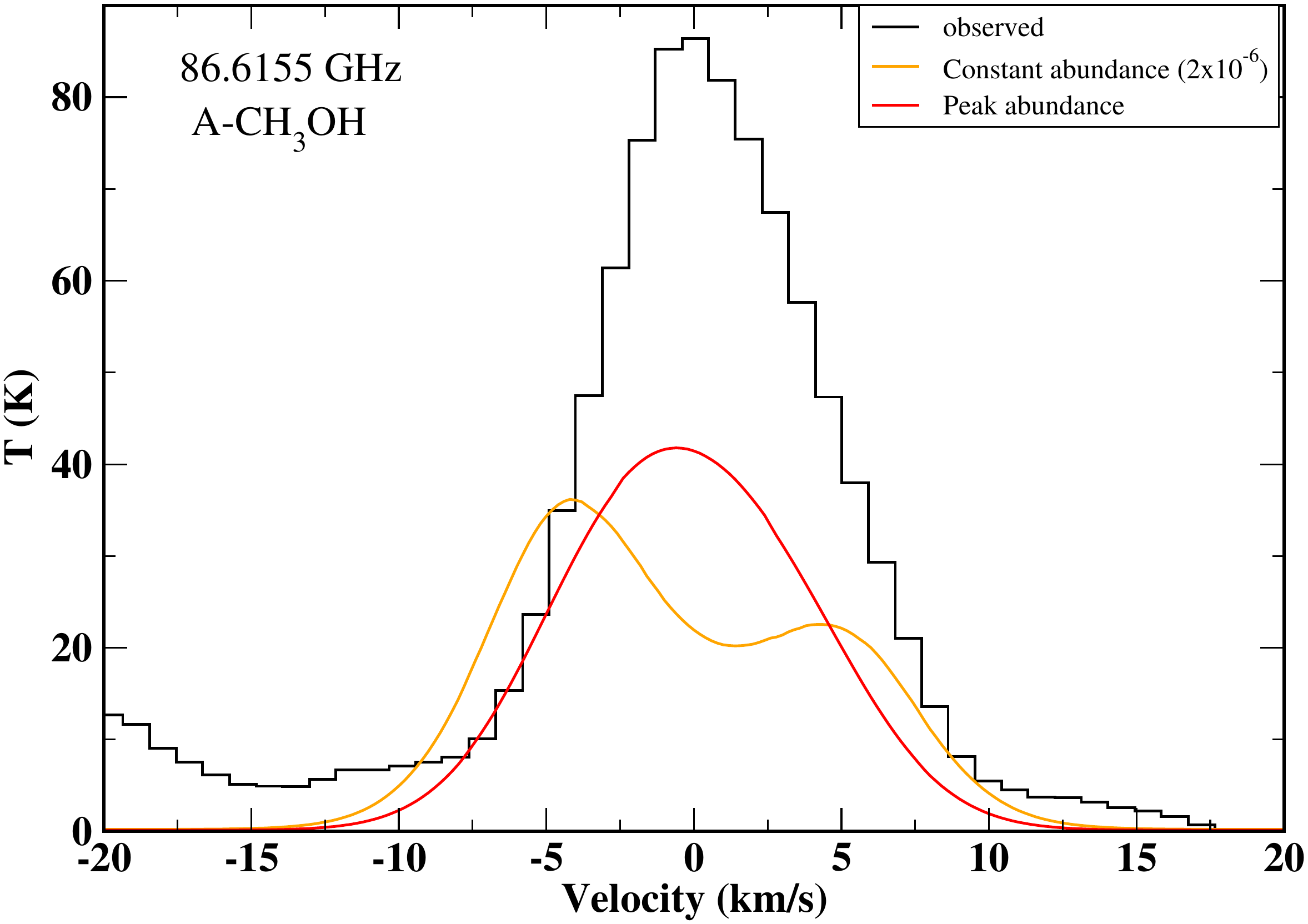}
\end{minipage}
\begin{minipage}{0.30\textwidth}
\includegraphics[width=\textwidth]{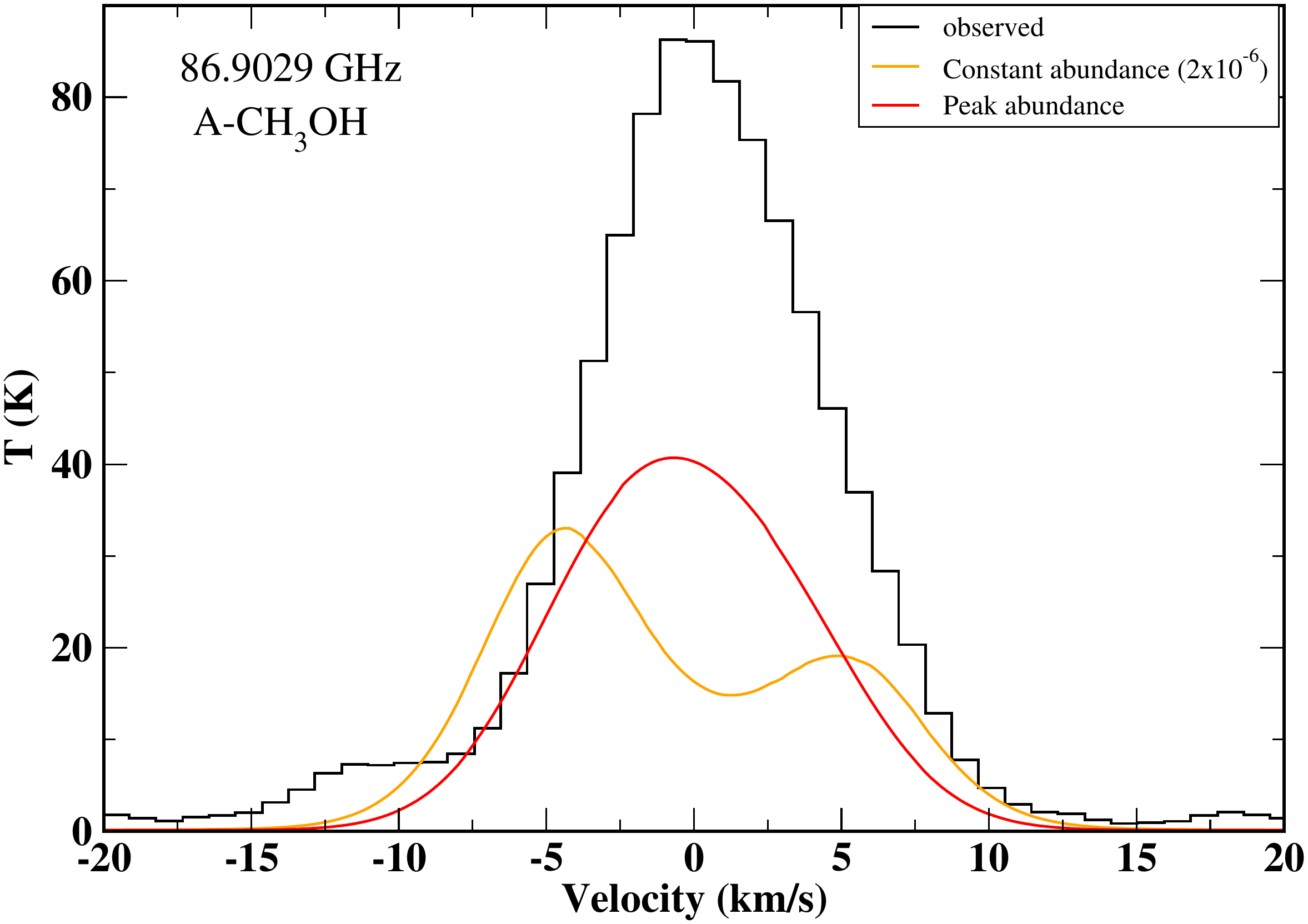}
\end{minipage}
 \begin{minipage}{0.30\textwidth}
 \includegraphics[width=\textwidth]{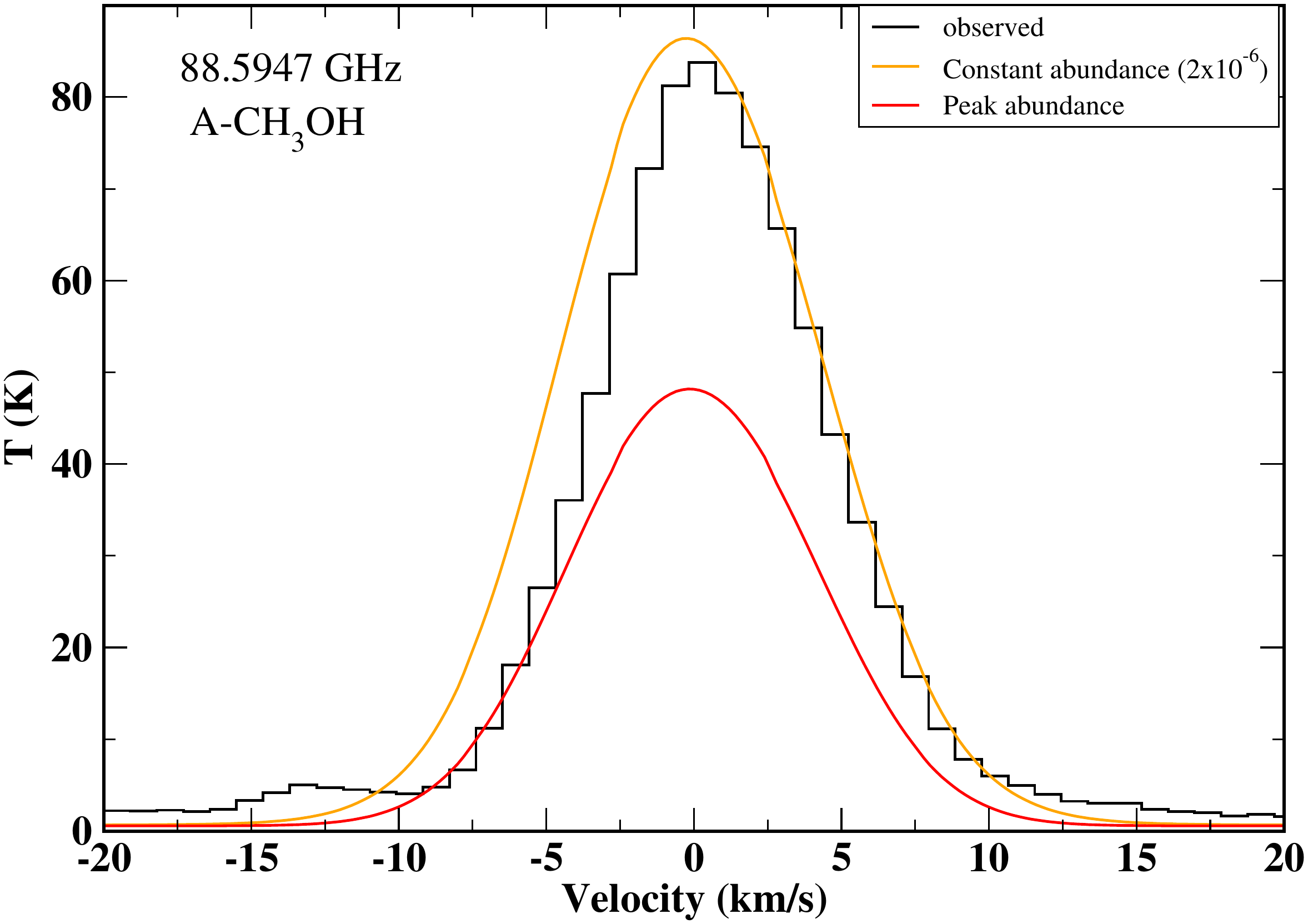}
 \end{minipage}
\hskip -0.9cm
\begin{minipage}{0.30\textwidth}
\includegraphics[width=\textwidth]{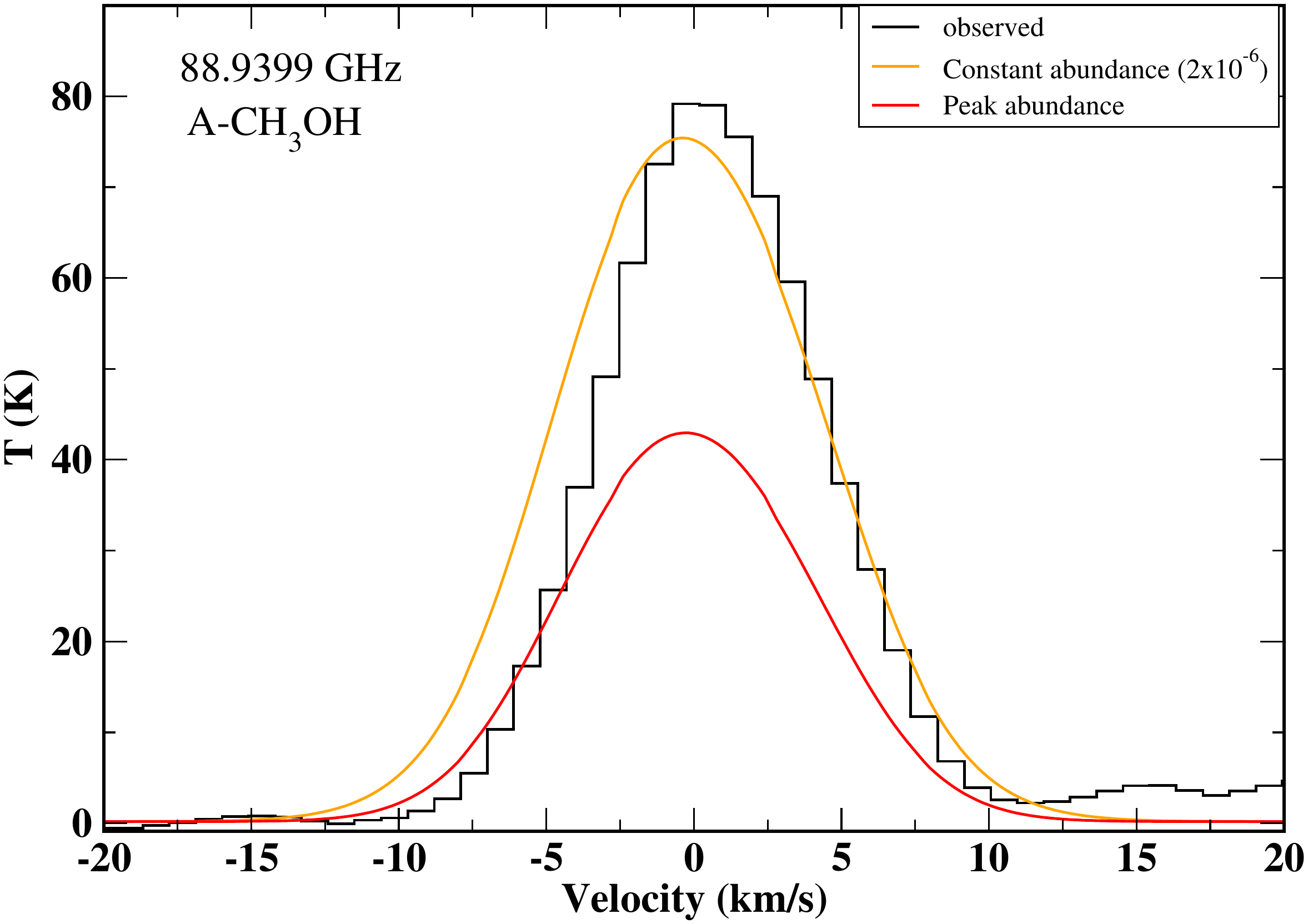}
\end{minipage}
\begin{minipage}{0.30\textwidth}
\includegraphics[width=\textwidth]{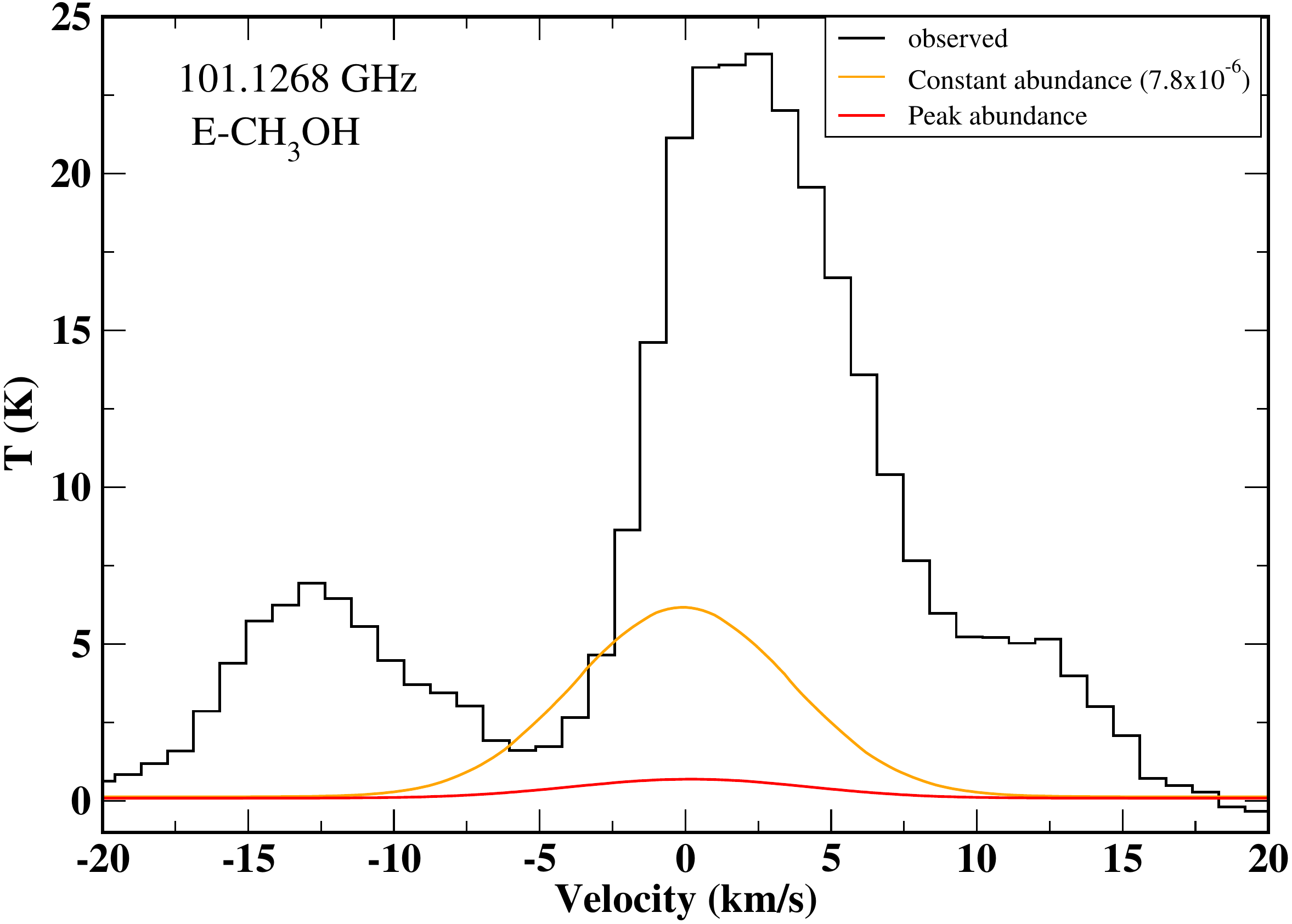}
\end{minipage}
\begin{minipage}{0.30\textwidth}
\includegraphics[width=\textwidth]{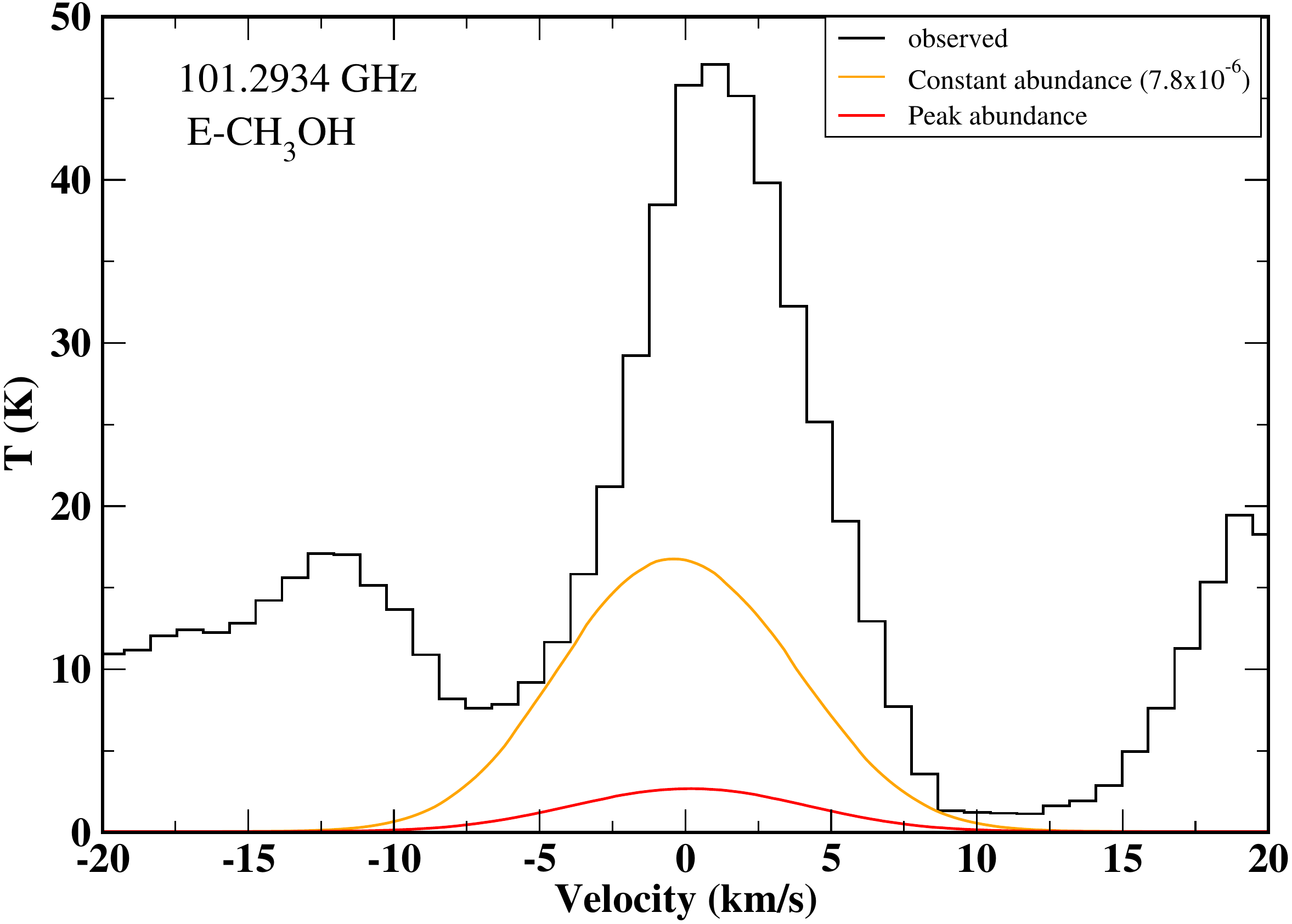}
\end{minipage}
\hskip -0.9 cm
\begin{minipage}{0.30\textwidth}
\includegraphics[width=\textwidth]{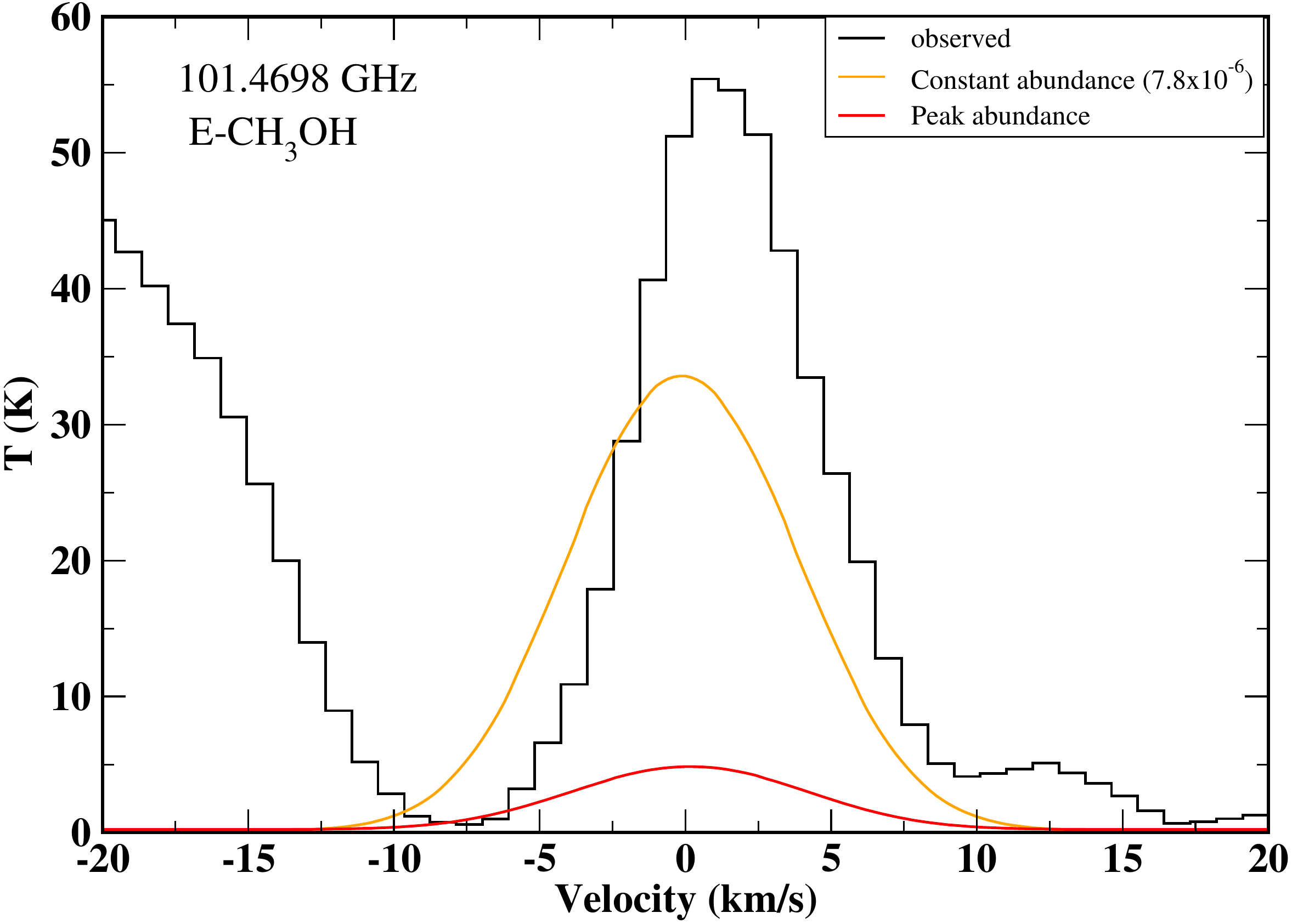}
\end{minipage}
\hskip 0.6cm
\caption{The modeled transitions of $\rm{A-CH_3OH}$ (first four transitions) and $\rm{E-CH_3OH}$ (last three transitions) along with their observed transitions (black lines) are shown. The line profile obtained with the constant abundance is displayed with the orange lines. The best fit for A-CH$_3$OH is obtained when a methanol abundance of $\sim 2 \times 10^{-6}$, an FWHM $\sim 8$ km/s, and $\beta=1.4$ are used. 
For E-CH$_3$OH, the FWHM and $\beta$ are kept the same as A-CH$_3$OH, but a constant abundance of $\sim 7.8 \times 10^{-6}$ is used. The red lines in the figure show the modeled line profiles when the peak spatial distribution of the methanol abundances from Fig. \ref{fig:abundance} is used.}
\label{fig:ch3oh-ratran}
\end{figure*}

\begin{figure*}
\begin{minipage}{0.30\textwidth}
\includegraphics[width=\textwidth]{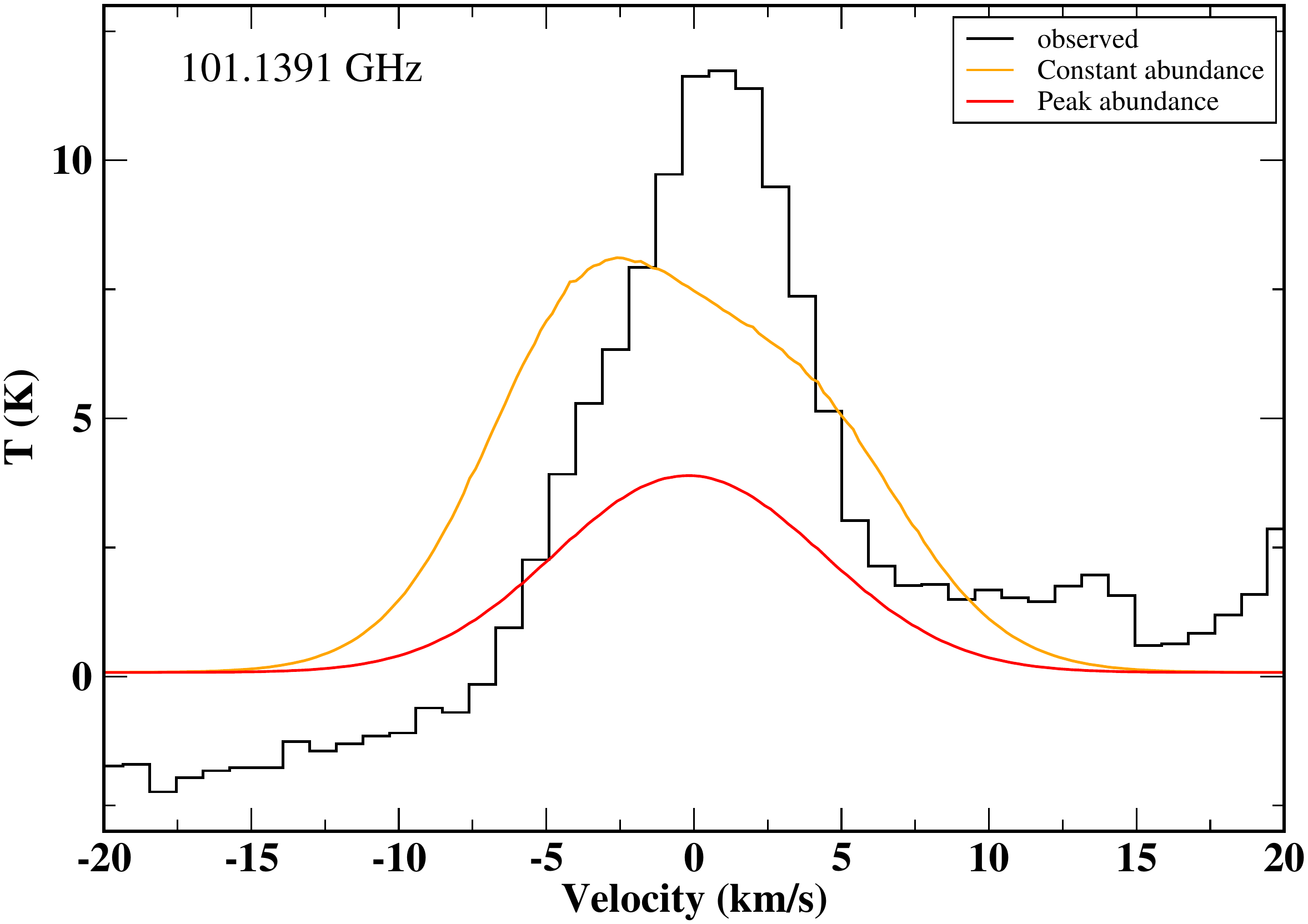}
\end{minipage}
\begin{minipage}{0.30\textwidth}
\includegraphics[width=\textwidth]{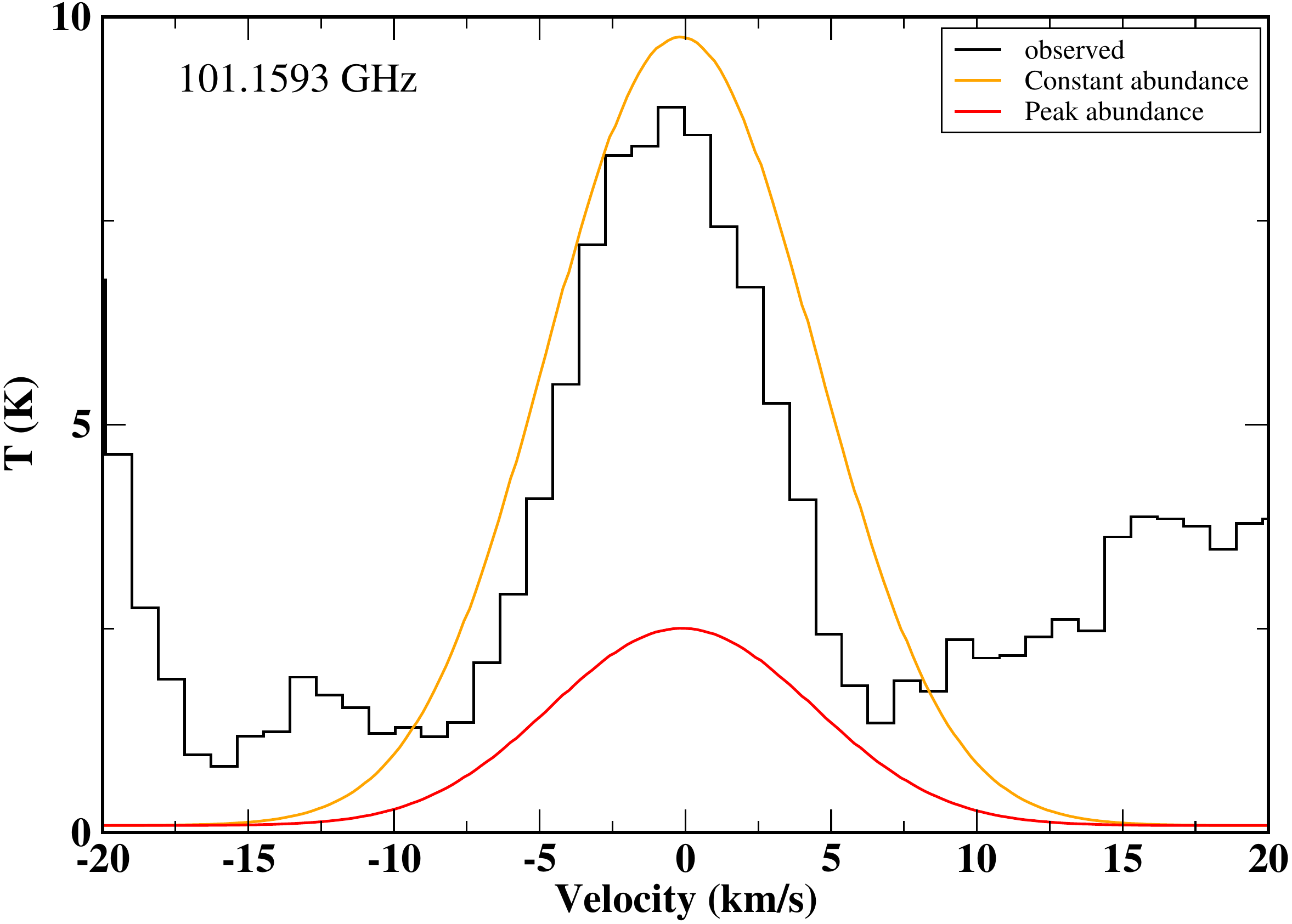}
\end{minipage}
 \begin{minipage}{0.30\textwidth}
 \includegraphics[width=\textwidth]{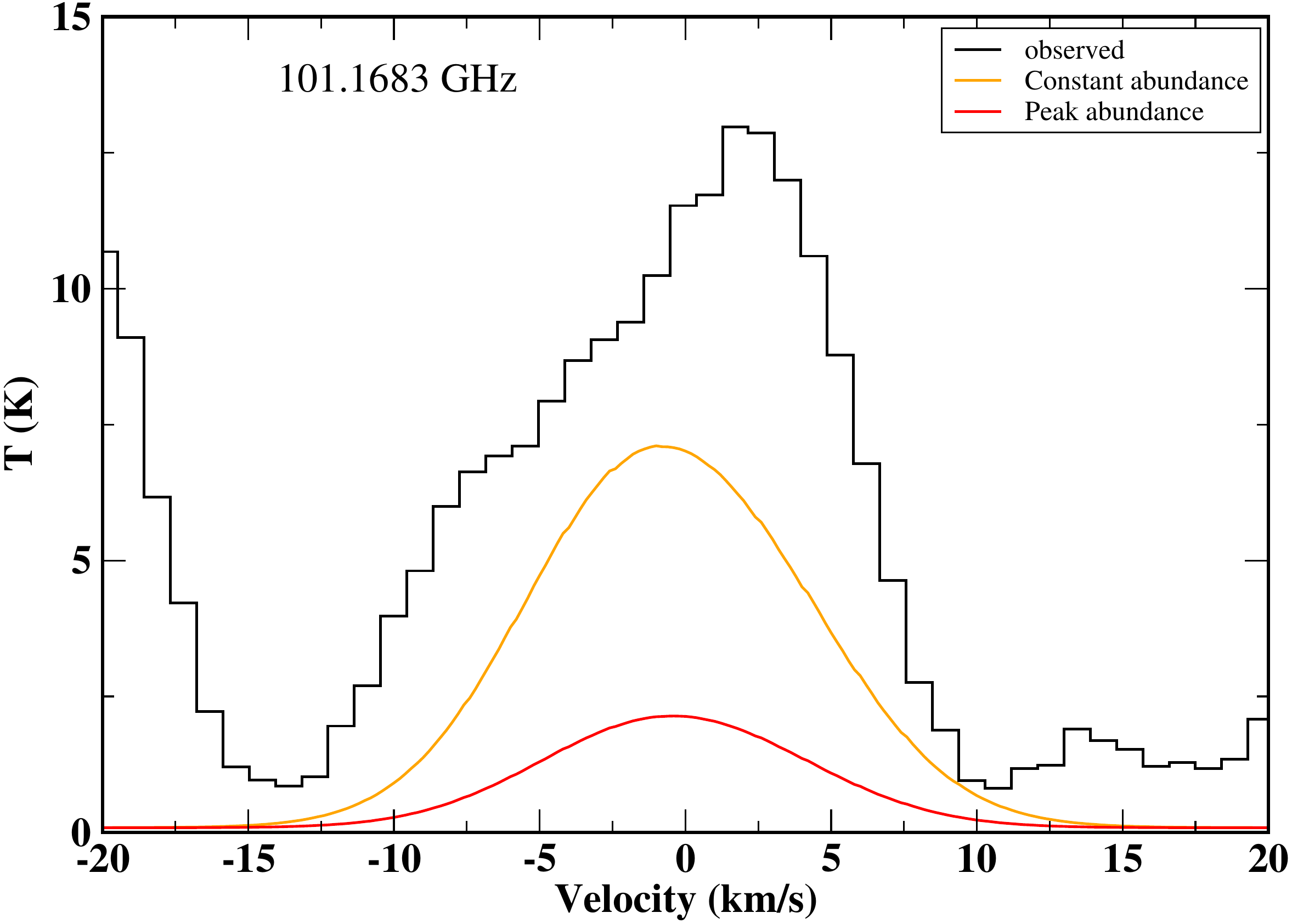}
 \end{minipage}
\hskip -0.9cm
\caption{The modeled line profiles of CH$_3$SH along with its observed line profiles are shown.
The best fit is obtained when a constant abundance $\sim 1.9 \times 10^{-8}$, an FWHM $\sim 9.45$ km/s, and $\beta=1.4$ are used. Observed line profiles are shown in black, whereas the modeled line profiles with the constant abundance are shown in orange. The line profiles obtained with the peak spatial distribution of the abundance profile are shown in red.  
}
\label{fig:ch3sh-ratran}
\end{figure*}

\begin{figure*}
\begin{minipage}{0.30\textwidth}
\includegraphics[width=\textwidth]{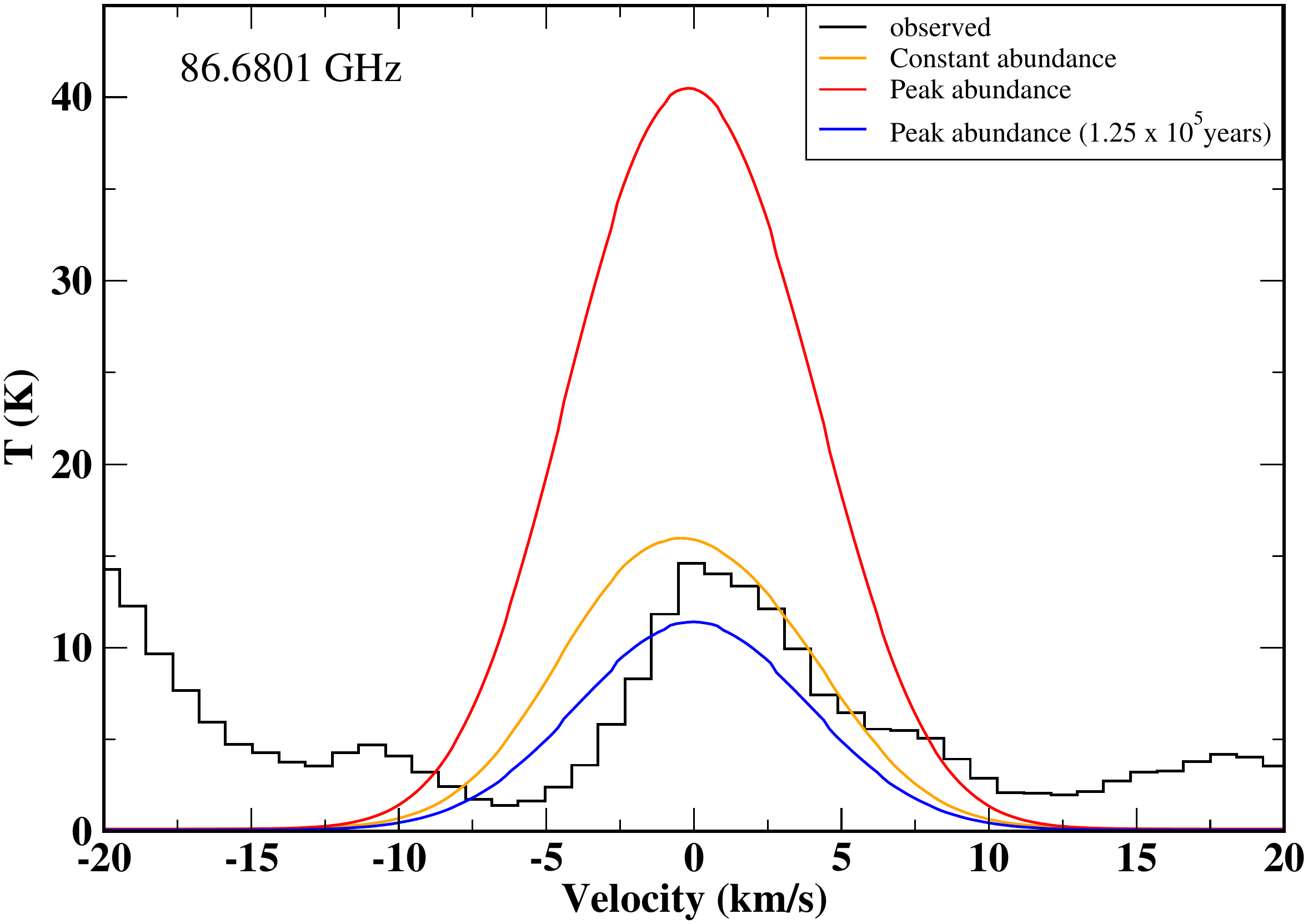}
\end{minipage}
\begin{minipage}{0.30\textwidth}
\includegraphics[width=\textwidth]{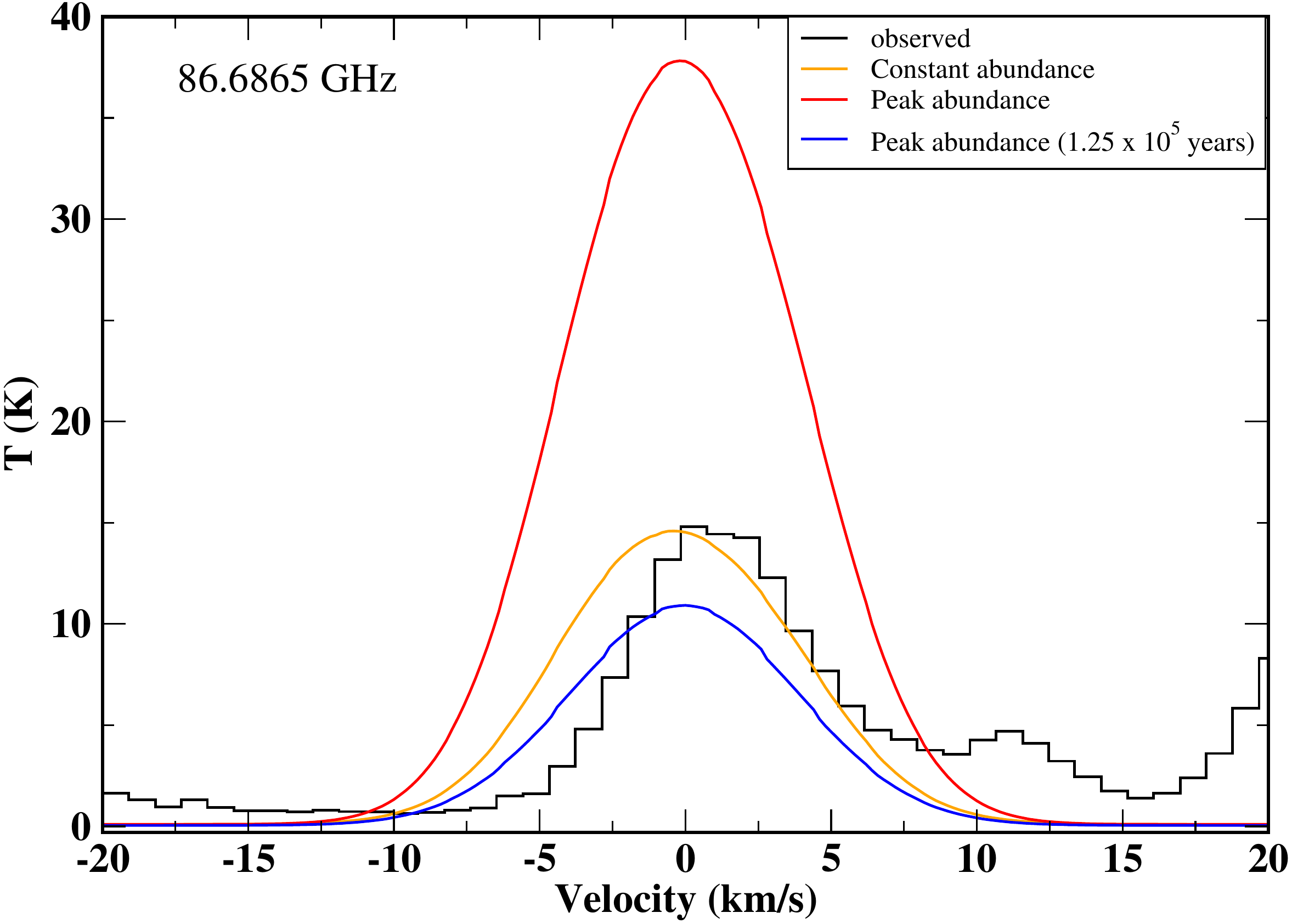}
\end{minipage}
 \begin{minipage}{0.30\textwidth}
 \includegraphics[width=\textwidth]{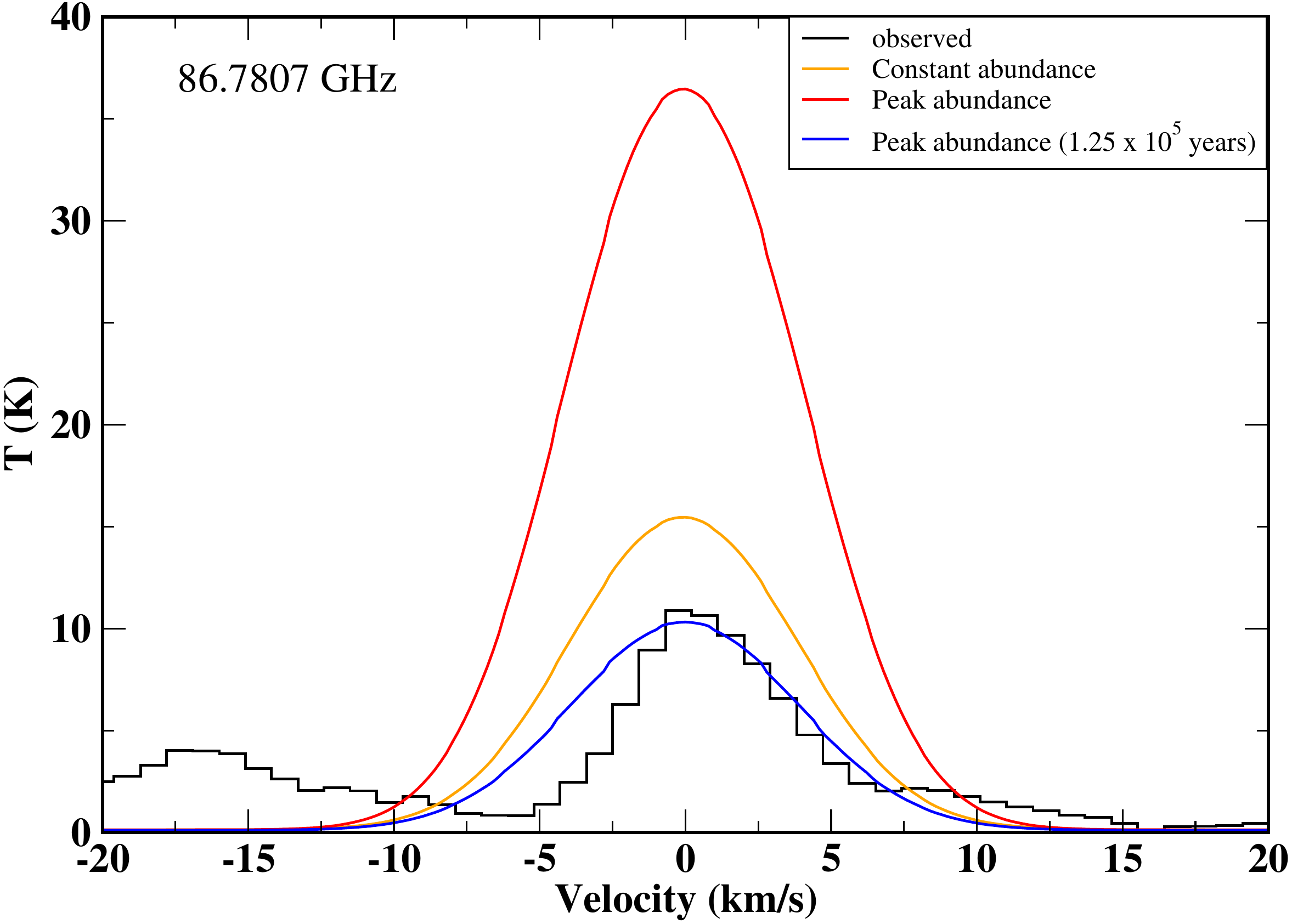}
 \end{minipage}
\hskip -0.9cm
\begin{minipage}{0.30\textwidth}
\includegraphics[width=\textwidth]{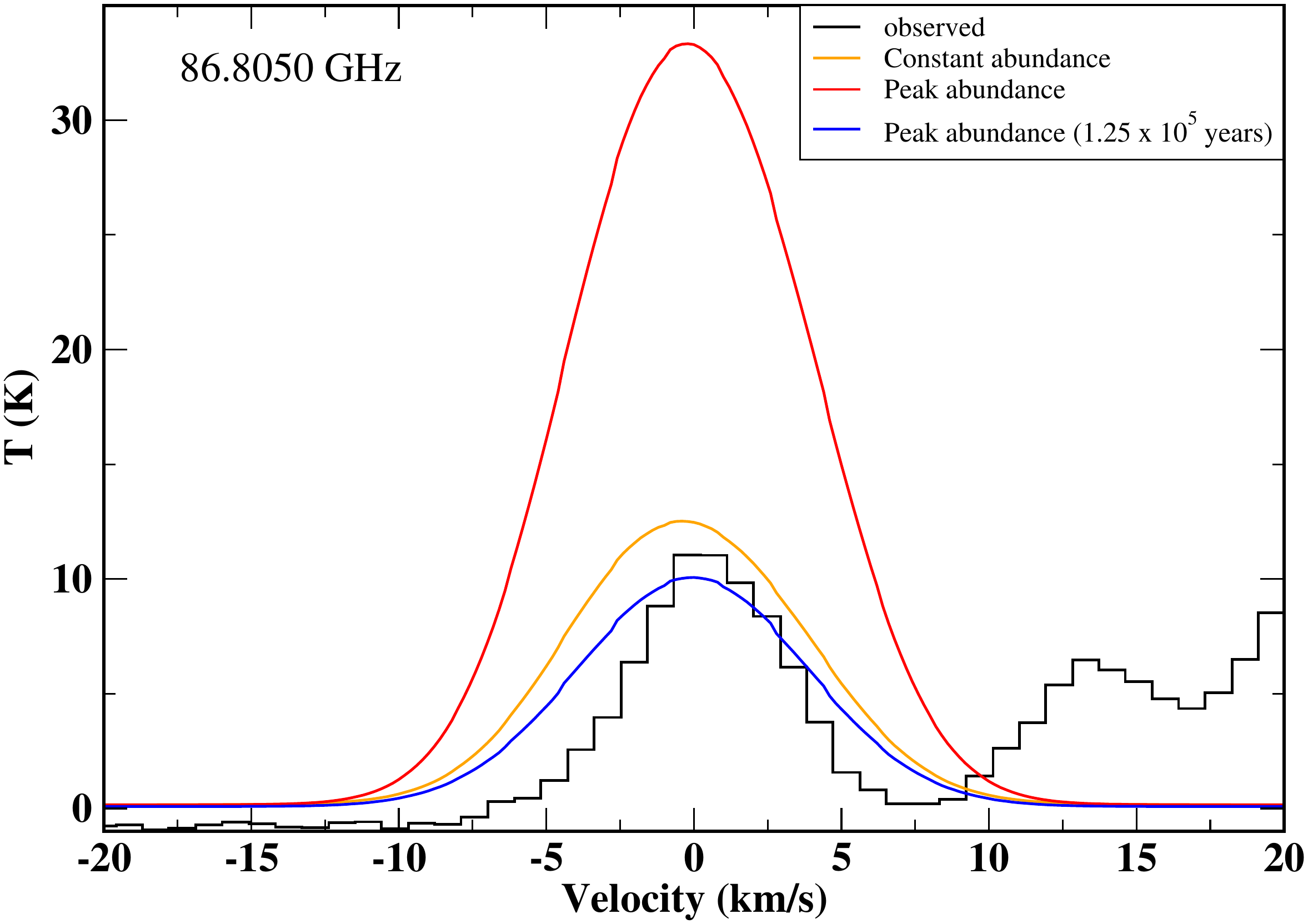}
\end{minipage}
\hskip 0.90cm
\begin{minipage}{0.30\textwidth}
\includegraphics[width=\textwidth]{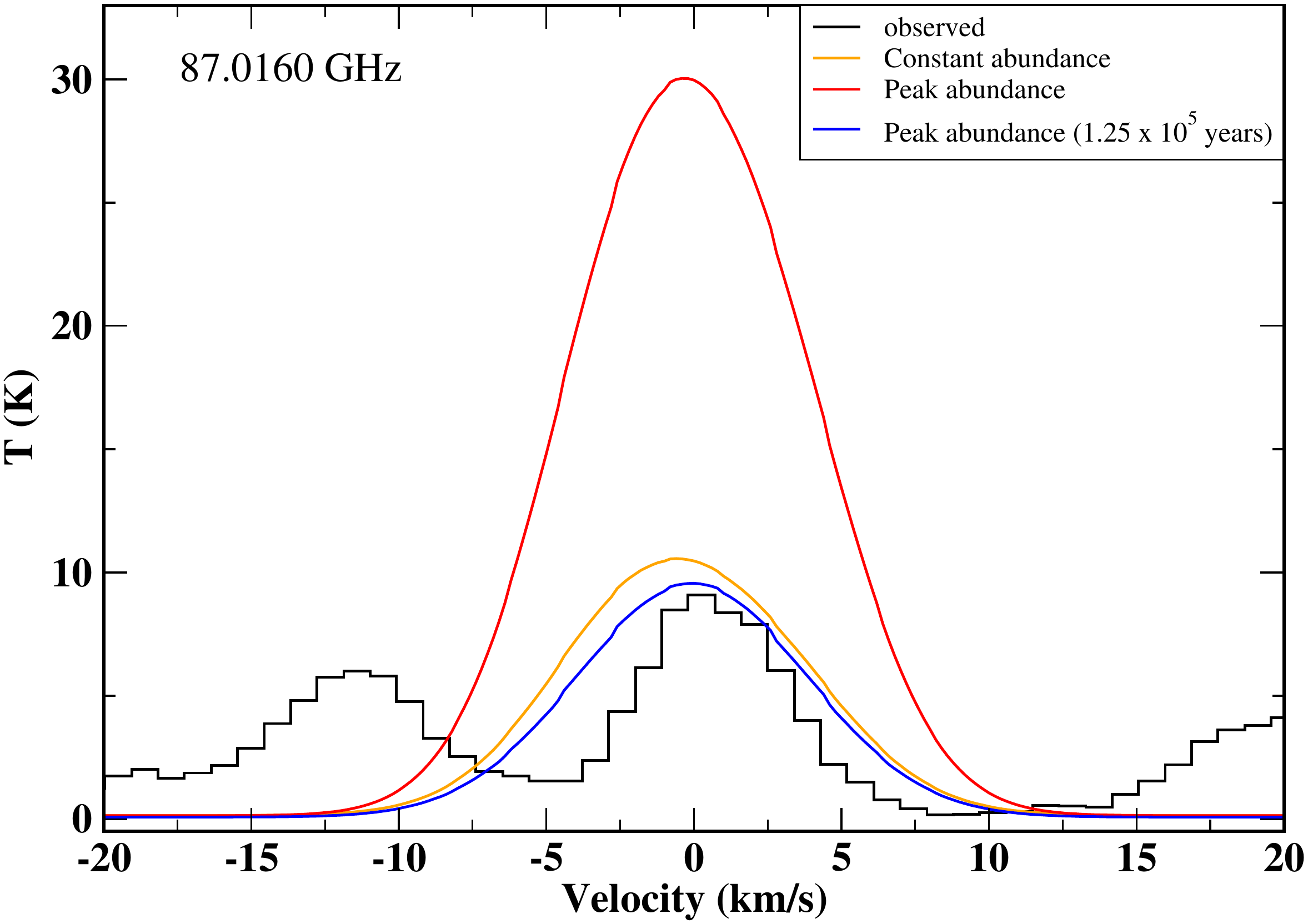}
\end{minipage}
\hskip -0.9 cm
\caption{A comparison between the observed (black line) CH$_3$NCO line profiles and the modeled line profiles is shown.  The modeled line profiles with the constant abundance are shown with the orange lines. The best fit is obtained when an FWHM of $\sim 7.5$ km/s, a constant abundance of $\sim 5 \times 10^{-9}$, and $\beta=1.4$ are used. The modeled line profiles with the peak values are shown with the red lines. It is noticed that our peak abundance profile over-predict the intensity of these transitions. An additional peak abundance profile (extracted at $1.25 \times 10^5$ years from our CMMC model) is used which shows a good fit (blue curve) with the observation.}
\label{fig:ch3nco-ratran}
\end{figure*}

\subsubsection{Complex organic molecules}
Recently, \cite{gora21} reported several transitions of some complex organic molecules in G31. 
Here, the observed emission features of $\rm{CH_3OH, CH_3SH}$, and CH$_3$NCO are modeled. 
\cite{ara08} observed the transition of $\rm{CH_3OH}$ in G31, and they used this molecule to understand the outflow present in this source. 
Figure \ref{fig:abundance} shows that the peak abundance of CH$_3$OH shows a jump at $\sim 10000$ AU and having a maximum value ($\sim 1.9 \times 10^{-6}$). 
The final abundance of methanol greatly differs from its peak abundance. In contrast to its peak value, it drastically decreased deep inside the cloud. 
\cite{gora21}
identified a transition of H$_2$CO and estimated its column density. In Figure \ref{fig:abundance}, the abundance variation of
H$_2$CO is shown. It also shows a jump profile like the methanol and has a peak abundance $1.2 \times 10^{-6}$. 
 This is the first exclusive modeling result in explaining the observed abundance of  CH$_3$SH and CH$_3$NCO in G31. Here, for the chemical evolution of CH$_3$NCO and CH$_3$SH, the pathways mentioned in \cite{gora20} and \cite{gor17a} are used. 
The peak abundance profile of CH$_3$SH shows a similar trend as the CH$_3$OH. It shows a jump at $10000$ AU from its minimum value and having
its maximum abundance of $\sim 1.4 \times 10^{-8}$. Like CH$_3$OH, the final abundance of CH$_3$SH drastically decreased in the inner region.
The peak and final abundance of CH$_3$NCO show similar value because of its formation during the simulation's late stages by the reaction between HNCO and CH$_3$ in the gas phase (with a rate $\sim 5 \times 10^{-11}$ cm$^3$ s$^{-1}$) cm$^3$ s$^{-1}$. 
Here, the activation barrier for the reaction between H and NH$_2$CHO  is used at its highest value ($3130$ K), as mentioned in
\cite{gora20}.
A peak abundance of $\sim 3.3 \times 10^{-8}$ for CH$_3$NCO is obtained at around $6000$ AU. 
A comparison between our simulated and observed (obtained from \cite{gora21} by the LTE fitting and rotation diagram and MCMC fitting carried out here) abundance is shown in Figure \ref{fig:comp}. It shows that the
obtained abundance of CH$_3$OH, H$_2$CO, CH$_3$SH, and CH$_3$NCO is consistent with the observation. In deriving the molecular abundance from the column density, a molecular hydrogen column density $\sim 1.53 \times 10^{25}$ cm$^{-2}$ \citep{gora21} is used.

 CH$_3$OH and CH$_3$SH are mainly formed on the interstellar grain surface \citep{das08a,das10,das11,gor17a}. These species can quickly transfer to the gas phase in the warmer region (temperature $>100$ K). Significant production of gas-phase CH$_3$NCO can be processed after the release of grain phase HNCO \citep{gora20}. So the gas-phase abundance of these three species depends mainly on the chemical process related to the grain. In our simulation, the entire cloud is divided into $23$ shells. For example, the outermost shell's temperature is allowed to evolve up to $23$ K. In the innermost shell, it is up to $1593$ K. So, in the case of the outer grids (beyond grid number $13$), it remains below $100$ K. For these shells, the peak value of these complex organic molecules appear during the end of the simulation. The peak value gradually appears in a shorter time for the inner grids (inside the $13^{th}$ grid). So, in brief, since the peak values which are noted in Table \ref{table:abundances} and plotted in Figure \ref{fig:abundance} is beyond the collapsing time scale, our time uncertainty in predicting the peak value is $\sim 1.5 \times 10^5$ years. There are many other uncertainties (physical and chemical parameters) that may influence these time scales. 

Initially, observed line profile of these COMs are modeled with the constant spatial abundances. As like the observation, all the transitions of these complex organic molecules are found in emission (see Figure \ref{fig:ch3oh-ratran}, 
\ref{fig:ch3sh-ratran}, and \ref{fig:ch3nco-ratran}). 
The collisional data file for the E-CH$_3$OH and A-CH$_3$OH with H$_2$ is taken from the
LAMDA database. 
For the methanol transitions shown in Figure \ref{fig:ch3oh-ratran}, a best-fit is obtained when an A-CH$_3$OH abundance of $\sim 2 \times 10^{-6}$ and an FWHM of 8 km/s is used. It is consistent with the methanol (A-CH$_3$OH) abundance obtained from the MCMC fit ($\sim 1.4 \times 10^{-6}$) of these transitions shown in Table \ref{table:mcmc_lte}.
The A-CH$_3$OH transitions at $88.594787$ GHz and $88.939971$ GHz shows a good fit with the observations. However, the transitions shown in the first two panels of Figure \ref{fig:ch3oh-ratran} do not fit well. With the peak abundance profile, the intensities of all the transitions are under produced.
Table \ref{table:mcmc_lte} shows a $3.9$ times higher column density is needed for E-CH$_3$OH than A-CH$_3$OH. Here, following this, an abundance $\sim 7.8 \times 10^{-6}$ is used for E-CH$_3$OH. 
As with A-CH$_3$OH, an FWHM of 8 km/s is used.
However, with the 1D RATRAN model, we cannot fit all the three
transitions (the last three panels of Figure \ref{fig:ch3oh-ratran}) of E-CH$_3$OH. As like A-CH$_3$OH, here also, with the peak abundance profile, the intensities of all the three transitions are under produced.
These three transitions (101.126857 GHz, 101.293415 GHz, and 101.469805 GHz) are slightly off-centered. Moreover, at 101.1269 GHz, 101.2927 GHz, and 101.2935 GHz, HOCH$_2$CN, CH$_3$OCN, and s-propanal may appear. Since \cite{gora21} did not find any other transitions of these species in the other part of their spectrum, they excluded these from their analysis.

 Only three out of the seven observed transitions of CH$_3$SH are shown in Figure \ref{fig:ch3sh-ratran} and five out of the six observed transitions of CH$_3$NCO shown in Figure \ref{fig:ch3nco-ratran}. It is because of the adopted collisional data files for these two species. The collisional data file is essential for the non-LTE calculation. 
The collisional data files were available for the modeled molecules for H$^{13}$CO$^+$, HCN, NH$_3$, SiO, and E-CH$_3$OH, A-CH$_3$OH. 
Since no such collisional data were available for CH$_3$NCO and CH$_3$SH, the collisional data of HNCO and $\rm{A-CH_3OH}$ with H$_2$ are considered in their place.
These files are prepared in the prescribed format of the RATRAN modeling. We must admit that these are very crude assumptions, but these are used here to check the educated guess of the line profiles. Since collisional rates for only a few levels were available, our file does not contain all the transitions of the CH$_3$SH and CH$_3$NCO. 

Figure \ref{fig:ch3sh-ratran} shows a comparison between the observed and modeled line profiles of CH$_3$SH. For the best-fitted model, a constant abundance of $\sim 1.9 \times 10^{-8}$ and an FWHM $9.45$ km/s for CH$_3$SH is used.
 Figure \ref{fig:ch3sh-ratran} shows that constant abundance does not fit well, and peak abundance profile underproduces the intensity. 
This is because three transitions of CH$_3$SH are blended among each other (see Figure \ref{fig:ch3sh-mcmc}).
Similarly, Figure \ref{fig:ch3nco-ratran} shows a comparison between the observed and modeled line profiles. For the best-fitted model, a constant abundance of $\sim 5.0 \times 10^{-9}$ and an FWHM $7.5$ km/s for CH$_3$NCO is used.
Further, the radial distribution of the abundance profiles (from Figure \ref{fig:abundance}) is used in our radiative transfer model to determine the simulated line
profiles of CH$_3$OH, CH$_3$SH, and CH$_3$NCO. The line profiles obtained with the peak values are shown in red. The intensity of the transitions of CH$_3$NCO is overestimated with the peak abundance profile. The peak CH$_3$NCO abundance appears during the latter stages of our simulation. Several studies with the RATRAN model are performed by considering various abundance profiles extracted at different times. The best fit is obtained when the peak abundance profile of CH$_3$NCO is extracted at $1.25 \times 10^5$ years (corresponding to a warmup time of $2.5 \times 10^4$ years after the isothermal collapse phase). 
 
\section{Conclusions}

\label{sec:conclusion}
 This paper has carried out an extensive chemical and radiative transfer model to explain several observed line profiles in G31. The physical conditions obtained in this region were coupled with our CMMC model to mimic a realistic situation. This coupled model has yielded a spatial distribution of abundances of some key interstellar species in G31. Our coupled model was successful in explaining the obtained quantities. 
  Various physical properties such as infall velocity, mass infall rate, FWHM, etc., are extracted from our radiative transfer models.
 A summary of our modeled and observed parameters is shown in Table \ref{table:RATRAN-best-fit}. {The major highlights of this work is reported below:\\
 
 $\bullet$
 With the LTE model, \cite{gora21} obtained an abundance $4.72 \times 10^{-8}$ and $2.7 \times 10^{-8}$, respectively for CH$_3$NCO and CH$_3$SH. With the rotational diagram analysis, they obtained $\sim 1.04 \times 10^{-9}$ and $1.86 \times 10^{-9}$, respectively for CH$_3$NCO and CH$_3$SH.
 The MCMC model reported here yields $7.84 \times 10^{-8}$ and $9.15 \times 10^{-9}$ abundance, respectively, for these two species.   Our CMMC model has reported a maximum peak abundance of CH$_3$NCO and CH$_3$SH $\sim 3.3 \times 10^{-8}$ and $1.4 \times 10^{-8}$, respectively.\\

 $\bullet$
 \cite{gora21} reported an infall velocity of $2.3$ km/s from the H$^{13}$CO$^+$ observation. Here, from the two slab model of H$^{13}$CO$^+$, an infall velocity of $2.5$ km/s is reported. Using this infall velocity, a mass infall rate of $1.3 \times 10^{-3}$ M$_\odot yr^{-1}$ is estimated by considering a distance of 3.7 kpc of G31. With the 1D RATRAN model, the best fit is obtained when an infall velocity of $4.9$ km/s at 1000 AU is used, which is consistent with what was obtained by \cite{oso09}.\\

 $\bullet$ It is noticed that dust emissivity plays a crucial role in defining the line profile. All the observed line profiles can be explained by the power-law emissivity relation of dust emission.   $\beta=1$ is needed to reproduce the observed line profiles of H$^{13}$CO$^+$, SiO, and CH$_3$CN, whereas $\beta=1.4$ is required in generating the observed line profiles of other species. In Table \ref{table:RATRAN-best-fit}, abundance, FWHM, and $\beta$ used to explain the observed line profile are summarized. \\

 $\bullet$ 1D RATRAN code is employed to explain the observed line profiles of H$^{13}$CO$^+$, HCN, SiO, NH$_3$, CH$_3$CN, CH$_3$OH, CH$_3$SH, and CH$_3$NCO.  An infalling envelope can successfully explain all the line profiles. Only for the SiO, it is required to include an additional outflow component in our model. \\
 
$\bullet$ The line profile of H$^{13}$CO$^+$ and CH$_3$CN is modeled for different resolutions. It is noticed that with the higher angular resolution, an inverse P-Cygni nature of CH$_3$CN and with the lower angular resolution, an inverse P-Cygni nature of the H$^{13}$CO$^+$ could be expected. These findings are consistent with the observation of \cite{gora21} and \cite{belt18}.}

\section{Acknowledgments}
This paper makes use of the  following ALMA  data:ADS/JAO.ALMA $2015.1.01193.S$. ALMA is a partner-ship of ESO (representing its member states), NSF (USA)and NINS (Japan), together with NRC (Canada), MOST and  ASIAA  (Taiwan),  and  KASI  (Republic  of  Korea), in  cooperation  with  the  Republic  of  Chile. The Joint ALMA  Observatory  is  operated  by  ESO, AUI/NRAO and NAOJ.

BB gratefully acknowledges the DST, India's Government, for providing financial assistance
through the DST-INSPIRE Fellowship [IF170046] scheme. AD acknowledges the ISRO respond project (Grant No. ISRO/RES/2/402/16-17)
for partial financial support. PG acknowledges CSIR extended SRF fellowship (Grant
No. 09/904 (0013) 2018 EMR-I) and Chalmers  Cosmic  Origins postdoctoral fellowship. SKM acknowledges CSIR  fellowship (Grant No. 09/904 (0014) 2018 EMR-I). This research was possible in part due to a Grant-In-Aid
from the Higher Education Department of the Government of West Bengal. We are thankful to the anonymous reviewers for providing insightful comments to improve this paper.


\bibliographystyle{model5-names}
\biboptions{authoryear}
\bibliography{ASR-G31}

\appendix
\setcounter{table}{0}
\renewcommand{\thetable}{A\arabic{table}}
\section{Abundances}
\clearpage

\label{sec:abundance-rad}
\begin{table*}
\tiny \setlength{\tabcolsep}{4pt}
\caption{Final and peak abundances at different radius obtained from our chemical model. The peak values are considered beyond the collapsing time scale. So our time uncertainty in the peak abundances is $1.5 \times 10^5$ years.} \label{table:abundances}
\begin{center}
\begin{adjustbox}{angle=90}
\begin{tabular}{|c|c|c|c|c|c|c|c|c|c|c|c|c|c|c|c|c|c|c|}
\hline
\hline
\multicolumn{3}{|c|}{HCO$^+$}&\multicolumn{2}{|c|}{HCN}&\multicolumn{2}{|c|}{SiO}&\multicolumn{2}{|c|}{NH$_3$}&\multicolumn{2}{|c|}{CH$_3$OH}&\multicolumn{2}{|c|}{CH$_3$SH}&\multicolumn{2}{|c|}{CH$_3$NCO}&\multicolumn{2}{|c|}{CH$_3$CN}&\multicolumn{2}{|c|}{H$_2$CO}\\
\hline
\hline
Radius (AU)&Peak value &Final value&Peak value&Final value&Peak value&Final value&Peak value&Final value&Peak value&Final value&Peak value&Final value&Peak value&Final value&Peak value&Final value&Peak value&Final value\\
\hline
\hline
1.56$\times10^{2}$&1.08$\times10^{-5}$ &4.76$\times10^{-16}$ &2.76$\times10^{-10}$&8.58$\times10^{-16}$&1.46$\times10^{-15}$ &4.66$\times10^{-17}$ &2.05$\times10^{-10}$ &4.91$\times10^{-13}$ &5.93$\times10^{-11}$&5.93 $\times10^{-11}$&2.41$\times10^{-16}$&8.45$\times10^{-23}$ &1.47$\times10^{-11}$ &1.37$\times10^{-11}$ &6.72$\times10^{-15}$&5.77$\times10^{-18}$&2.71$\times10^{-9}$&1.88$\times10^{-17}$\\
2.12$\times10^{2}$&4.62$\times10^{-6}$&3.56$\times10^{-15}$&2.39$\times10^{-9}$&1.88$\times10^{-15}$&2.21$\times10^{-14}$ &2.21$\times10^{-14}$ &3.04$\times10^{-10}$ &4.83$\times10^{-11}$ &1.10$\times10^{-10}$&1.38$\times10^{-12}$ &1.35$\times10^{-15}$&3.10$\times10^{-25}$ &2.06$\times10^{-10}$ &1.93$\times10^{-10}$ &1.73$\times10^{-13}$&2.29$\times10^{-17}$&1.52$\times10^{-8}$&1.36$\times10^{-14}$\\
2.87$\times10^{2}$&1.74$\times10^{-6}$&1.58$\times10^{-14}$&1.15$\times10^{-7}$&4.54$\times10^{-16}$&2.33$\times10^{-10}$ &1.08$\times10^{-14}$ &2.73$\times10^{-7}$ &9.50$\times10^{-12}$ &4.44$\times10^{-8}$& 5.47$\times10^{-15}$&1.06$\times10^{-14}$&4.14$\times10^{-39}$ &9.35$\times10^{-10}$ &8.83$\times10^{-10}$ &1.79$\times10^{-9}$&7.76$\times10^{-19}$&1.91$\times10^{-7}$&1.98$\times10^{-17}$\\
3.88$\times10^{2}$&6.24$\times10^{-7}$&2.07$\times10^{-13}$&9.96$\times10^{-8}$ & 3.94$\times10^{-14}$&2.92$\times10^{-10}$ &2.30$\times10^{-10}$ &5.89$\times10^{-7}$ &4.57 $\times10^{-8}$&7.18$\times10^{-8}$&3.37$\times10^{-11}$ &2.03$\times10^{-10}$&9.59$\times10^{-34}$ &2.81$\times10^{-9}$ &2.62$\times10^{-9}$ &2.44$\times10^{-8}$&6.73$\times10^{-16}$&2.57$\times10^{-7}$&3.67$\times10^{-16}$\\
5.24$\times10^{2}$&2.37$\times10^{-7}$&5.83$\times10^{-13}$&9.47$\times10^{-8}$ &5.67$\times10^{-14}$&3.84$\times10^{-10}$ &3.31$\times10^{-10}$ &5.98$\times10^{-7}$ &3.30$\times10^{-7}$ &1.37$\times10^{-7}$&1.85$\times10^{-11}$ &4.89$\times10^{-10}$&1.18$\times10^{-31}$ &3.99$\times10^{-9}$ &3.69$\times10^{-9}$ &2.64$\times10^{-8}$&2.70$\times10^{-15}$&3.62$\times10^{-7}$&5.88$\times10^{-15}$\\
7.09$\times10^{2}$&9.71$\times10^{-8}$&1.55$\times10^{-12}$&1.00$\times10^{-7}$ &9.45$\times10^{-14}$&5.10$\times10^{-10}$ &4.67$\times10^{-10}$ &4.90$\times10^{-7}$ &3.68$\times10^{-7}$ &2.52$\times10^{-7}$&9.69$\times10^{-14}$ &1.47$\times10^{-9}$&3.58 $\times10^{-20}$&6.41$\times10^{-9}$ &5.71$\times10^{-9}$ &1.67$\times10^{-8}$&3.34$\times10^{-15}$&4.39$\times10^{-7}$&1.31$\times10^{-13}$\\
9.58$\times10^{2}$&2.81$\times10^{-8}$&3.79$\times10^{-12}$&9.85$\times10^{-8}$ &1.52$\times10^{-13}$ &6.07$\times10^{-10}$ &5.97$\times10^{-10}$ &5.02$\times10^{-7}$ &3.51$\times10^{-7}$ &5.21$\times10^{-7}$&4.90$\times10^{-15}$ &3.28$\times10^{-9}$&2.82$\times10^{-16}$ &8.92$\times10^{-9}$ &7.96$\times10^{-9}$ &1.03$\times10^{-8}$&1.33$\times10^{-15}$&5.67$\times10^{-7}$&1.75$\times10^{-12}$\\
1.29$\times10^{3}$&1.15$\times10^{-8}$&8.98$\times10^{-12}$&1.14$\times10^{-7}$ &2.26$\times10^{-13}$ &5.84$\times10^{-10}$ &5.65$\times10^{-10}$ &5.86$\times10^{-7}$ &5.37$\times10^{-7}$ &7.96$\times10^{-7}$&4.25$\times10^{-14}$ &4.19$\times10^{-9}$&3.73$\times10^{-13}$ &1.05$\times10^{-8}$ &9.72$\times10^{-9}$ &1.16$\times10^{-8}$&1.50$\times10^{-15}$&7.36$\times10^{-7}$&3.87$\times10^{-11}$\\
1.75$\times10^{3}$&7.29$\times10^{-9}$&1.05$\times10^{-11}$&1.44$\times10^{-7}$ &5.14$\times10^{-13}$ &1.08$\times10^{-9}$ &1.08$\times10^{-9}$ &6.41$\times10^{-7}$ &4.26$\times10^{-7}$ &1.09$\times10^{-6}$&1.07$\times10^{-10}$ &4.65$\times10^{-9}$&1.51$\times10^{-11}$ &1.28$\times10^{-8}$ &1.12$\times10^{-8}$ &1.56$\times10^{-8}$&7.76$\times10^{-16}$&8.16$\times10^{-7}$&1.93$\times10^{-9}$\\
2.37$\times10^{3}$&5.45$\times10^{-9}$& 7.33$\times10^{-11}$&1.85$\times10^{-7}$ &1.23$\times10^{-12}$ &1.14$\times10^{-9}$ &1.14$\times10^{-9}$ &5.68$\times10^{-7}$ &2.85$\times10^{-7}$ &1.42$\times10^{-6}$&3.20$\times10^{-10}$ &4.90$\times10^{-9}$&4.42$\times10^{-11}$ &1.72$\times10^{-8}$ &1.51$\times10^{-8}$ &2.07$\times10^{-8}$&3.64$\times10^{-15}$&7.26$\times10^{-7}$&7.92$\times10^{-8}$\\
3.20$\times10^{3}$&4.88$\times10^{-9}$&7.55$\times10^{-11}$&2.41$\times10^{-7}$ &3.11$\times10^{-12}$ &1.19$\times10^{-9}$ &6.92$\times10^{-10}$ &5.21$\times10^{-7}$ &4.22$\times10^{-7}$ &1.71$\times10^{-6}$&1.29 $\times10^{-9}$&4.54$\times10^{-9}$&3.78$\times10^{-10}$ &2.02$\times10^{-8}$ &2.00$\times10^{-8}$ &2.65$\times10^{-8}$&1.06$\times10^{-14}$&8.33$\times10^{-7}$&1.25$\times10^{-7}$\\
4.32$\times10^{3}$&4.75$\times10^{-9}$& 5.58$\times10^{-11}$&3.11$\times10^{-7}$&5.32$\times10^{-12}$ &1.51$\times10^{-9}$ &2.71$\times10^{-10}$ &7.53$\times10^{-7}$ &6.77$\times10^{-7}$ &1.82$\times10^{-6}$&3.39$\times10^{-9}$ &4.84$\times10^{-9}$&1.14$\times10^{-9}$ &2.24$\times10^{-8}$ &2.24$\times10^{-8}$ &3.25$\times10^{-8}$&1.55$\times10^{-14}$&9.73$\times10^{-7}$&1.04$\times10^{-7}$\\
5.84$\times10^{3}$&4.43$\times10^{-9}$& 7.64$\times10^{-11}$&4.02$\times10^{-7}$ &1.27$\times10^{-8}$& 2.01$\times10^{-9}$&6.05$\times10^{-10}$ &3.18$\times10^{-7}$ &2.87$\times10^{-7}$ &1.88$\times10^{-6}$&2.13 $\times10^{-8}$&5.49$\times10^{-9}$&1.09$\times10^{-12}$ &3.27$\times10^{-8}$ &3.27$\times10^{-8}$ &3.81$\times10^{-8}$&3.16$\times10^{-10}$&1.15$\times10^{-6}$&2.05$\times10^{-8}$\\
7.90$\times10^{3}$&4.53$\times10^{-9}$& 1.14$\times10^{-9}$& 5.35$\times10^{-7}$&8.35$\times10^{-8}$&4.39$\times10^{-9}$ &4.39$\times10^{-9}$ &3.40$\times10^{-8}$ &1.67$\times10^{-8}$ &9.92$\times10^{-7}$&4.16$\times10^{-9}$ &6.32$\times10^{-9}$&1.00$\times10^{-12}$ &1.36$\times10^{-9}$ &1.10$\times10^{-12}$ &3.27$\times10^{-8}$&1.07$\times10^{-9}$&1.18$\times10^{-6}$&1.32$\times10^{-8}$\\
1.07$\times10^{4}$& 7.56$\times10^{-9}$& 7.50$\times10^{-9}$& 4.66$\times10^{-7}$&7.36$\times10^{-8}$& 3.17$\times10^{-9}$&3.14$\times10^{-9}$ &4.96$\times10^{-8}$ &1.70$\times10^{-8}$ &1.78$\times10^{-7}$&9.47$\times10^{-10}$ &1.39$\times10^{-8}$&7.15$\times10^{-9}$ &6.16$\times10^{-10}$ &1.99$\times10^{-12}$ &7.13$\times10^{-9}$&2.96$\times10^{-10}$&4.61$\times10^{-7}$&7.21$\times10^{-9}$\\
1.44$\times10^{4}$& 1.87$\times10^{-8}$& 1.87$\times10^{-8}$&7.82$\times10^{-8}$ &4.11$\times10^{-8}$ &2.18$\times10^{-9}$ &2.10$\times10^{-9}$ &6.89$\times10^{-8}$ &2.44$\times10^{-8}$ &8.21$\times10^{-9}$&3.13 $\times10^{-9}$&6.02$\times10^{-10}$&6.02$\times10^{-10}$ &1.57$\times10^{-10}$ &4.44$\times10^{-12}$ &1.67$\times10^{-10}$&1.04$\times10^{-10}$&2.41$\times10^{-8}$&1.13$\times10^{-8}$\\
1.95$\times10^{4}$&2.63$\times10^{-8}$&2.63$\times10^{-8}$& 2.45$\times10^{-8}$&1.85$\times10^{-8}$& 1.31$\times10^{-9}$&1.47$\times10^{-11}$ &9.73$\times10^{-8}$ &4.24$\times10^{-8}$ &3.91$\times10^{-10}$&3.19 $\times10^{-10}$&3.18$\times10^{-10}$&8.67$\times10^{-11}$ &5.09$\times10^{-11}$ &1.25$\times10^{-12}$ &7.55$\times10^{-11}$&4.26$\times10^{-11}$&1.04$\times10^{-8}$&2.39$\times10^{-9}$\\
2.64$\times10^{4}$&2.01$\times10^{-8}$&2.01$\times10^{-8}$&2.55$\times10^{-8}$ &2.07$\times10^{-8}$&2.88$\times10^{-9}$ &2.40$\times10^{-11}$ &1.48$\times10^{-7}$ &8.30$\times10^{-8}$ &3.94$\times10^{-10}$&3.08$\times10^{-10}$ &4.15$\times10^{-10}$&3.85 $\times10^{-11}$& 3.19$\times10^{-11}$&2.63 $\times10^{-12}$&1.72$\times10^{-10}$&1.23$\times10^{-10}$&1.28$\times10^{-8}$&2.12$\times10^{-9}$\\
3.56$\times10^{4}$&1.33$\times10^{-8}$& 1.15$\times10^{-8}$&3.78$\times10^{-8}$ &3.68$\times10^{-8}$ &4.86$\times10^{-9}$ &1.48$\times10^{-10}$ &2.12$\times10^{-7}$ &1.63$\times10^{-7}$ &5.71$\times10^{-10}$&4.12 $\times10^{-10}$&4.53$\times10^{-10}$& 9.38$\times10^{-11}$& 3.40$\times10^{-11}$&9.18$\times10^{-12}$ &3.67$\times10^{-10}$&3.67$\times10^{-10}$&1.85$\times10^{-8}$&5.46$\times10^{-9}$\\
4.82$\times10^{4}$&1.05$\times10^{-8}$&9.78$\times10^{-9}$&4.83$\times10^{-8}$ &4.83$\times10^{-8}$ &6.94$\times10^{-9}$ &8.37$\times10^{-10}$ &2.30$\times10^{-7}$ &2.30$\times10^{-7}$ &7.83$\times10^{-10}$&6.10$\times10^{-10}$ &4.43$\times10^{-10}$& 4.43$\times10^{-10}$& 9.02$\times10^{-11}$&9.02$\times10^{-11}$ &6.77$\times10^{-10}$&5.36$\times10^{-10}$&2.79$\times10^{-8}$&1.29$\times10^{-8}$\\
6.51$\times10^{4}$&1.02$\times10^{-8}$&9.81$\times10^{-9}$&2.02$\times10^{-8}$ &2.02$\times10^{-8}$ &7.62$\times10^{-9}$ &1.88$\times10^{-9}$ &1.97$\times10^{-7}$ &1.97$\times10^{-7}$ &7.97$\times10^{-10}$&6.02$\times10^{-10}$ &1.94$\times10^{-10}$&1.88$\times10^{-10}$ &2.79$\times10^{-11}$ &2.79$\times10^{-11}$ &1.05$\times10^{-9}$&1.99$\times10^{-10}$&3.84$\times10^{-8}$&7.42$\times10^{-9}$\\
8.80$\times10^{4}$& 1.02$\times10^{-8}$& 1.02$\times10^{-8}$&3.72$\times10^{-8}$ &2.26$\times10^{-8}$ &8.49$\times10^{-9}$ &4.12$\times10^{-9}$ &1.99$\times10^{-7}$ &1.99$\times10^{-7}$ &9.53$\times10^{-10}$&9.53$\times10^{-10}$ &1.17$\times10^{-10}$&1.17$\times10^{-10}$ &3.02$\times10^{-11}$ &2.84$\times10^{-11}$ &2.07$\times10^{-9}$&3.51$\times10^{-10}$&5.99$\times10^{-8}$&1.52$\times10^{-8}$\\
1.19$\times10^{5}$&1.01$\times10^{-8}$&1.01$\times10^{-8}$&4.57$\times10^{-8}$ &1.93$\times10^{-8}$ &9.22$\times10^{-9}$ &6.32$\times10^{-9}$ &1.69$\times10^{-7}$ &1.69$\times10^{-7}$ &1.73$\times10^{-9}$&1.67$\times10^{-9}$ &2.83$\times10^{-11}$&2.83$\times10^{-11}$ &2.19$\times10^{-11}$ &1.38$\times10^{-11}$ &2.83$\times10^{-9}$&3.97$\times10^{-10}$&7.35$\times10^{-8}$&1.63$\times10^{-8}$\\
\hline
\hline
\end{tabular}
\end{adjustbox}
\end{center}
\end{table*}

 \begin{table*}
\centering
{\scriptsize
 \caption{ Abundance, linewidth, and $\beta$ obtained from our best-fitted RATRAN model are noted. 
For the comparison, we have reported the abundances obtained by the other methods (i.e., MCMC method discussed in section \ref{sec:MCMC}, and LTE and rotational method carried out by \cite{gora21}). Moreover, the abundances obtained from our chemical model are also noted (peak abundance noted in Table \ref{table:abundances}). \label{table:RATRAN-best-fit}.}
\begin{tabular}{|l|l|l|l|l|l|l|l|}
  \hline
  \hline
  Species & \multicolumn{5}{|c|}{Abundance}& Line width &$\beta$\\
  &RATRAN&LTE$^g$&Rotational diagram$^g$&MCMC$^h$&Chemical model&(km.s$^{-1}$)&\\
  \hline
  \hline
 H$^{13}$CO$^+$& 7.08$\times$10$^{-11}$&-&-&-&1.16$\times$10$^{-10}$-1.66$\times$10$^{-7}$&1.42&1.0\\
 HCN& 7.6$\times$10$^{-8}$&-&-&-&2.76$\times$10$^{-10}$-5.35$\times$10$^{-7}$&10.00&1.4\\
 SiO&9.5$\times$10$^{-10}$ &9.54$\times$10$^{-11}$&-&-&1.46$\times$10$^{-15}$-9.22$\times$10$^{-9}$&4.67&1.0\\
 A-CH$_3$OH&2.0$\times$10$^{-6}$&1.2$\times$10$^{-6}$&1.92$\times$10$^{-6}$&1.37$\times$10$^{-6}$&5.93$\times$10$^{-11}$-1.88$\times$10$^{-6}$&8.00&1.4\\
 E-CH$_3$OH&7.8$\times$10$^{-6}$&1.2$\times$10$^{-6}$&1.92$\times$10$^{-6}$&5.36$\times$10$^{-6}$&5.93$\times$10$^{-11}$-1.88$\times$10$^{-6}$&7.67&1.4\\
 CH$_3$SH&1.9$\times$10$^{-8}$&2.7$\times$10$^{-8}$&1.86$\times$10$^{-9}$&9.15$\times$10$^{-9}$&2.41$\times$10$^{-16}$-1.39$\times$10$^{-8}$&9.45&1.4\\
 CH$_3$NCO&5.0$\times$10$^{-9}$&4.72$\times$10$^{-8}$&1.04$\times$10$^{-9}$&7.84$\times$10$^{-8}$&9.02$\times$10$^{-11}$-3.27$\times$10$^{-8}$&7.49&1.4\\
 NH$_3$ (100M)&2.0$\times$10$^{-9}$-1.0$\times10^{-7}$&-&-&-&3.04$\times$10$^{-10}$-7.53$\times$10$^{-7}$&4.90-8.33&1.4\\
 NH$_3$ (VLA)&1.0$\times$10$^{-7}$-1.6$\times$10$^{-7}$&-&-&-&3.04$\times$10$^{-10}$-7.53$\times$10$^{-7}$&4.90-8.33&1.4\\
 CH$_3$CN&6.0$\times$10$^{-8}$&&-&-&6.72$\times$10$^{-15}$-3.81$\times$10$^{-8}$&2.5&1.0\\
  \hline
  \hline
 \end{tabular}}\\
 {\noindent $^g$ \cite{gora21}\\$^h$ Using N$_{H2}$ = 1.53$\times$10$^{25}$ from \cite{gora21}}
 \end{table*}
\end{document}